\documentclass[pdftex,twocolumn,epjc3]{svjour3}          

\RequirePackage{amsmath}
\RequirePackage{multirow}
\RequirePackage{subfigure}
\RequirePackage{widetext}

\allowdisplaybreaks[3]

\RequirePackage[T1]{fontenc}

\smartqed  

\RequirePackage{graphicx}
\RequirePackage{mathptmx}      
\RequirePackage{flushend}
\RequirePackage[numbers,sort&compress]{natbib}
\RequirePackage[colorlinks,citecolor=blue,urlcolor=blue,linkcolor=blue]{hyperref}

\journalname{Eur. Phys. J. C}

\begin{document}

\title{Decay properties of $P$-wave bottom baryons within light-cone sum rules}

\author{Hui-Min Yang\thanksref{addr1} \and Hua-Xing Chen\thanksref{addr1,addr2} \and Er-Liang Cui\thanksref{addr3} \and Atsushi Hosaka\thanksref{addr4,addr5} \and Qiang Mao\thanksref{addr6}
}

\institute{
School of Physics, Beihang University, Beijing 100191, China\label{addr1}
\and
School of Physics, Southeast University, Nanjing 210094, China\label{addr2}
\and
College of Science, Northwest A{\rm \&}F University, Yangling 712100, China\label{addr3}
\and
Research Center for Nuclear Physics (RCNP), Osaka University, Ibaraki 567-0047, Japan\label{addr4}
\and
Advanced Science Research Center, Japan Atomic Energy Agency (JAEA), Tokai 319-1195, Japan\label{addr5}
\and
Department of Electrical and Electronic Engineering, Suzhou University, Suzhou 234000, China\label{addr6}
}

\date{Received: date / Accepted: date}

\maketitle

\begin{abstract}
We use the method of light-cone sum rules to study decay properties of $P$-wave bottom baryons belonging to the $SU(3)$ flavor $\mathbf{6}_F$ representation. In Ref.~\cite{Cui:2019dzj} we have studied their mass spectrum and pionic decays, and found that the $\Sigma_{b}(6097)$ and $\Xi_{b}(6227)$ can be well interpreted as $P$-wave bottom baryons of $J^P = 3/2^-$. In this paper we further study their decays into ground-state bottom baryons and vector mesons. We propose to search for a new state $\Xi_b({5/2}^-)$, that is the $J^P = 5/2^-$ partner state of the $\Xi_{b}(6227)$, in the $\Xi_b({5/2}^-) \to \Xi_b^{*}\rho \to \Xi_b^{*}\pi\pi$ decay process. Its mass is $12 \pm 5$~MeV larger than that of the $\Xi_{b}(6227)$.
\end{abstract}

%
%
%
\section{Introduction}\label{sec:intro}
%

In the past years important progress has been made in the field of heavy baryons, and many heavy baryons were observed in various experiments~\cite{pdg,Yelton:2016fqw,Kato:2016hca,Aaij:2017nav,Aaij:2017vbw,Aaij:2018yqz,Aaij:2018tnn,Aaij:2019amv}. These heavy baryons are interesting in a theoretical point of view~\cite{Korner:1994nh,Manohar:2000dt,Klempt:2009pi}: the light degrees of freedom (light quarks and gluons) circle around the nearly static heavy quark, so that the whole system behaves as the QCD analogue of the hydrogen bounded by electromagnetic interaction.
To understand them, various phenomenological models have been applied, such as
the relativized potential quark model~\cite{Capstick:1986bm},
the relativistic quark model~\cite{Ebert:2007nw},
the constituent quark model~\cite{Ortega:2012cx},
the chiral quark model~\cite{Zhong:2007gp},
the heavy hadron chiral perturbation theory~\cite{Cheng:2015naa},
the hyperfine interaction~\cite{Copley:1979wj,Karliner:2008sv},
the Feynman-Hellmann theorem~\cite{Roncaglia:1995az},
the combined expansion in $1/m_Q$ and $1/N_c$~\cite{Jenkins:1996de},
the pion induced reactions~\cite{Kim:2014qha},
the variational approach~\cite{Roberts:2007ni},
the relativistic flux tube model~\cite{Chen:2014nyo},
the Faddeev approach~\cite{Garcilazo:2007eh},
the Regge trajectory~\cite{Guo:2008he},
the extended local hidden gauge approach~\cite{Liang:2014eba},
the unitarized dynamical model~\cite{GarciaRecio:2012db},
the unitarized chiral perturbation theory~\cite{Lu:2014ina},
and QCD sum rules~\cite{Chen:2016phw,Mao:2017wbz,Bagan:1991sg,Neubert:1991sp,Broadhurst:1991fc,Huang:1994zj,Dai:1996yw,Colangelo:1998ga,Groote:1996em,Zhu:2000py,Lee:2000tb,Huang:2000tn,Wang:2003zp,Duraes:2007te,Zhou:2014ytp,Zhou:2015ywa,Chen:2019bip}, etc.
We refer to reviews~\cite{Chen:2016spr,Albuquerque:2018jkn,Cheng:2015iom,Crede:2013sze,Bianco:2003vb} for their recent progress.

In Refs.~\cite{Chen:2015kpa,Mao:2015gya} we have systematically applied the method of QCD sum rules~\cite{Shifman:1978bx,Reinders:1984sr} to study $P$-wave heavy baryons within the heavy quark effective theory (HQET)~\cite{Grinstein:1990mj,Eichten:1989zv,Falk:1990yz}, where we systematically constructed all the $P$-wave heavy baryon interpolating fields, and applied them to study the mass spectrum of $P$-wave heavy baryons. Later in Ref.~\cite{Chen:2017sci} we further studied their decay properties using light-cone sum rules, including:
\begin{itemize}

\item $S$-wave decays of flavor $\mathbf{\bar 3}_F$ $P$-wave heavy baryons into ground-state heavy baryons and pseudoscalar mesons;

\item $S$-wave decays of flavor $\mathbf{6}_F$ $P$-wave heavy baryons into ground-state heavy baryons and pseudoscalar mesons;

\item $S$-wave decays of flavor $\mathbf{\bar 3}_F$ $P$-wave heavy baryons into ground-state heavy baryons and vector mesons.

\end{itemize}
Very quickly, one notices that in order to make a complete study of $P$-wave heavy baryons, we still need to study:
\begin{itemize}

\item $S$-wave decays of flavor $\mathbf{6}_F$ $P$-wave heavy baryons into ground-state heavy baryons and vector mesons.

\end{itemize}
Besides them, we also need to systematically study $D$-wave and radiative decay properties of $P$-wave heavy baryons.

In the present study we will study $S$-wave decays of flavor $\mathbf{6}_F$ $P$-wave heavy baryons into ground-state heavy baryons and vector mesons. We will further use the obtained results to investigate the $\Sigma_{b}(6097)$ and $\Xi_{b}(6227)$ recently observed by LHCb~\cite{Aaij:2018yqz,Aaij:2018tnn}.
Our previous sum rule study in Ref.~\cite{Cui:2019dzj} suggests that they can be well interpreted as $P$-wave bottom baryons of $J^P = 3/2^-$. This conclusion is supported by Refs.~\cite{Chen:2018orb,Chen:2018vuc,Yang:2018lzg,Wang:2018fjm,Aliev:2018vye,Aliev:2018lcs,Azizi:2015ksa}, and we refer to Refs.~\cite{Chua:2018lfa,Karliner:2018bms,Huang:2018bed,Yu:2018yxl,Jia:2019bkr,Liang:2019aag,Guo:2019kdc,Padmanath:2013bla,Padmanath:2017lng,Cheng:2006dk,Nagahiro:2016nsx} for more relevant discussions. The result of Ref.~\cite{Cui:2019dzj} also suggests that they belong to the bottom baryon doublet $[\mathbf{6}_F, 2, 1, \lambda]$, whose definition will be given below. This doublet contains six bottom baryons, $\Sigma_b({3\over2}^-/{5\over2}^-)$, $\Xi^\prime_b({3\over2}^-/{5\over2}^-)$, and $\Omega_b({3\over2}^-/{5\over2}^-)$. We predicted the mass and decay width of the $\Omega_b(3/2^-)$ state to be
\begin{eqnarray*}
M_{\Omega_b(3/2^-)} &=& 6.46 \pm 0.12 {~\rm GeV} \, ,
\\ \Gamma_{\Omega_b(3/2^-)} &=& 58{^{+65}_{-33}} {~\rm MeV} \, ,
\end{eqnarray*}
and masses of the three $J^P = 5/2^-$ states to be
\begin{eqnarray*}
&& M_{\Sigma_b(5/2^-)} = 6.11 \pm 0.12~{\rm GeV} \, ,
\\ && M_{\Sigma_b(5/2^-)} -  M_{\Sigma_b(3/2^-)} = 13 \pm 5~{\rm MeV} \, ,
\\ && M_{\Xi_b^{\prime}(5/2^-)} = 6.29 \pm 0.11~{\rm GeV} \, ,
\\ && M_{\Xi^\prime_b(5/2^-)} -  M_{\Xi^\prime_b(3/2^-)} = 12 \pm 5~{\rm MeV} \, ,
\\ && M_{\Omega_b(5/2^-)} = 6.47 \pm 0.12~{\rm GeV} \, ,
\\ && M_{\Omega_b(5/2^-)} -  M_{\Omega_b(3/2^-)} = 11 \pm 5~{\rm MeV} \, .
\end{eqnarray*}
The three $J^P = 5/2^-$ states are probably quite narrow, because their $S$-wave decays into ground-state bottom baryons and pseudoscalar mesons can not happen, and widths of the following $D$-wave decays are extracted to be zero in Ref.~\cite{Cui:2019dzj}:
\begin{eqnarray}
\nonumber &(w^\prime)& \Gamma_{\Sigma_b^{-}({5/2}^-) \rightarrow \Lambda_b^{0} \pi^-} = 0 \, ,
\\
\nonumber &(x^\prime)& \Gamma_{\Xi_b^{\prime-}({5/2}^-) \rightarrow \Xi_b^{0} \pi^-} = 0 \, ,
\\
\nonumber &(y^\prime)& \Gamma_{\Xi_b^{\prime-}({5/2}^-) \rightarrow \Lambda_b^{0} K^-} = 0 \, ,
\\
\nonumber &(z^\prime)& \Gamma_{\Omega_b^{-}({5/2}^-) \rightarrow \Xi_b^{0} K^-} = 0 \, .
\end{eqnarray}
To further study their decay properties, in this paper we will investigate their $S$-wave decays into ground-state bottom baryons together with vector mesons $\rho$ and $K^*$.

This paper is organized as follows. In Sec.~\ref{sec:sdecay} we study $S$-wave decays of flavor $\mathbf{6}_F$ $P$-wave bottom baryons into ground-state bottom baryons and vector mesons, separately in several subsections for the four bottom baryon multiplets, $[\mathbf{6}_F, 1, 0, \rho]$, $[\mathbf{6}_F, 0, 1, \lambda]$, $[\mathbf{6}_F, 1, 1, \lambda]$, and $[\mathbf{6}_F, 2, 1, \lambda]$. A short summary is given in Sec.~\ref{sec:summary}. Some relevant parameters and formulae are given in \ref{sec:sbottom} and \ref{sec:othersumrule}.

%
\section{Decay properties of $P$-wave bottom baryons}\label{sec:sdecay}
%

\begin{figure*}[hbt]
\begin{center}
\scalebox{0.6}{\includegraphics{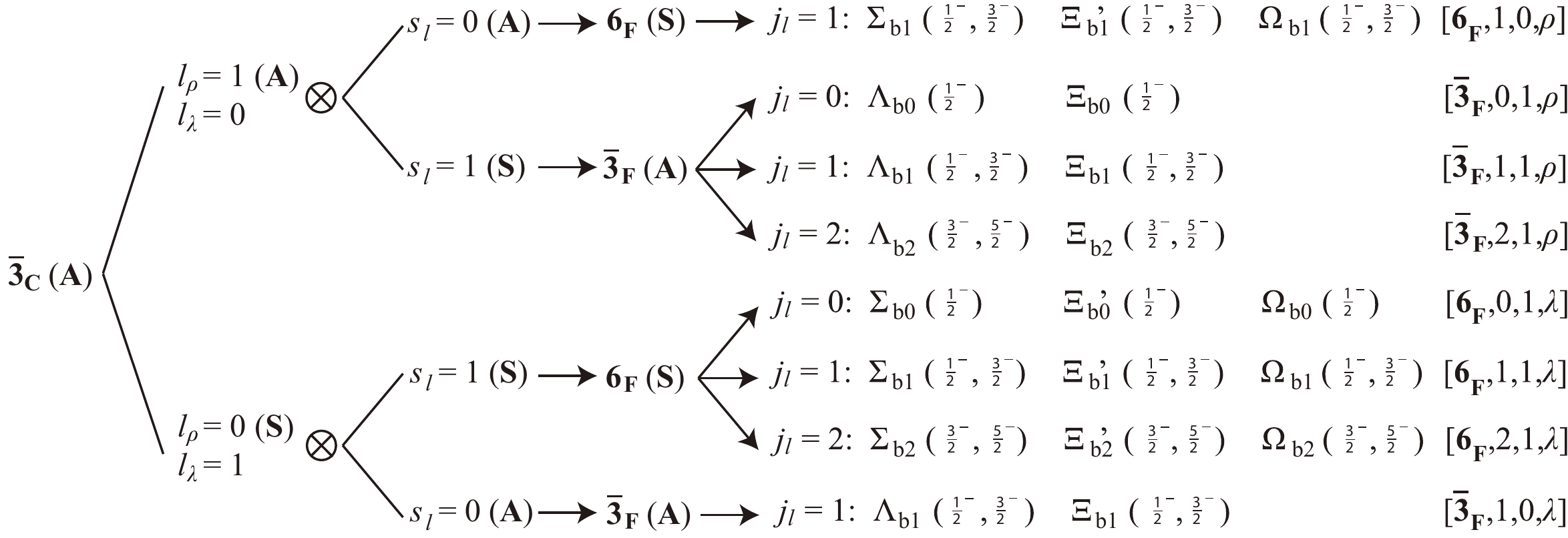}}
\caption{Categorization of $P$-wave bottom baryons. Taken from Ref.~\cite{Cui:2019dzj}.
\label{fig:pwave}}
\end{center}
\end{figure*}

At the beginning let us briefly introduce our notations. A $P$-wave bottom baryon ($bqq$) consists of one bottom quark ($b$) and two light quarks ($qq$). Its orbital excitation can be either between the two light quarks ($l_\rho = 1$) or between the bottom quark and the two-light-quark system ($l_\lambda = 1$), so there are $\rho$-type bottom baryons ($l_\rho = 1$ and $l_\lambda = 0$) and $\lambda$-type ones ($l_\rho = 0$ and $l_\lambda = 1$). Altogether its internal symmetries are as follows:
\begin{itemize}

\item The color structure of the two light quarks is antisymmetric ($\mathbf{\bar 3}_C$).

\item The $SU(3)$ flavor structure of the two light quarks is either antisymmetric ($\mathbf{\bar 3}_F$) or symmetric ($\mathbf{6}_F$).

\item The spin structure of the two light quarks is either antisymmetric ($s_l \equiv s_{qq} = 0$) or symmetric ($s_l = 1$).

\item The orbital structure of the two light quarks is either antisymmetric ($l_\rho = 1$) or symmetric ($l_\rho = 0$).

\item Due to the Pauli principle, the total symmetry of the two light quarks is antisymmetric.

\end{itemize}
According to the above symmetries, one can categorize the $P$-wave bottom baryons into eight baryon multiplets, as shown in Fig.~\ref{fig:pwave}. We denote these multiplets as $[F({\rm flavor}), j_l, s_l, \rho/\lambda]$, with $j_l$ the total angular momentum of the light components ($j_l = l_\lambda \otimes l_\rho \otimes s_l$). Every multiplet contains one or two bottom baryons, whose total angular momenta are $j = j_l \otimes s_b = | j_l \pm 1/2 |$, with $s_b$ the spin of the bottom quark. Especially, the heavy quark effective theory tells that the bottom baryons inside the same doublet with $j = j_l - 1/2$ and $j = j_l + 1/2$ have similar masses.

\begin{widetext}
In this section we investigate $S$-wave decays of flavor $\mathbf{6}_F$ $P$-wave bottom baryons into ground-state bottom baryons and vector mesons. To do this we use the method of light-cone sum rules within HQET, and investigate the following decay channels (the coefficients at right hand sides are isospin factors):
\begin{eqnarray}
&(a1)& {\Gamma\Big[} \Sigma_b[1/2^-] \rightarrow \Lambda_b + \rho \rightarrow\Lambda_b+\pi+\pi{\Big ]}
= { \Gamma\Big[} \Sigma_b^{-}[1/2^-] \rightarrow \Lambda_b^{0} +\pi^0+ \pi^- {\Big ]} \, ,
\label{eq:couple1}
\\ &(a2)& { \Gamma\Big[}\Sigma_b[1/2^-] \rightarrow \Sigma_b + \rho\rightarrow\Sigma_b+\pi+\pi{\Big ]}
= 2 \times { \Gamma\Big[}\Sigma_b^{-}[1/2^-] \rightarrow \Sigma_b^{0} +\pi^0+ \pi^-{\Big ]} \, ,
\\ &(a3)&{ \Gamma\Big[}\Sigma_b[1/2^-] \rightarrow \Sigma_b^{*} + \rho\rightarrow\Sigma_b^{*}+\pi+\pi{\Big ]}
= 2 \times { \Gamma\Big[}\Sigma_b^{-}[1/2^-] \rightarrow \Sigma_b^{*0} +\pi^0+ \pi^-{\Big ]} \, ,
\\ &(b1)& { \Gamma\Big[}\Xi_b^{\prime}[1/2^-] \rightarrow \Xi_b + \rho\rightarrow\Xi_b+\pi+\pi{\Big ]}
= {3\over2} \times { \Gamma\Big[}\Xi_b^{\prime -}[1/2^-] \rightarrow \Xi_b^{0} + \pi^0+\pi^-{\Big ]} \, ,
\\ &(b2)& { \Gamma\Big[} \Xi_b^{\prime}[1/2^-] \rightarrow \Lambda_b + K^*\rightarrow\Lambda_b+K+\pi {\Big ]}
={3\over2}\times { \Gamma\Big[} \Xi_b^{\prime -}[1/2^-] \rightarrow \Lambda_b^{0} + K^0+\pi^- {\Big ]} \, ,
\\ &(b3)& { \Gamma\Big[}\Xi_b^{\prime}[1/2^-] \rightarrow \Xi_b^{\prime} + \rho\rightarrow\Xi_b^{\prime}+\pi+\pi{\Big ]}
= {3\over2} \times  { \Gamma\Big[}\Xi_b^{\prime -}[1/2^-] \rightarrow \Xi_b^{\prime0} + \pi^0+\pi^-{\Big ]} \, ,
\\ &(b4)& { \Gamma\Big[}\Xi_b^{\prime}[1/2^-] \rightarrow \Sigma_b + K^*\rightarrow\Sigma_b+K+\pi{\Big ]}
= {9\over2}\times { \Gamma\Big[}\Xi_b^{\prime -}[1/2^-] \rightarrow \Sigma_b^{0} + K^0+\pi^-{\Big ]} \, ,
\\&(b5)&{ \Gamma\Big[}\Xi_b^{\prime}[1/2^-] \rightarrow \Xi_b^{*} + \rho\rightarrow\Xi_b^{*}+\pi+\pi{\Big ]}
= {3\over2} \times  { \Gamma\Big[}\Xi_b^{\prime -}[1/2^-] \rightarrow \Xi_b^{*0} + \pi^0+\pi^-{\Big ]} \, ,
\\&(b6)&{ \Gamma\Big[}\Xi_b^{\prime}[1/2^-] \rightarrow \Sigma_b^{*} + K^*\rightarrow\Sigma_b^{*}+K+\pi{\Big ]}
= {9\over2} \times  { \Gamma\Big[}\Xi_b^{\prime -}[1/2^-] \rightarrow \Sigma_b^{*0} + K^0+\pi^-{\Big ]} \, ,
\\ &(c1)& { \Gamma\Big[}\Omega_b[1/2^-] \rightarrow \Xi_b + K^*\rightarrow\Xi_b+K+\pi{\Big ]}
= 3 \times { \Gamma\Big[}\Omega_b^{-}[1/2^-] \rightarrow \Xi_b^{0} + K^0+\pi^-{\Big ]} \, ,
\\ &(c2)& { \Gamma\Big[}\Omega_b[1/2^-] \rightarrow \Xi_b^{\prime} + K^*\rightarrow\Xi_b^{\prime}+K+\pi{\Big ]}
= 3 \times { \Gamma\Big[}\Omega_b^{-}[1/2^-] \rightarrow \Xi_b^{\prime0} + K^0+\pi^-{\Big ]} \, ,
\\ &(c3)&{ \Gamma\Big[}\Omega_b[1/2^-] \rightarrow \Xi_b^{*} + K^*\rightarrow\Xi_b^{*}+K+\pi{\Big ]}
= 3 \times { \Gamma\Big[}\Omega_b^{-}[1/2^-] \rightarrow \Xi_b^{*0} + K^0+\pi^-{\Big ]} \, ,
\\ &(d1)&{ \Gamma\Big[} \Sigma_b[3/2^-] \rightarrow \Lambda_b + \rho \rightarrow\Lambda_b+\pi+\pi{\Big ]}
= { \Gamma\Big[} \Sigma_b^{-}[3/2^-] \rightarrow \Lambda_b^{0} +\pi^0+ \pi^- {\Big ]} \, ,
\\ &(d2)& { \Gamma\Big[}\Sigma_b[3/2^-] \rightarrow \Sigma_b + \rho\rightarrow\Sigma_b+\pi+\pi{\Big ]}
= 2 \times { \Gamma\Big[}\Sigma_b^{-}[3/2^-] \rightarrow \Sigma_b^{0} +\pi^0+ \pi^-{\Big ]} \, ,
\\&(d3)& { \Gamma\Big[}\Sigma_b[3/2^-] \rightarrow \Sigma_b^{*} + \rho\rightarrow\Sigma_b^{*}+\pi+\pi {\Big ]}
= 2 \times { \Gamma\Big[}\Sigma_b^{-}[3/2^-] \rightarrow \Sigma_b^{*0} + \pi^0+\pi^- {\Big ]} \, ,
\\ &(e1)& { \Gamma\Big[}\Xi_b^{\prime}[3/2^-] \rightarrow \Xi_b + \rho\rightarrow\Xi_b+\pi+\pi{\Big ]}
= {3\over2} \times { \Gamma\Big[}\Xi_b^{\prime -}[3/2^-] \rightarrow \Xi_b^{0} + \pi^0+\pi^-{\Big ]} \, ,
\\ &(e2)& { \Gamma\Big[} \Xi_b^{\prime}[3/2^-] \rightarrow \Lambda_b + K^*\rightarrow\Lambda_b+K+\pi {\Big ]}
={3\over2}\times { \Gamma\Big[} \Xi_b^{\prime -}[3/2^-] \rightarrow \Lambda_b^{0} + K^0+\pi^- {\Big ]} \, ,
\\ &(e3)& { \Gamma\Big[}\Xi_b^{\prime}[3/2^-] \rightarrow \Xi_b^{\prime} + \rho\rightarrow\Xi_b^{\prime}+\pi+\pi{\Big ]}
= {3\over2} \times  { \Gamma\Big[}\Xi_b^{\prime -}[3/2^-] \rightarrow \Xi_b^{\prime0} + \pi^0+\pi^-{\Big ]} \, ,
\\ &(e4)& { \Gamma\Big[}\Xi_b^{\prime}[3/2^-] \rightarrow \Sigma_b^{\prime} + K^*\rightarrow\Sigma_b^{\prime}+K+\pi{\Big ]}
= {9\over2} \times  { \Gamma\Big[}\Xi_b^{\prime -}[3/2^-] \rightarrow \Sigma_b^{\prime0} + K^0+\pi^-{\Big ]} \, ,
\\ &(e5)& { \Gamma\Big[}\Xi_b^{\prime}[3/2^-] \rightarrow \Xi_b^{*} + \rho\rightarrow\Xi_b^{*}+\pi+\pi {\Big ]}
= {3\over2} \times { \Gamma\Big[}\Xi_b^{\prime-}[3/2^-] \rightarrow \Xi_b^{*0} + \pi^0+\pi^- {\Big ]} \, ,
\\ &(e6)& { \Gamma\Big[}\Xi_b^{\prime}[3/2^-] \rightarrow \Sigma_b^{*} + K^* \rightarrow \Sigma_b^* + K + \pi{\Big ]}
 = {9\over2} \times { \Gamma\Big[}\Xi_b^{\prime-}[3/2^-] \rightarrow \Sigma_b^{*0} + K^0+\pi^- {\Big ]} \, ,
\\  &(f1)& { \Gamma\Big[}\Omega_b[3/2^-] \rightarrow \Xi_b + K^*\rightarrow\Xi_b +K+\pi {\Big ]} = 3 \times { \Gamma\Big[}\Omega_b^{-}[3/2^-] \rightarrow \Xi_b^0 + K^0+\pi^- {\Big ]} \, ,
\\  &(f2)& { \Gamma\Big[}\Omega_b[3/2^-] \rightarrow \Xi_b^{\prime} + K^*\rightarrow\Xi_b^{\prime}+K+\pi {\Big ]} = 3 \times { \Gamma\Big[}\Omega_b^{-}[3/2^-] \rightarrow \Xi_b^{\prime0} + K^0+\pi^- {\Big ]} \, ,
\\ &(f3)& { \Gamma\Big[}\Omega_b[3/2^-] \rightarrow \Xi_b^{*} + K^*\rightarrow\Xi_b^{*}+K+\pi {\Big ]} = 3 \times { \Gamma\Big[}\Omega_b^{-}[3/2^-] \rightarrow \Xi_b^{*0} + K^0+\pi^- {\Big ]} \, ,
\\ &(g1)& { \Gamma\Big[}\Sigma_b^{}[5/2^-] \rightarrow \Sigma_b^{*}+\rho\rightarrow\Sigma_b^{*}+\pi\pi {\Big]}
= 2\times{ \Gamma\Big[}\Sigma_b^{-}[5/2^-]\rightarrow\Sigma_b^{*0}+\pi^0+\pi^-{\Big]} \, ,
\\ &(h1)& { \Gamma\Big[} \Xi_b^{\prime}[5/2^-]\rightarrow\Sigma_b^{*}+K^*\rightarrow\Sigma_b^{*}+K+\pi{\Big]}
={9\over2}\times{\Gamma\Big[}\Xi_b^{\prime-}[5/2^-]\rightarrow\Sigma_b^{*0}+K^0+\pi^-{\Big]}\, ,
\\&(h2)&  { \Gamma\Big[}\Xi_b^{\prime}[5/2^-]\rightarrow\Xi_b^{*}+\rho\rightarrow\Xi_b^{*}+\pi+\pi{\Big]}
={3\over2}\times{\Gamma\Big[}\Xi_b^{\prime-}[5/2^-]\rightarrow\Xi_b^{*0}+\pi^0+\pi^-{\Big]} \, ,
\\&(i1)& {\Gamma\Big[}\Omega_b[5/2^-]\rightarrow\Xi_b^{*}+K^*\rightarrow\Xi_b^{*}+K+\pi{\Big]}
=3\times{\Gamma\Big[}\Omega_b^-[5/2^-]\rightarrow\Xi_b^{*0}+K^0+\pi^-{\Big]} \, .
\label{eq:couple28}
\end{eqnarray}
\end{widetext}
We can calculate their decay widths through the following Lagrangians
\begin{eqnarray*}
\mathcal{L}_{X_b({1/2}^-) \rightarrow Y_b({1/2}^+) V} &=& g {\bar X_b}(1/2^-) \gamma_\mu \gamma_5 Y_b(1/2^+) V^\mu \, ,
\\ \mathcal{L}_{X_b({1/2}^-) \rightarrow Y_b({3/2}^+) V} &=& g {\bar X_{b}}(1/2^-) Y_{b}^{\mu}(3/2^+) V_\mu \, ,
\\ \mathcal{L}_{X_b({3/2}^-) \rightarrow Y_b({1/2}^+) V} &=& g {\bar X_{b}^{\mu}}(3/2^-) Y_{b}(1/2^+) V_\mu \, ,
\\ \mathcal{L}_{X_b({3/2}^-) \rightarrow Y_b({3/2}^+) V} &=& g {\bar X_b}^{\nu}(3/2^-) \gamma_\mu \gamma_5 Y_{b\nu}(3/2^+) V^\mu \, ,
\\ \mathcal{L}_{X_b({5/2}^-) \rightarrow Y_b({3/2}^+) V} &=& g {\bar X_{b}^{\mu\nu}}(5/2^-) Y_{b\mu}(3/2^+) V_\nu
\\ \nonumber && ~ + g {\bar X_{b}^{\nu\mu}}(5/2^-) Y_{b\mu}(3/2^+) V_\nu \, ,
\end{eqnarray*}
where $X_b^{(\mu\nu)}$, $Y_b^{(\mu)}$, and $V^\mu$ denotes the $P$-wave bottom baryon, ground-state bottom baryon, and vector meson, respectively.

As an example, we study the $S$-wave decay of the $\Sigma_b^-({1/2}^-)$ belonging to $[\mathbf{6}_F, 1, 0, \rho]$ into $\Lambda_b^0(1/2^+)$ and $\rho^-(1^-)$ in the next subsection, and investigate the four bottom baryon multiplets, $[\mathbf{6}_F, 1, 0, \rho]$, $[\mathbf{6}_F, 0, 1, \lambda]$, $[\mathbf{6}_F, 1, 1, \lambda]$, and $[\mathbf{6}_F, 2, 1, \lambda]$, separately in the following subsections.

\subsection{$\Sigma_b^-({1/2}^-)$ of $[\mathbf{6}_F, 1, 0, \rho]$ decaying into $\Lambda_b^0(1/2^+)$ and $\rho^-(1^-)$}

In this subsection we study the $S$-wave decay of the $\Sigma_b^-({1/2}^-)$ belonging to $[\mathbf{6}_F, 1, 0, \rho]$ into $\Lambda_b^0(1/2^+)$ and $\rho^-(1^-)$. To do this we consider the following three-point correlation function:
\begin{eqnarray}
\nonumber \Pi(\omega, \, \omega^\prime) &=& \int d^4 x~e^{-i k \cdot x}~\langle 0 | J_{1/2,-,\Sigma_b^-,1,0,\rho}(0) \bar J_{\Lambda_b^{0}}(x) | \rho^-(q) \rangle
\\ &=& {1+v\!\!\!\slash\over2} G_{\Sigma_b^-[{1\over2}^-] \rightarrow \Lambda_b^{0}\rho^-} (\omega, \omega^\prime) \, ,
\end{eqnarray}
where $J_{1/2,-,\Sigma_b^-,1,0,\rho}$ and $J_{\Lambda_b^{0}}$ are the interpolating fields coupling to $\Sigma_b^-({1/2}^-)$ and $\Lambda_b^0$:
\begin{eqnarray}
&& J_{1/2,-,\Sigma_b^-,1,0,\rho}
\\ \nonumber && ~~~~~~ = i \epsilon_{abc} \Big ( [\mathcal{D}_t^{\mu} d^{aT}] C \gamma_5 d^b -  d^{aT} C \gamma_5 [\mathcal{D}_t^{\mu} d^b] \Big ) \gamma_t^{\mu} \gamma_5 h_v^c \, ,
\\ && J_{\Lambda_b^0} = \epsilon_{abc} [u^{aT} C\gamma_{5} d^{b}] h_{v}^{c} \, ,
\end{eqnarray}
We refer to Refs.~\cite{Liu:2007fg,Chen:2015kpa}, where we systematically constructed all the $S$- and $P$-wave heavy baryon interpolating fields. In the above expressions $h_v(x)$ is the heavy quark field; $k^\prime = k + q$, with $k^\prime$, $k$, and $q$ the momenta of the $\Sigma_b^-({1/2}^-)$, $\Lambda_b^0$, and $\rho^-$, respectively; $\omega = v \cdot k$ and $\omega^\prime = v \cdot k^\prime$. Note that the definitions of $\omega$ and $\omega^\prime$ in the present study are the same as those used in Refs.~\cite{Chen:2017sci,Cui:2019dzj}, but different from those used in Refs.~\cite{Chen:2015kpa,Mao:2015gya}.

$\\$

\begin{widetext}
At the hadronic level, we write $G_{\Sigma_b^-[{1\over2}^-] \rightarrow \Lambda_b^{0}\rho^-}$ as:
\begin{eqnarray}
G_{\Sigma_b^-[{1\over2}^-] \rightarrow \Lambda_b^{0}\rho^-} (\omega, \omega^\prime) &=& g_{\Sigma_b^-[{1\over2}^-] \rightarrow \Lambda_b^{0}\rho^-} \times { f_{\Sigma_b^-[{1\over2}^-]} f_{\Lambda_b^{0}} \over (\bar \Lambda_{\Sigma_b^-[{1\over2}^-]} - \omega^\prime) (\bar \Lambda_{\Lambda_b^{0}} - \omega)} \, . \label{G0C}
\end{eqnarray}

At the quark and gluon level, we calculate $G_{\Sigma_b^-[{1\over2}^-] \rightarrow \Lambda_b^{0}\rho^-}$ using the method of operator product expansion (OPE):
\begin{eqnarray}
\label{eq:sumrule}
&& G_{\Sigma_b^-[{1\over2}^-] \rightarrow \Lambda_b^0\rho^-} (\omega, \omega^\prime)
\\ \nonumber &=& \int_0^\infty dt \int_0^1 du e^{i (1-u) \omega^\prime t} e^{i u \omega t} \times 8 \times \Big (
\frac{if_\rho^{\bot}m_\rho^2}{48}\langle\bar q q\rangle\psi_{3;\rho}^{||}(u)
+ \frac{if_\rho^{||}m_\rho^3}{8\pi^2 t^4(v \cdot q)^2}\phi_{2;\rho}^{||}(u)
\\ \nonumber &&
- \frac{i f_\rho^{||} m_\rho^3}{4\pi^2t^4(v \cdot q)^2}\phi_{3;\rho}^\bot (u)+\frac{i f_\rho^{||} m_\rho^3}{8 \pi^2 t^4 (v \cdot q)^2} \psi_{4;\rho}^{||}(u)-\frac{i f_\rho^{||} m_\rho}{4 \pi^2 t^4}\phi_{2;\rho}^{||}(u)+\frac{i f_\rho^{||} m_\rho}{2 \pi^2 t^4}\phi_{3;\rho}^\bot(u)
\\ \nonumber &&
+ \frac{i f_\rho^\bot m_\rho^2 t^2}{768} \langle g_s \bar q \sigma  G q\rangle \psi_{3;\rho}^{||}(u)-\frac{i f_\rho^{||} m_\rho^3}{64 \pi^2 t^2}\phi_{4;\rho}^{||}(u) \Big )
\\ \nonumber &-&
\int_0^\infty dt \int_0^1 du \int \mathcal{D} \underline{\alpha} e^{i \omega^{\prime} t(\alpha_2 + u \alpha_3)} e^{i \omega t(1 - \alpha_2 - u \alpha_3)} \times \Big (\frac{i f_\rho^{||} m_\rho^3}{8 \pi^2 t^2}\Phi_{4;\rho}^{||}(\underline{\alpha})
+ \frac{i f_\rho^{||} m_\rho^3}{8 \pi^2 t^2}\widetilde{\Phi}_{4;\rho}^{||}(\underline{\alpha})
\\ \nonumber &&
+\frac{i f_\rho^{||} m_\rho^3}{16 \pi^2 t^2}\Psi_{4;\rho}^{||}(\underline{\alpha})+\frac{i f_\rho^{||} m_\rho^3}{16 \pi^2 t^2}\widetilde{\Psi}_{4;\rho}^{||}(\underline{\alpha})+\frac{i f_\rho^{||} m_\rho^3 u}{4 \pi^2 t^2}\Phi_{4;\rho}^{||}(\underline{\alpha})+\frac{i f_\rho^{||} m_\rho^3 u}{8 \pi^2 t^2}\Psi_{4;\rho}^{||}(\underline{\alpha})\Big)\, .
\end{eqnarray}

After Wick rotations and making double Borel transformation with the variables $\omega$ and $\omega^\prime$ to be $T_1$ and $T_2$, we obtain
\begin{eqnarray}
&& g_{\Sigma_b^-[{1\over2}^-] \rightarrow \Lambda_b^{0}\rho^-} f_{\Sigma_b^-[{1\over2}^-]} f_{\Lambda_b^{0}} e^{- {\bar \Lambda_{\Sigma_b^-[{1\over2}^-]} \over T_1}} e^{ - {\bar \Lambda_{\Lambda_b^{0}} \over T_2}}
\\ \nonumber &=& 8 \times \Big ( \frac{- f_\rho^\bot m_\rho^2}{48}\langle \bar q q \rangle T f_0({\omega_c \over T})\psi_{3;\rho}^{||}(u_0)-\frac{f_\rho^{||} m_\rho^3}{8\pi^2} T^3 f_2({\omega_c \over T}) \int_0^{u_0} du_1 \int_0^{u_1} du_2\phi_{2;\rho}^{||}(u_2)
\\ \nonumber &&
+\frac{f_\rho^{||} m_\rho^3}{4\pi^2} T^3 f_2({\omega_c \over T}) \int_0^{u_0} du_1 \int_0^{u_1} du_2 \phi_{3;\rho}^\bot(u_2)-\frac{f_\rho^{||} m_\rho^3}{8\pi^2} T^3 f_2({\omega_c \over T}) \int_0^{u_0} du_1 \int_0^{u_1} du_2 \psi_{4;\rho}^{||}(u_2)
\\ \nonumber &&
+ \frac{f_\rho^{||} m_\rho}{4\pi^2} T^5 f_4({\omega_c \over T}) \phi_{2;\rho}^{||}(u_0) - \frac{f_\rho^{||} m_\rho}{2 \pi^2} T^5 f_4({\omega_c \over T}) \phi_{3;\rho}^\bot(u_0)+\frac{f_\rho^\bot m_\rho^2}{768} \langle g_s \bar q \sigma  G q \rangle {1 \over T} \psi_{3;\rho}^{||}(u_0) - \frac{f_\rho^{||} m_\rho^3}{64} T^3 f_2({\omega_c \over T})\phi_{4;\rho}^{||}(u_0) \Big )
\\ \nonumber &-&
\Big(\frac{f_\rho^{||} m_\rho^3}{8\pi^2} T^3 f_2({\omega_c \over T}) \int_0^{1 \over 2} d\alpha_2 \int_{{1 \over 2}-\alpha_2}^{1-\alpha_2} d\alpha_3 {1 \over \alpha_3} \Phi_{4;\rho}^{||}(\underline{\alpha}) + \frac{f_\rho^{||} m_\rho^3}{8\pi^2} T^3 f_2({\omega_c \over T}) \int_0^{1 \over 2} d\alpha_2 \int_{{1 \over 2}-\alpha_2}^{1-\alpha_2} d\alpha_3 {1 \over \alpha_3} \widetilde{\Phi}_{4;\rho}^{||}(\underline{\alpha})
\\ \nonumber &&
+\frac{f_\rho^{||} m_\rho^3}{16 \pi^2} T^3 f_2({\omega_c \over T}) \int_0^{1 \over 2} d\alpha_2 \int_{{1 \over 2}-\alpha_2}^{1-\alpha_2} d\alpha_3 {1 \over \alpha_3} \Psi_{4;\rho}^{||}(\underline{\alpha})+ \frac{f_\rho^{||} m_\rho^3}{16 \pi^2} T^3 f_2({\omega_c \over T}) \int_0^{1 \over 2} d\alpha_2 \int_{{1 \over 2}-\alpha_2}^{1-\alpha_2} d\alpha_3 {1 \over \alpha_3} \widetilde{\Psi}_{4;\rho}^{||}(\underline{\alpha})
\\ \nonumber &&
+ \frac{f_\rho^{||} m_\rho^3 u}{4\pi^2} T^3 f_2({\omega_c \over T}) \int_0^{1 \over 2} d\alpha_2 \int_{{1 \over 2}-\alpha_2}^{1-\alpha_2} d\alpha_3 {1\over\alpha_3} \Phi_{4;\rho}^{||}(\underline{\alpha}) + \frac{f_\rho^{||} m_\rho^3 u}{8\pi^2} T^3 f_2({\omega_c \over T}) \int_0^{1 \over 2} d\alpha_2 \int_{{1 \over 2}-\alpha_2}^{1-\alpha_2} d\alpha_3 {1\over\alpha_3} \Psi_{4;\rho}^{||}(\underline{\alpha}) \Big) \, .
\end{eqnarray}
Here $u_0 = {T_1 \over T_1 + T_2}$, $T = {T_1 T_2 \over T_1 + T_2}$, and $f_n(x) = 1 - e^{-x} \sum_{k=0}^n {x^k \over k!}$. Explicit forms of the light-cone distribution amplitudes contained in the above expression can be found in Refs.~\cite{Ball:1998je,Ball:2006wn,Ball:2004rg,Ball:1998kk,Ball:1998sk,Ball:1998ff,Ball:2007rt,Ball:2007zt}, and we work at the renormalization scale 2~GeV for the parameters involved. More sum rule examples can be found in \ref{sec:othersumrule}.
\end{widetext}

In the present study we work at the symmetric point $T_1 = T_2 = 2T$ so that $u_0 = {1\over2}$. We use the following values for the bottom quark mass and various quark and gluon condensates~\cite{pdg,Yang:1993bp,Hwang:1994vp,Ovchinnikov:1988gk,Jamin:2002ev,Ioffe:2002be,Narison:2002pw,Gimenez:2005nt,colangelo}:
%
\begin{eqnarray}
\nonumber && m_b = 4.66 \pm 0.03~{\rm GeV} \, ,
\\ \nonumber && \langle \bar qq \rangle = - (0.24 \pm 0.01 \mbox{ GeV})^3 \, ,
\\ \nonumber && \langle \bar ss \rangle = (0.8\pm 0.1)\times \langle\bar qq \rangle \, ,
\\ && \langle g_s \bar q \sigma G q \rangle = M_0^2 \times \langle \bar qq \rangle\, ,
\label{eq:condensates}
\\ \nonumber && \langle g_s \bar s \sigma G s \rangle = M_0^2 \times \langle \bar ss \rangle\, ,
\\ \nonumber && M_0^2= 0.8 \mbox{ GeV}^2\, ,
\\ \nonumber && \langle g_s^2GG\rangle =(0.48\pm 0.14) \mbox{ GeV}^4\, .
\end{eqnarray}

Now the coupling constant $g_{\Sigma_b^-[{1\over2}^-] \rightarrow \Lambda_b^{0}\rho^-}$ only depends on two free parameters, the threshold value $\omega_c$ and the Borel mass $T$. We choose $\omega_c = 1.$ 485GeV to be the average of the threshold values of the $\Sigma_b(1/2^-)$ and $\Lambda_b^{0}$ mass sum rules (see \ref{sec:sbottom} and Ref.~\cite{Cui:2019dzj} for the parameters of the $\Sigma_b(1/2^-)$ and $\Lambda_b^{0}$), and extract the coupling constant $g_{\Sigma_b^-[{1\over2}^-] \rightarrow \Lambda_b^{0}\rho^-}$ to be
\begin{eqnarray}
&& g_{\Sigma_b^-[{1\over2}^-] \rightarrow \Lambda_b^{0}\rho^-}
\\ \nonumber &=& 0.17~{^{+0.24}_{-0.17}} =0.17~{^{+0.01}_{-0.00}}~{^{+0.06}_{-0.05}}~{^{+0.06}_{-0.05}}~{^{+0.20}_{-0.17}}~{^{+0.10}_{-0.11}} \, ,
\end{eqnarray}
where the uncertainties are due to the Borel mass, the parameters of the $\Lambda_b^{0}$, the parameters of the $\Sigma_b^-({1/2}^-)$, the light-cone distribution amplitudes of vector mesons~\cite{Ball:1998je,Ball:2006wn,Ball:2004rg,Ball:1998kk,Ball:1998sk,Ball:1998ff,Ball:2007rt,Ball:2007zt}, and various quark masses and condensates listed in Eq.~(\ref{eq:condensates}), respectively. Besides these statistical uncertainties, there is another (theoretical) uncertainty, which comes from the scale dependence. In the present study we do not consider this, and simply work at the renormalization scale 2~GeV, since $\sqrt{M_{\Sigma_b^-({1/2}^-)}^2 - M_{\Lambda_b^{0}}^2} = 2.4$~GeV. However, it is useful to give a rough estimation on this. Since the largest uncertainty comes from the light-cone distribution amplitudes of vector mesons, we choose the values for the parameters contained in these amplitudes to be at the renormalization scale 1~GeV (see Tables~1 and 2 of Ref.~\cite{Ball:2007zt}), and redo the above calculations to obtain:
\begin{equation}
g_{\Sigma_b^-[{1\over2}^-] \rightarrow \Lambda_b^{0}\rho^-}(1~\mbox{GeV}) = 0.42 \, .
\end{equation}
Hence, the scale dependence leads to a significant uncertainty, and the total uncertainty of our results can be even larger.

For completeness, we show $g_{\Sigma_b^-[{1\over2}^-] \rightarrow \Lambda_b^{0}\rho^-}$ in Fig.~\ref{fig:610rho}(a) as a function of the Borel mass $T$, and find that it only slightly depends on the Borel mass, where the working region for $T$ has been evaluated in the $\Sigma_b(1/2^-)$ mass sum rules~\cite{Chen:2015kpa,Mao:2015gya,Chen:2017sci,Cui:2019dzj} to be $0.31$ GeV $<T<0.34$ GeV (we also summarize this in Table~\ref{tab:pwaveparameter}). Note that the definitions of $\omega$ and $\omega^\prime$ in this paper are the same as those used in Refs.~\cite{Chen:2017sci,Cui:2019dzj}, but different from those used in Refs.~\cite{Chen:2015kpa,Mao:2015gya}, so the Borel windows used in this paper are also the same/similar as those used in Refs.~\cite{Chen:2017sci,Cui:2019dzj}, but just about half of those used in Refs.~\cite{Chen:2015kpa,Mao:2015gya}.

The two-body decay $\Sigma_b^-({1/2}^-) \rightarrow \Lambda_b^{0}\rho^-$ is kinematically forbidden, but the three-body decay process $\Sigma_b^-({1/2}^-) \rightarrow \Lambda_b^{0}\rho^- \rightarrow \Lambda_b^{0}\pi^0\pi^-$ is kinematically allowed, whose decay amplitude is
\begin{eqnarray}
&& \mathcal{M} \left( 0 \rightarrow 3 + 4 \rightarrow 3 + 2 + 1 \right)
\\ \nonumber &\equiv& \mathcal{M} \left( \Sigma_b^-({1/2}^-) \rightarrow \Lambda_b^{0} + \rho^- \rightarrow \Lambda_b^{0} + \pi^0 + \pi^- \right)
\\ \nonumber &=& g_{0 \rightarrow 3 + 4} \times g_{4 \rightarrow 2 + 1} \times \bar u_{0} \gamma_\mu \gamma_5 u_3 \times \left( g_{\mu\nu} - { p_{4,\mu} p_{4,\nu} \over m_4^2} \right)
\\ \nonumber && ~~~~~ \times { 1 \over p_4^2 - m_4^2 + i m_4 \Gamma_4 } \times \left( p_{1,\nu} - p_{2,\nu} \right) \, .
\end{eqnarray}
Here $0$ denotes the initial state $\Sigma_b^-({1/2}^-)$; 4 denotes the intermediate state $\rho^{-}$; $1$, $2$ and $3$ denote the finial states $\pi^-$, $\pi^0$ and $\Lambda_b^{0}$, respectively.

This amplitude can be used to further calculate its decay width
\begin{eqnarray}
&& \Gamma \left( 0 \rightarrow 3 + 4 \rightarrow 3 + 2 + 1 \right)
\\ \nonumber &\equiv& \Gamma \left( \Sigma_b^-({1/2}^-) \rightarrow \Lambda_b^{0} + \rho^- \rightarrow \Lambda_b^{0} + \pi^0 + \pi^- \right)
\\ \nonumber &=& {1 \over (2\pi)^3} {1 \over 32 m_0^3} \times g^2_{0 \rightarrow 3 + 4} \times g^2_{4 \rightarrow 2 + 1} \times \int d m_{12} d m_{23}
\\ \nonumber && ~~~~~ \times {1\over2}~{\rm Tr}\Big[ \left( p\!\!\!\slash_3 + m_3 \right) \gamma_{\mu^\prime} \gamma_5 \left( p\!\!\!\slash_0 + m_0 \right) \gamma_{\mu} \gamma_5 \Big]
\\ \nonumber && ~~~~~ \times \left( g_{\mu\nu} - { p_{4,\mu} p_{4,\nu} \over m_4^2} \right)\left( g_{\mu^\prime\nu^\prime} - { p_{4,\mu^\prime} p_{4,\nu^\prime} \over m_4^2} \right)
\\ \nonumber && ~~~~~ \times {\left( p_{1,\nu} - p_{2,\nu} \right) \left( p_{1,\nu^\prime} - p_{2,\nu^\prime} \right) \over |p_4^2 - m_4^2 + i m_4 \Gamma_4|^2 } \, ,
\end{eqnarray}
so that the width of the $\Sigma_b^-({1/2}^-) \rightarrow \Lambda_b^{0}\rho^- \rightarrow \Lambda_b^{0}\pi^0\pi^-$ decay is evaluated to be
\begin{eqnarray}
&&\Gamma_{\Sigma_b^-[{1\over2}^-] \rightarrow \Lambda_b^{0}\rho^- \rightarrow \Lambda_b^{0}\pi^0\pi^-}
\\ \nonumber &=& 8.9~{^{+32.5}_{-8.9}}{\rm~keV} = 8.9~{^{+1.1}_{-0.0}}~{^{+6.7}_{-5.0}}~{^{+7.0}_{-4.5}}~{^{+27.9}_{-8.9}}~{^{+13.6}_{-7.5}}{\rm~keV} \, .
\end{eqnarray}

In the following subsections we apply the same procedures to separately study the four bottom baryon multiplets, $[\mathbf{6}_F, 1, 0, \rho]$, $[\mathbf{6}_F, 0, 1, \lambda]$, $[\mathbf{6}_F, 1, 1, \lambda]$, and $[\mathbf{6}_F, 2, 1, \lambda]$.

\subsection{The bottom baryon doublet $[\mathbf{6}_F, 1, 0, \rho]$}

\begin{table*}[hbt]
\begin{center}
\renewcommand{\arraystretch}{1.5}
\caption{$S$-wave decays of $P$-wave bottom baryons belonging to the doublet $[\mathbf{6}_F, 1, 0, \rho]$ into ground-state bottom baryons and vector mesons.}
\begin{tabular}{ c | c | c | c}
\hline\hline
 ~~~Decay channels~~~ & ~Coupling constant $g$ ~ & Partial width  & Total width
\\ \hline\hline
(a1) $\Sigma_b({1\over2}^-)\to \Lambda_b({1\over2}^+) \rho\to\Lambda_b({1\over2}^+)\pi\pi$  & $0.17$  & $ 8.9${\rm~keV} & \multirow{2}{*}{$8.9${\rm~MeV}}
\\
(a2) $\Sigma_b({1\over2}^-)\to \Sigma_b({1\over2}^+) \rho\to\Sigma_b({1\over2}^+)\pi\pi$ & $0.35$ & $2.2\times 10^{-3}${\rm~keV}
\\ \hline
(b1) $\Xi_b^{\prime}({1\over2}^-)\to \Xi_b({1\over2}^+) \rho\to\Xi_b({1\over2}^+)\pi\pi$ & $0.03$ & $ 0.2${\rm~keV} & \multirow{2}{*}{$0.2${\rm~keV}}
\\
(b3) $\Xi_b^{\prime}({1\over2}^-) \to \Xi_b^{\prime}({1\over2}^+) \rho \to \Xi_b^{\prime}({1 \over 2}^+) \pi\pi$ & $0.24$ & $5.1\times 10^{-3}${\rm~keV}
\\ \hline
(d1) $\Sigma_b({3\over2}^-)\to\Lambda_b({1\over2}^+)\rho\to\Lambda_b({1\over2}^+)\pi\pi$ &$0.29$&$4.1${\rm~keV}&
\multirow{2}{*}{$4.1${\rm~keV}}
\\
(d2) $\Sigma_b({3\over2}^-)\to\Sigma_b({1\over2}^+)\rho\to\Sigma_b({1\over2}^+)\pi\pi$ &$0.19$&$2.1\times10^{-4}${\rm~keV}
\\ \hline
(e1) $\Xi_b^{\prime}({3\over2}^-)\to \Xi_b({1\over2}^+) \rho\to\Xi_b({1\over2}^+)\pi\pi$ & $0.06$ & $0.2${\rm~keV} & \multirow{2}{*}{$0.2${\rm~keV}}
\\
(e3) $\Xi_b^{\prime}({3\over2}^-) \to \Xi_b^{\prime}({1\over2}^+) \rho \to \Xi_b^{\prime}({1 \over 2}^+) \pi\pi$ & $0.14$ & $5.8\times 10^{-4}${\rm~keV}
\\ \hline\hline
\end{tabular}
\label{tab:decay610rho}
\end{center}
\end{table*}

The bottom baryon doublet $[\mathbf{6}_F, 1, 0, \rho]$ consists of six members: $\Sigma_b({1\over2}^-/{3\over2}^-)$, $\Xi^\prime_b({1\over2}^-/{3\over2}^-)$, and $\Omega_b({1\over2}^-/{3\over2}^-)$. We use the method of light-cone sum rules within HQET to study their decays into ground-state bottom baryons and vector mesons.

\begin{figure*}[htb]
\begin{center}
\subfigure[]{
\scalebox{0.42}{\includegraphics{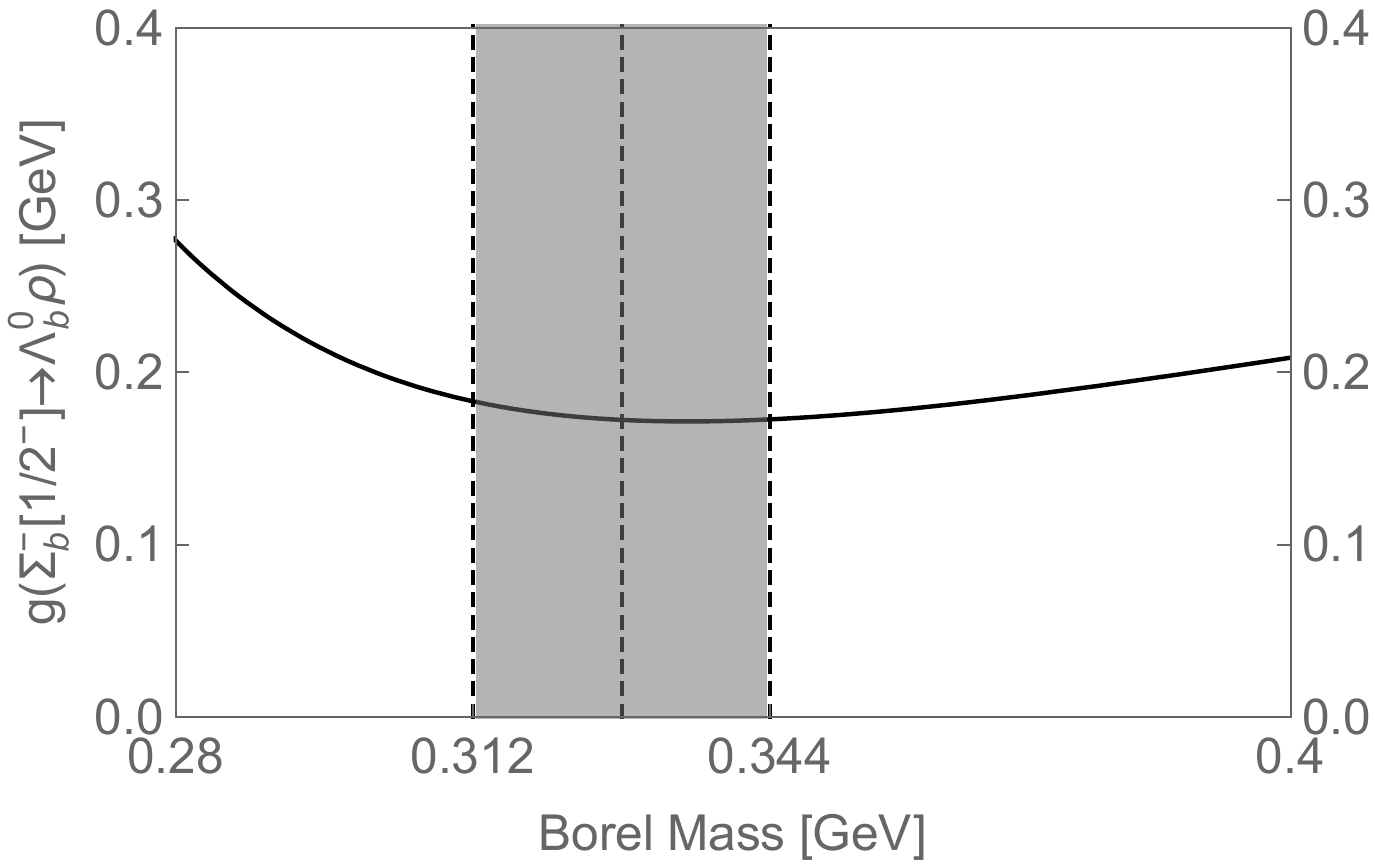}}}
\subfigure[]{
\scalebox{0.42}{\includegraphics{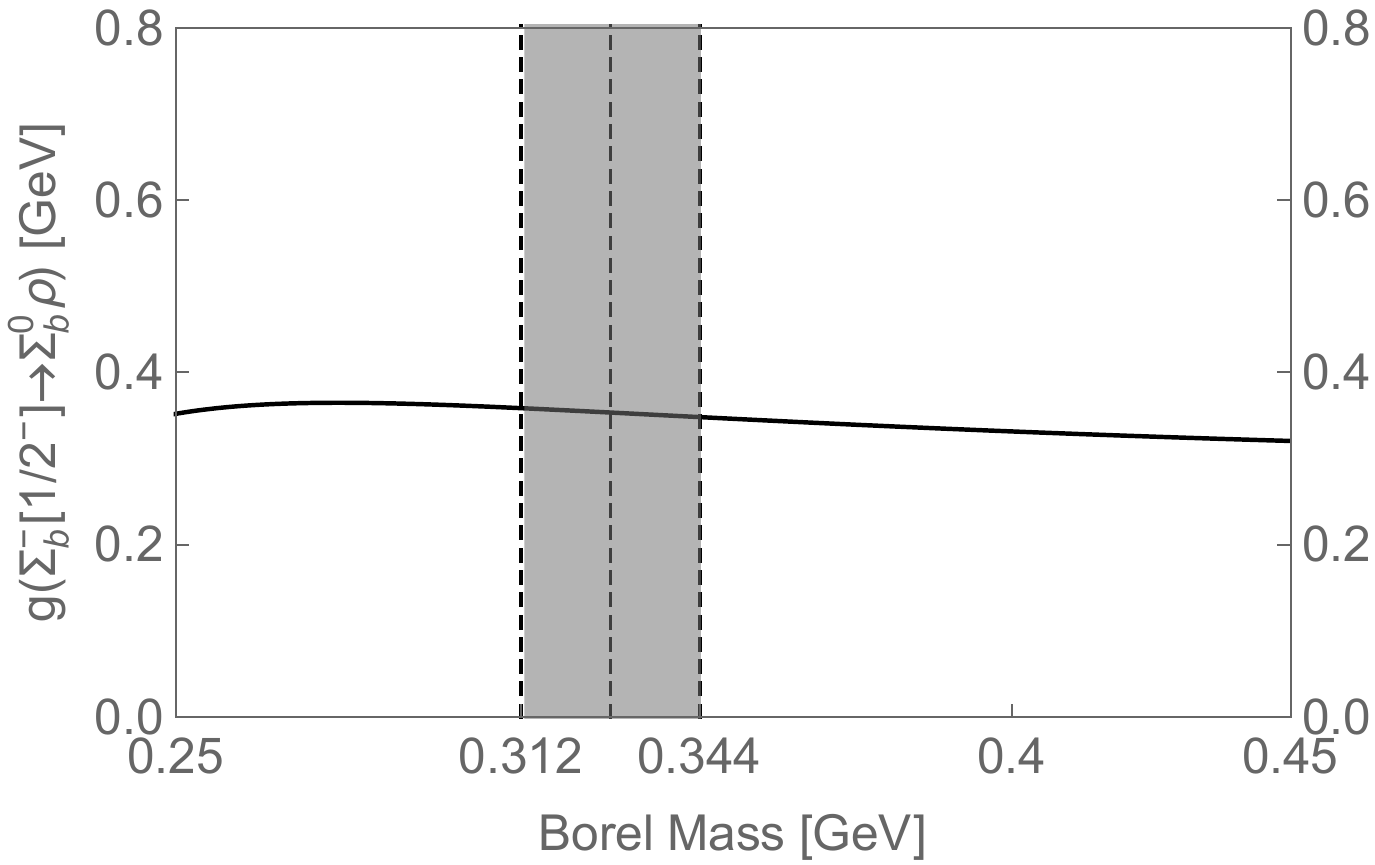}}}
\subfigure[]{
\scalebox{0.44}{\includegraphics{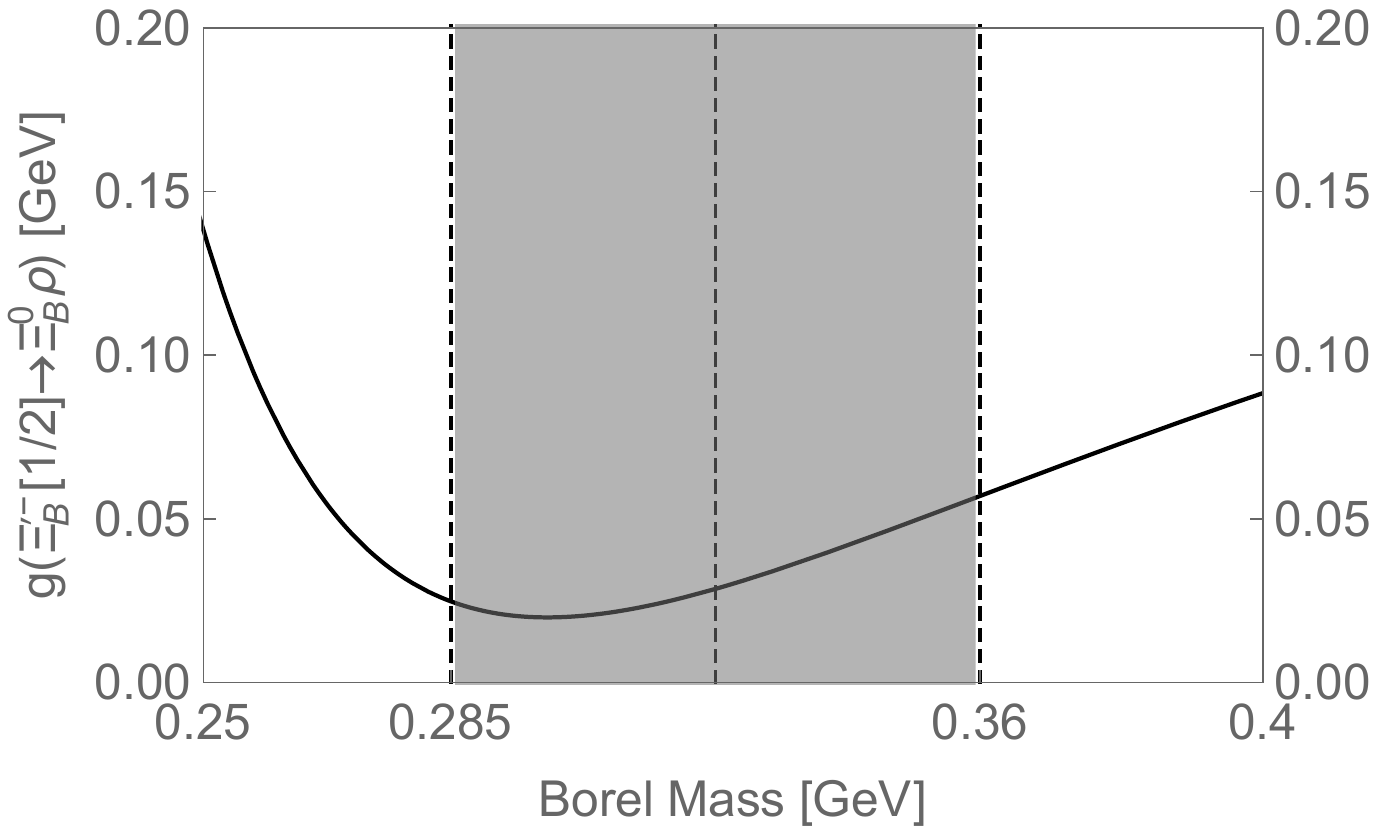}}}
\subfigure[]{
\scalebox{0.42}{\includegraphics{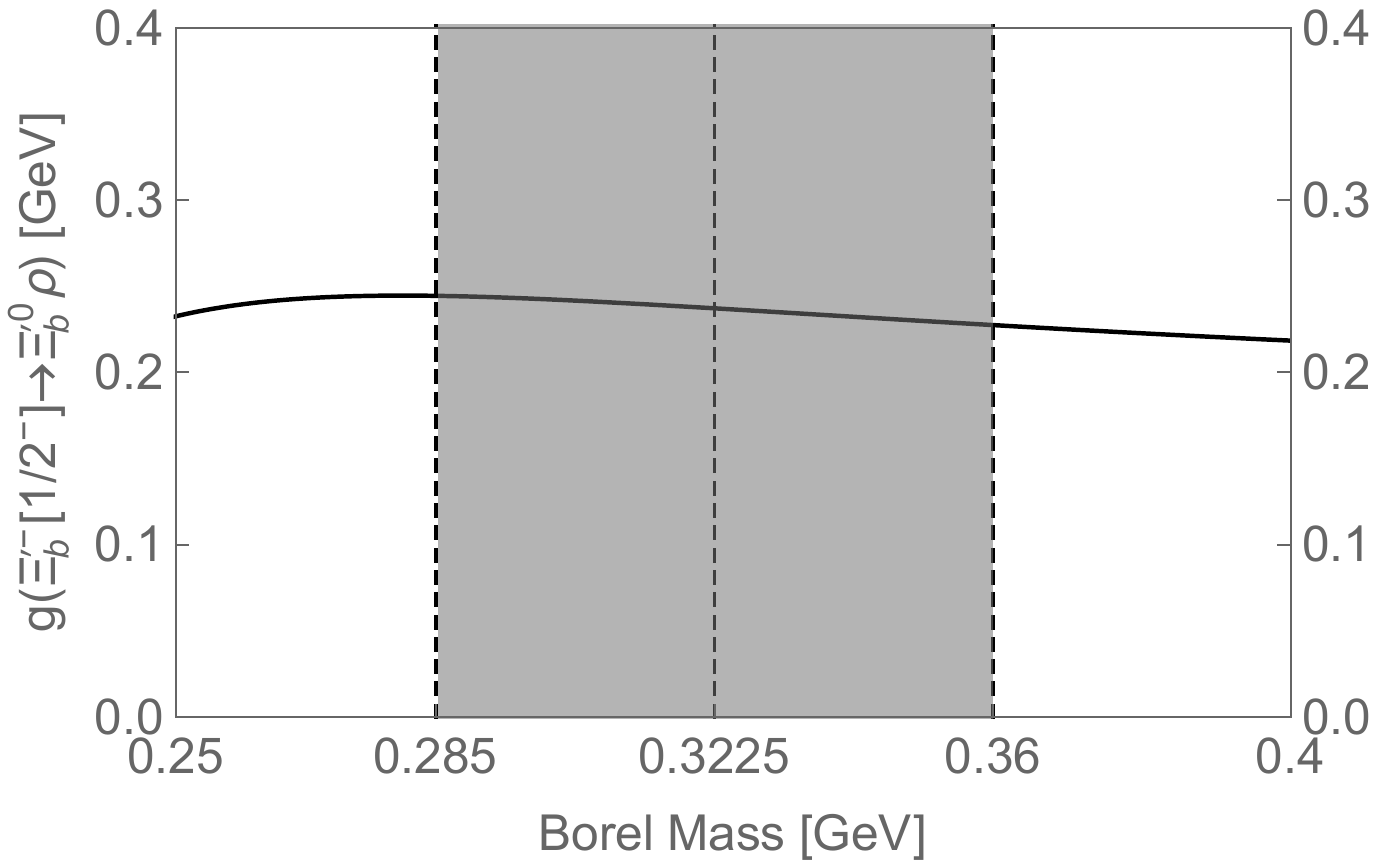}}}
\subfigure[]{
\scalebox{0.42}{\includegraphics{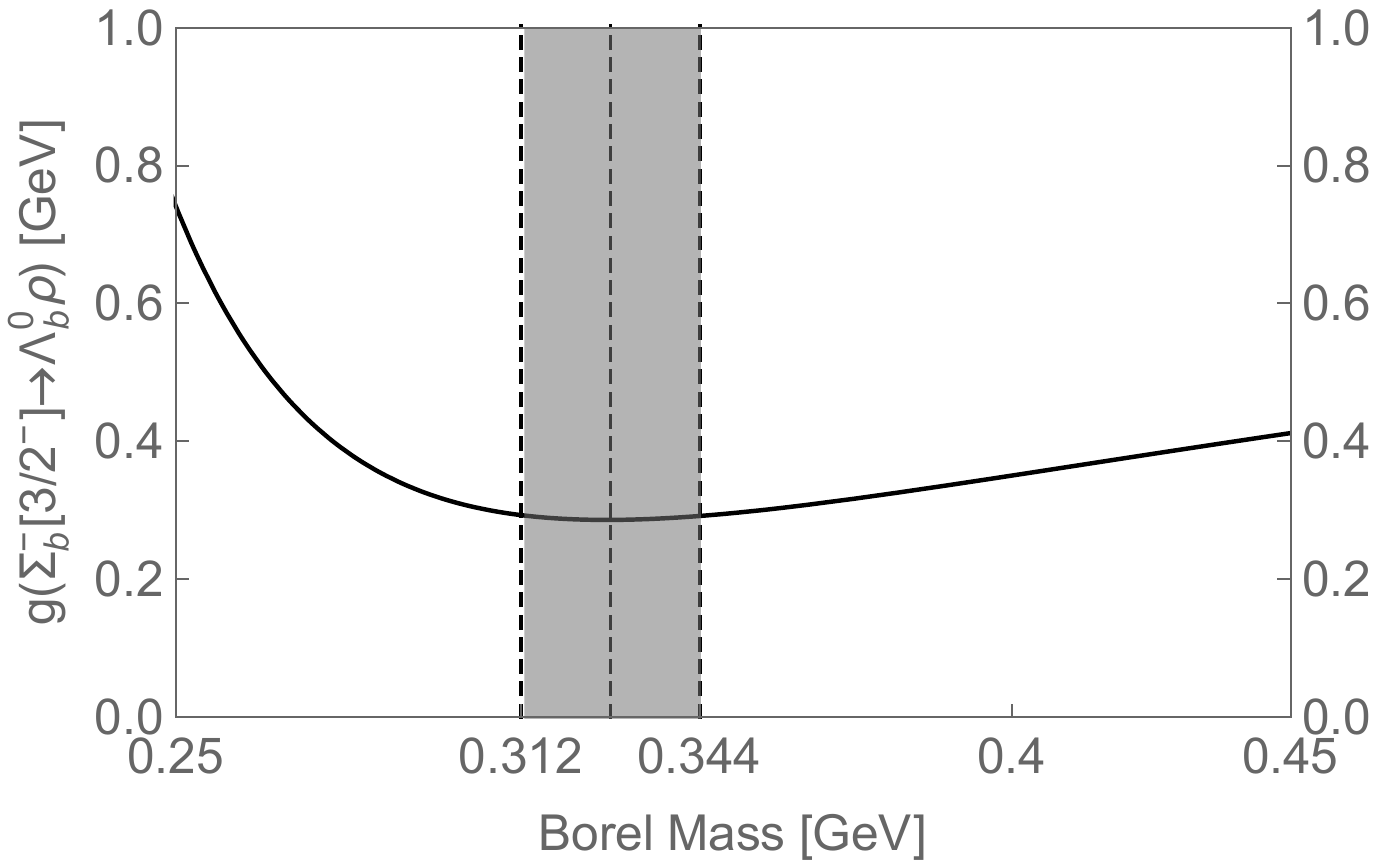}}}
\subfigure[]{
\scalebox{0.42}{\includegraphics{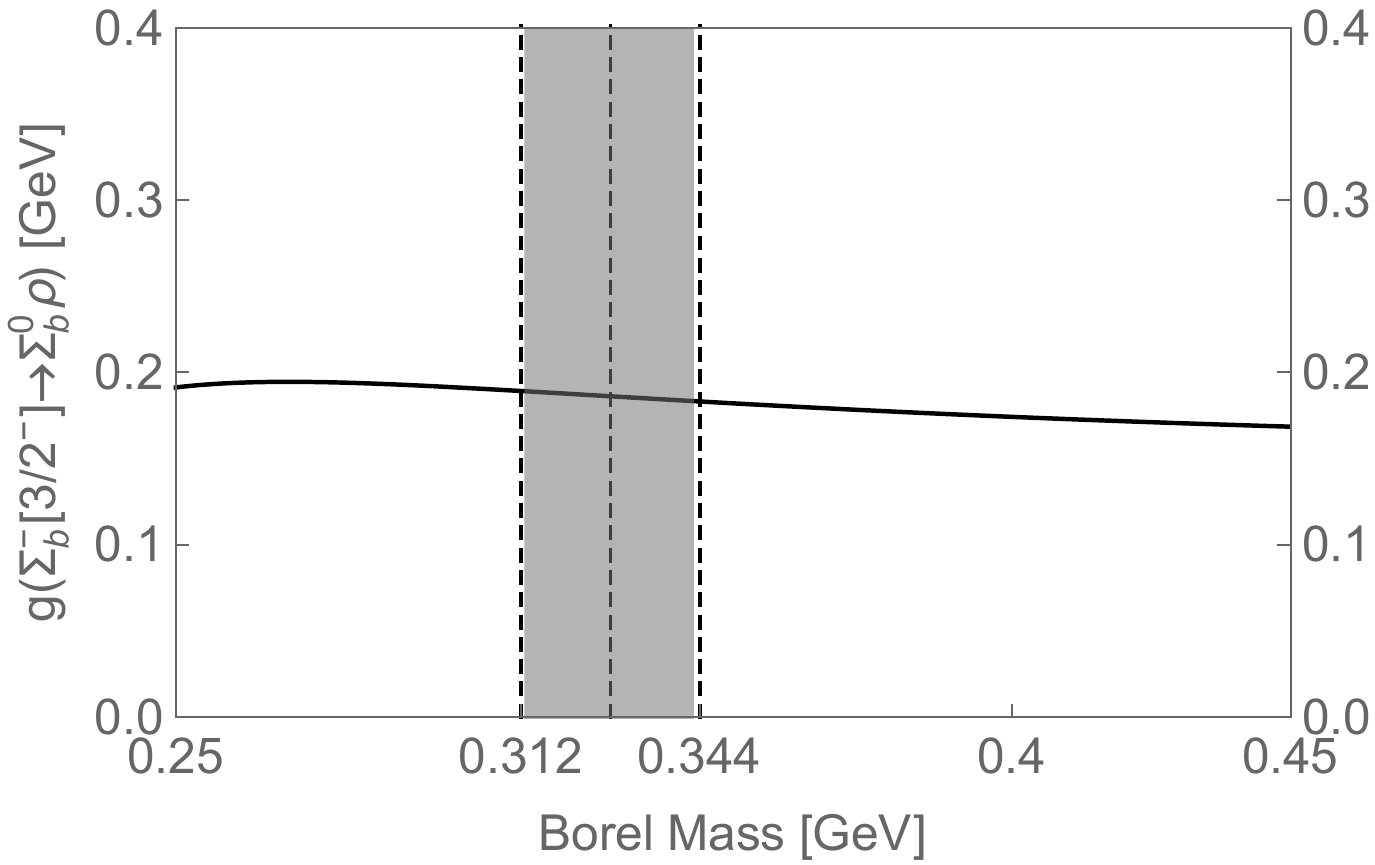}}}
\subfigure[]{
\scalebox{0.44}{\includegraphics{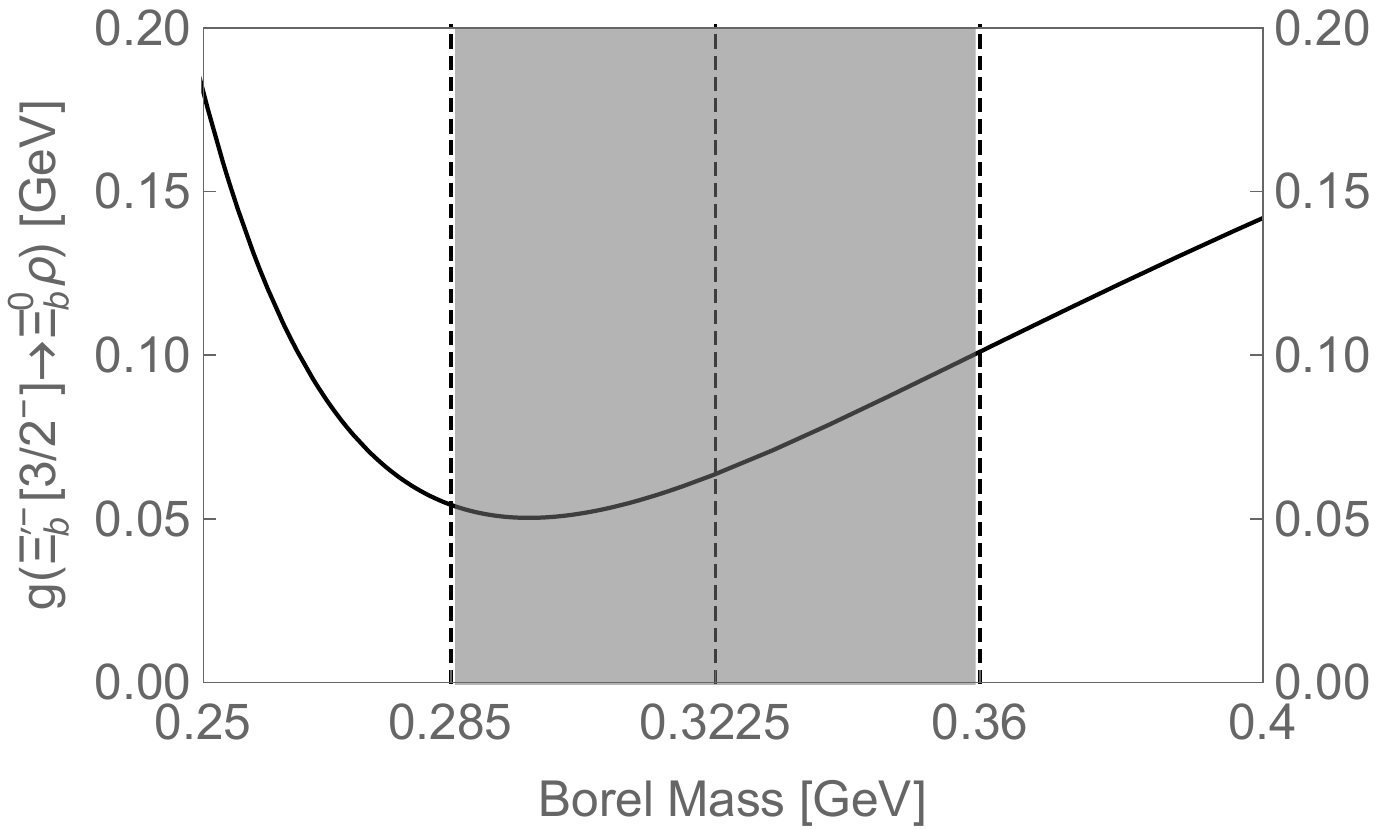}}}
\subfigure[]{
\scalebox{0.42}{\includegraphics{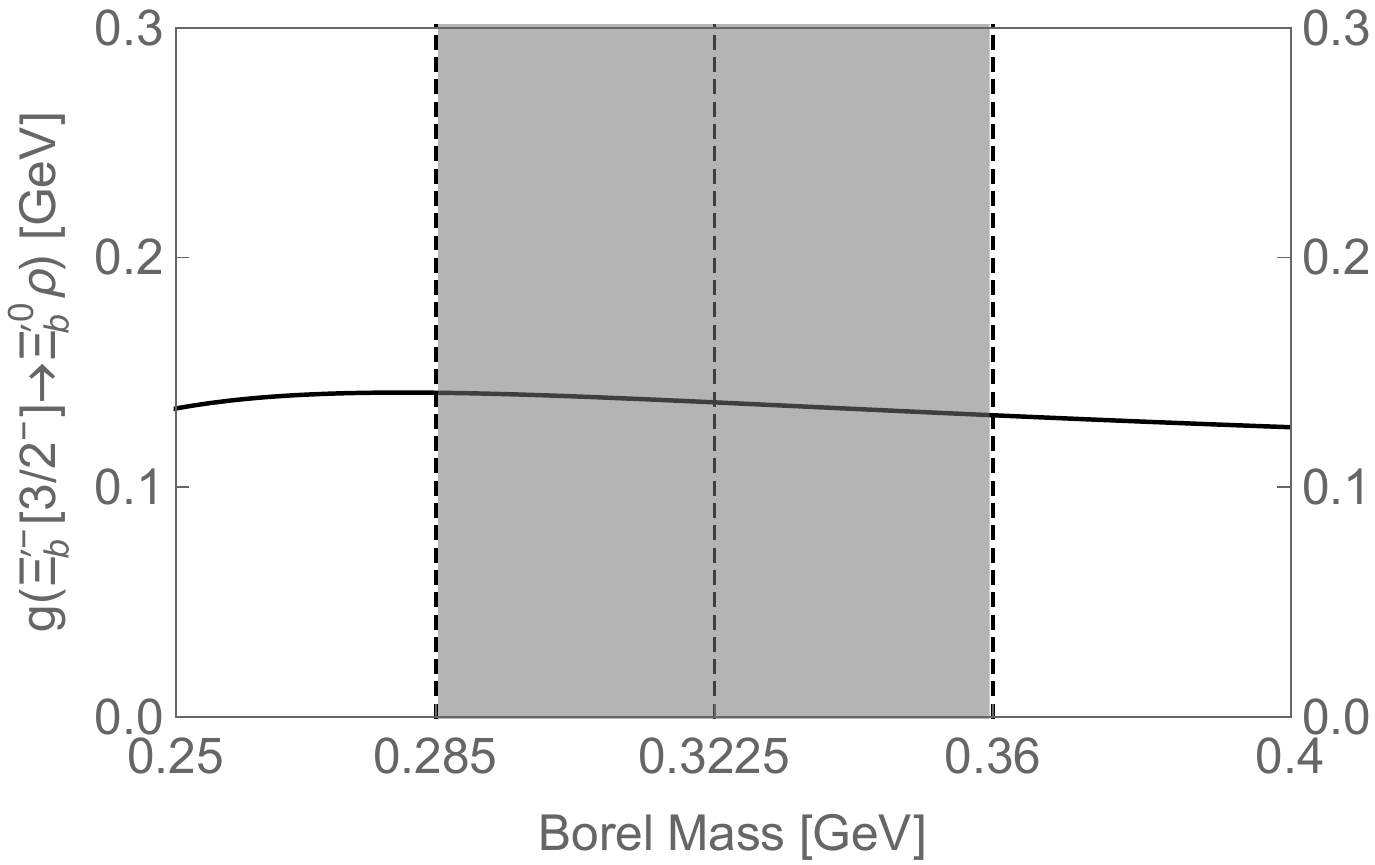}}}
\end{center}
\caption{The coupling constants (a) $g_{\Sigma_b^-[{1\over2}^-] \rightarrow \Lambda_b^0 \rho^-}$, (b) $g_{\Sigma_b^-[{1\over2}^-] \rightarrow \Sigma_b^0\rho^-}$, (c) $g_{\Xi_b^{\prime-}[{1\over2}^-] \rightarrow \Xi_b^0 \rho^-}$, (d) $g_{\Xi_b^{\prime-}[{1\over2}^-] \rightarrow \Xi_b^{\prime0} \rho^-}$, (e) $g_{\Sigma_b^-[{3\over2}^-] \rightarrow \Lambda_b^0 \rho^-}$, (f) $g_{\Sigma_b^-[{3\over2}^-] \rightarrow \Sigma_b^0 \rho^-}$, (g) $g_{\Xi_b^{\prime-}[{3\over2}^-] \rightarrow \Xi_b^0 \rho^-}$, and (h) $g_{\Xi_b^{\prime-}[{3\over2}^-] \rightarrow \Xi_b^{\prime0} \rho^-}$ as functions of the Borel mass $T$. Here the baryons $\Sigma_b({1\over2}^-/{3\over2}^-)$ and $\Xi^\prime_b({1\over2}^-/{3\over2}^-)$ belong to the bottom baryon doublet $[\mathbf{6}_F, 1, 0, \rho]$, and the working regions for $T$ have been evaluated in mass sum rules~\cite{Chen:2015kpa,Mao:2015gya,Chen:2017sci,Cui:2019dzj} and summarized in Table~\ref{tab:pwaveparameter}.
\label{fig:610rho}}
\end{figure*}

There are altogether twenty-four non-vanishing decay channels, whose coupling constants are extracted to be
\begin{eqnarray}
\nonumber &(a1)& g_{\Sigma_b[{1\over2}^-]\to \Lambda_b[{1\over2}^+] \rho} = 0.17 \, ,
\\
\nonumber &(a2)& g_{\Sigma_b[{1\over2}^-]\to \Sigma_b[{1\over2}^+] \rho} = 0.35 \, ,
\\
\nonumber &(a3)& g_{\Sigma_b[{1\over2}^-]\to \Sigma_b^{*}[{3\over2}^+] \rho}= 0.20 \, ,
\\
\nonumber &(b1)& g_{\Xi_b^{\prime}[{1\over2}^-]\to \Xi_b[{1\over2}^+] \rho} = 0.03 \, ,
\\
\nonumber &(b2)& g_{\Xi_b^{\prime}[{1\over2}^-]\to \Lambda_b[{1\over2}^+] K^{*}}= 0.31 \, ,
\\
\nonumber &(b3)& g_{\Xi_b^{\prime}[{1\over2}^-]\to \Xi_b^{\prime}[{1\over2}^+] \rho } = 0.24\, ,
\\
\nonumber &(b4)& g_{\Xi_b^{\prime}[{1\over2}^-]\to \Sigma_b[{1\over2}^+]K^{*}}= 0.38 \, ,
\\
\nonumber &(b5)& g_{\Xi_b^{\prime}[{1\over2}^-]\to \Xi_b^{*}[{3\over2}^+]\rho}= 0.14 \, ,
\\
\nonumber &(b6)& g_{\Xi_b^{\prime}[{1\over2}^-]\to\Sigma_b^{*}[{3\over2}^+]K^{*}}= 0.22 \, ,
\\
\nonumber &(c1)& g_{\Omega_b[{1\over2}^-]\to \Xi_b[{1\over2}^+]K^{*}}= 0.24 \, ,
\\
\nonumber &(c2)& g_{\Omega_b[{1\over2}^-]\to \Xi_b^{\prime}[{1\over2}^+]K^{*}}= 0.49 \, ,
\\
\nonumber &(c3)& g_{\Omega_b[{1\over2}^-]\to \Xi_b^{*}[{3\over2}^+]K^{*}}= 0.28 \, ,
\\        &(d1)& g_{\Sigma_b[{3\over2}^-]\to\Lambda_b[{1\over2}^+]\rho}=0.29 \, ,
\label{coupling610rho}
\\
\nonumber &(d2)& g_{\Sigma_b[{3\over2}^-]\to\Sigma_b[{1\over2}^+]\rho}=0.19 \, ,
\\
\nonumber &(d3)& g_{\Sigma_b[{3\over2}^-]\to\Sigma_b^{*}[{3\over2}^+]\rho}= 0.24 \, ,
\\
\nonumber &(e1)& g_{\Xi_b^{\prime}[{3\over2}^-]\to \Xi_b[{1\over2}^+] \rho}=0.06 \, ,
\\
\nonumber &(e2)& g_{\Xi_b^{\prime}[{3\over2}^-]\to\Lambda_b[{1\over2}^+]K^{*}}=0.42 \, ,
\\
\nonumber &(e3)& g_{\Xi_b^{\prime}[{3\over2}^-] \to \Xi_b^{\prime}[{1\over2}^+] \rho }=0.14 \, ,
\\
\nonumber &(e4)& g_{\Xi_b^{\prime}[{3\over2}^-]\to\Sigma_b[{1\over2}^+]K^{*}}=0.22 \, ,
\\
\nonumber &(e5)& g_{\Xi_b^{\prime}[{3\over2}^-]\to\Xi_b^{*}[{3\over2}^+]\rho}=0.16 \, ,
\\
\nonumber &(e6)& g_{\Xi_b^{\prime}[{3\over2}^-]\to\Sigma_b^{*}[{3\over2}^+]K^{*}}=0.25 \, ,
\\
\nonumber &(f1)& g_{\Omega_b[{3\over2}^-]\to\Xi_b[{1\over2}^+]K^{*}}=0.34 \, ,
\\
\nonumber &(f2)& g_{\Omega_b[{3\over2}^-]\to\Xi_b^{\prime}[{1\over2}^+]K^{*}}=0.28 \, ,
\\
\nonumber &(f3)& g_{\Omega_b[{3\over2}^-]\to\Xi_b^{*}[{3\over2}^+]K^{*}}=0.33 \, .
\end{eqnarray}
Then we compute the three-body decay widths, which are kinematically allowed:
\begin{eqnarray}
\nonumber &(a1)& \Gamma_{\Sigma_b[{1\over2}^-]\to \Lambda_b[{1\over2}^+] \rho\to\Lambda_b[{1\over2}^+]\pi\pi} =8.9 {\rm~keV} \, ,
\\
\nonumber &(a2)& \Gamma_{\Sigma_b[{1\over2}^-]\to \Sigma_b[{1\over2}^+] \rho\to\Sigma_b[{1\over2}^+]\pi\pi} = 2.2\times10^{-3} {\rm~keV} \, ,
\\
\nonumber &(b1)& \Gamma_{\Xi_b^{\prime}[{1\over2}^-]\to \Xi_b[{1\over2}^+] \rho\to\Xi_b[{1\over2}^+]\pi\pi} = 0.2 {\rm~keV} \, ,
\\
\nonumber &(b3)& \Gamma_{\Xi_b^{\prime}[{1\over2}^-]\to \Xi_b^{\prime}[{1\over2}^+] \rho \to \Xi_b^{\prime}[{1\over2}^+] \pi \pi}=5.1\times10^{-3} {\rm~keV} \, ,
\\ &(d1)& \Gamma_{\Sigma_b[{3\over2}^-]\to\Lambda_b[{1\over2}^+]\rho\to\Lambda_b[{1\over2}^+]\pi\pi} = 4.1 {\rm~keV} \, ,
\label{width610lambda}
\\
\nonumber &(d2)& \Gamma_{\Sigma_b[{3\over2}^-]\to\Sigma_b[{1\over2}^+]\rho\to\Sigma_b[{1\over2}^+]\pi\pi}=2.1\times10^{-4}  {\rm~keV } \, ,
\\
\nonumber &(e1)& \Gamma_{\Xi_b^{\prime}[{3\over2}^-]\to \Xi_b[{1\over2}^+] \rho\to\Xi_b[{1\over2}^+]\pi\pi}=0.2  {\rm~keV} \, ,
\\
\nonumber &(e3)& \Gamma_{\Xi_b^{\prime}[{3\over2}^-] \to \Xi_b^{\prime}[{1\over2}^+] \rho \to \Xi_b^{\prime}[{1 \over 2}^+]\pi\pi}=5.8\times10^{-4} {\rm~keV} \, .
\end{eqnarray}
We summarize these results in Table~\ref{tab:decay610rho}. For completeness, we also show the coupling constants as functions of the Borel mass $T$ in Fig.~\ref{fig:610rho}.

\subsection{The bottom baryon singlet $[\mathbf{6}_F, 0, 1, \lambda]$}

\begin{table*}[hbt]
\begin{center}
\renewcommand{\arraystretch}{1.5}
\caption{$S$-wave decays of $P$-wave bottom baryons belonging to the doublet $[\mathbf{6}_F, 0, 1, \lambda]$ into ground-state bottom baryons and vector mesons.}
\begin{tabular}{ c | c | c | c}
\hline\hline
 ~~~Decay channels~~~ & ~Coupling constant $g$ ~ & Partial width  & Total width
\\ \hline\hline
(a2) $\Sigma_b({1\over2}^-)\to \Sigma_b({1\over2}^+) \rho\to\Sigma_b({1\over2}^+)\pi\pi$ & $2.24$ & $5.9\times10^{-2}${\rm~keV}&$5.9\times10^{-2}${\rm~keV}
\\ \hline
(b3) $\Xi_b^{\prime}({1\over2}^-) \to \Xi_b^{\prime}({1\over2}^+) \rho \to \Xi_b^{\prime}({1 \over 2}^+) \pi\pi$ & $1.58$ & $0.2${\rm~keV}&$0.2${\rm~keV}
\\ \hline\hline
\end{tabular}
\label{tab:decay601lambda}
\end{center}
\end{table*}

The bottom baryon doublet $[\mathbf{6}_F, 1, 0, \rho]$ consists of three members: $\Sigma_b({1\over2}^-)$, $\Xi^\prime_b({1\over2}^-)$, and $\Omega_b({1\over2}^-)$. We use the method of light-cone sum rules within HQET to study their decays into ground-state bottom baryons and vector mesons.

\begin{figure*}[htb]
\begin{center}
\subfigure[]{
\scalebox{0.45}{\includegraphics{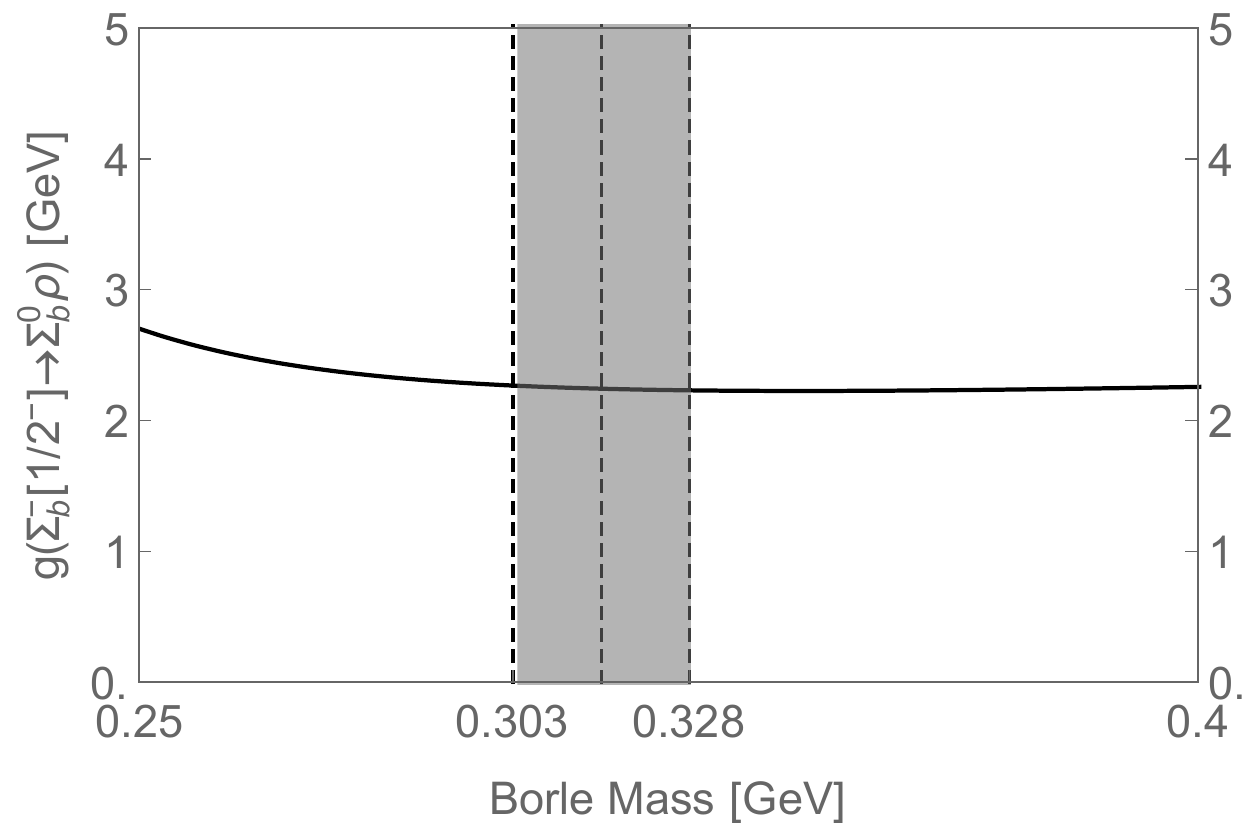}}}
\subfigure[]{
\scalebox{0.45}{\includegraphics{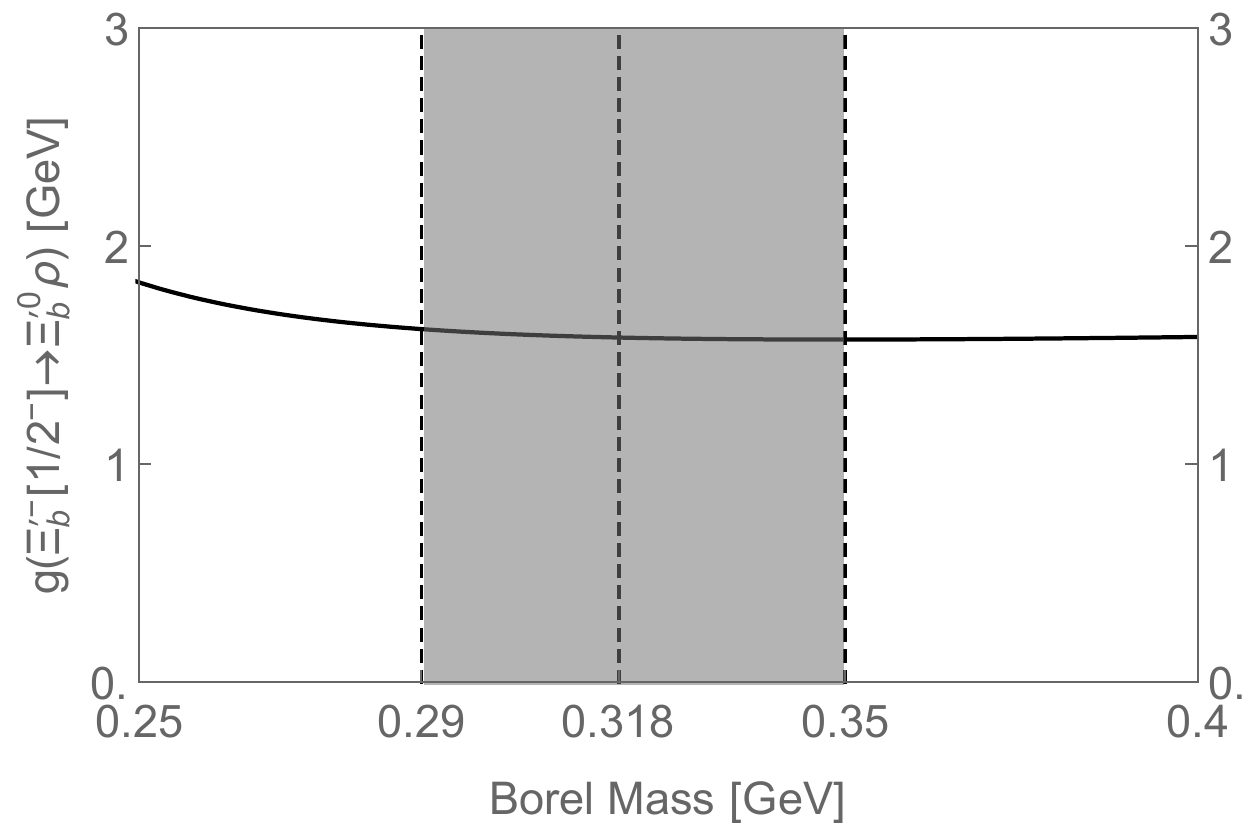}}}
\end{center}
\caption{The coupling constants (a) $g_{\Sigma_b^-[{1\over2}^-] \rightarrow \Sigma_b^0 \rho^-}$ and (b) $g_{\Xi_b^{\prime-}[{1\over2}^-] \rightarrow \Xi_b^{\prime0} \rho^-}$ as functions of the Borel mass $T$. Here the baryons $\Sigma_b({1\over2}^-)$ and $\Xi^\prime_b({1\over2}^-)$ belong to the bottom baryon singlet $[\mathbf{6}_F, 0, 1, \lambda]$, and the working regions for $T$ have been evaluated in mass sum rules~\cite{Chen:2015kpa,Mao:2015gya,Chen:2017sci,Cui:2019dzj} and summarized in Table~\ref{tab:pwaveparameter}.
\label{fig:601lambda}}
\end{figure*}

There are altogether eight non-vanishing decay channels, whose coupling constants are extracted to be
\begin{eqnarray}
\nonumber &(a2)& g_{\Sigma_b[{1\over2}^-]\to \Sigma_b[{1\over2}^+] \rho} = 2.25 \, ,
\\
\nonumber &(a3)& g_{\Sigma_b[{1\over2}^-]\to \Sigma_b^{*}[{3\over2}^+] \rho}= 7.74 \, ,
\\
\nonumber &(b3)& g_{\Xi_b^{\prime}[{1\over2}^-]\to \Xi_b^{\prime}[{1\over2}^+] \rho } = 1.58 \, ,
\\
&(b4)& g_{\Xi_b^{\prime}[{1\over2}^-]\to \Sigma_b[{1\over2}^+]K^{*}}= 1.77 \, ,
\label{coupling601lambda}
\\
\nonumber &(b5)& g_{\Xi_b^{\prime}[{1\over2}^-]\to \Xi_b^{*}[{3\over2}^+]\rho}= 5.70 \, ,
\\
\nonumber &(b6)& g_{\Xi_b^{\prime}[{1\over2}^-]\to\Sigma_b^{*}[{3\over2}^+]K^{*}}= 5.77 \, ,
\\
\nonumber &(c2)& g_{\Omega_b[{1\over2}^-]\to \Xi_b^{\prime}[{1\over2}^+]K^{*}}= 2.37 \, ,
\\
\nonumber &(c3)& g_{\Omega_b[{1\over2}^-]\to \Xi_b^{*}[{3\over2}^+]K^{*}}= 8.18 \, .
\end{eqnarray}
Then we compute the three-body decay widths, which are kinematically allowed:
\begin{eqnarray}
&(a2)& \Gamma_{\Sigma_b[{1\over2}^-]\to \Sigma_b[{1\over2}^+] \rho\to\Sigma_b[{1\over2}^+]\pi\pi} = 5.9\times10^{-2} {\rm~keV} \, ,
\label{width601lambda}
\\
\nonumber &(b3)& \Gamma_{\Xi_b^{\prime}[{1\over2}^-]\to \Xi_b^{\prime}[{1\over2}^+] \rho \to \Xi_b^{\prime}[{1\over2}^+] \pi \pi}=0.2 {\rm~keV} \, .
\end{eqnarray}
We summarize these results in Table~\ref{tab:decay601lambda}. For completeness, we also show the coupling constants as functions of the Borel mass $T$ in Fig.~\ref{fig:601lambda}.

\subsection{The bottom baryon doublet $[\mathbf{6}_F, 1, 1, \lambda]$}

\begin{table*}[hbt]
\begin{center}
\renewcommand{\arraystretch}{1.5}
\caption{$S$-wave decays of $P$-wave bottom baryons belonging to the doublet $[\mathbf{6}_F, 1, 1, \lambda]$ into ground-state bottom baryons and vector mesons.}
\begin{tabular}{ c | c | c | c}
\hline\hline
 ~~~Decay channels~~~ & ~Coupling constant $g$ ~ & Partial width  & Total width
\\ \hline\hline
(a1) $\Sigma_b({1\over2}^-)\to \Lambda_b({1\over2}^+) \rho\to\Lambda_b({1\over2}^+)\pi\pi$  & $0.83$  & $ 207.5${\rm~keV} & \multirow{2}{*}{$207.4${\rm~keV}}
\\
(a2) $\Sigma_b({1\over2}^-)\to \Sigma_b({1\over2}^+) \rho\to\Sigma_b({1\over2}^+)\pi\pi$ & $2.21$ & $5.7\times10^{-2}${\rm~keV}
\\ \hline
(b1) $\Xi_b^{\prime}({1\over2}^-)\to \Xi_b({1\over2}^+) \rho\to\Xi_b({1\over2}^+)\pi\pi$ & $0.59$ & $ 61.0${\rm~keV} & \multirow{2}{*}{$61.2${\rm~keV}}
\\
(b3) $\Xi_b^{\prime}({1\over2}^-) \to \Xi_b^{\prime}({1\over2}^+) \rho \to \Xi_b^{\prime}({1 \over 2}^+) \pi\pi$ & $1.45$ & $0.2${\rm~keV}
\\ \hline
(d1) $\Sigma_b({3\over2}^-)\to\Lambda_b({1\over2}^+)\rho\to\Lambda_b({1\over2}^+)\pi\pi$ &$1.62$&$261.1${\rm~keV}&
\multirow{2}{*}{$261.1{\rm~keV}$}
\\
(d2) $\Sigma_b({3\over2}^-)\to\Sigma_b({1\over2}^+)\rho\to\Sigma_b({1\over2}^+)\pi\pi$ &$0.90$&$3.2\times10^{-3}${\rm~keV}
\\ \hline
(e1) $\Xi_b^{\prime}({3\over2}^-)\to \Xi_b({1\over2}^+) \rho\to\Xi_b({1\over2}^+)\pi\pi$ & $0.57$ & $18.8$ {\rm~keV}& \multirow{2}{*}{$18.8${\rm~keV}}
\\
(e3) $\Xi_b^{\prime}({3\over2}^-) \to \Xi_b^{\prime}({1\over2}^+) \rho \to \Xi_b^{\prime}({1 \over 2}^+) \pi\pi$ & $0.84$ & $2.1\times 10^{-2}${\rm~keV}
\\ \hline\hline
\end{tabular}
\label{tab:decay611lambda}
\end{center}
\end{table*}

The bottom baryon doublet $[\mathbf{6}_F, 1, 1, \lambda]$ consists of six members: $\Sigma_b({1\over2}^-/{3\over2}^-)$, $\Xi^\prime_b({1\over2}^-/{3\over2}^-)$, and $\Omega_b({1\over2}^-/{3\over2}^-)$. We use the method of light-cone sum rules within HQET to study their decays into ground-state bottom baryons and vector mesons.

\begin{figure*}[htb]
\begin{center}
\subfigure[]{
\scalebox{0.37}{\includegraphics{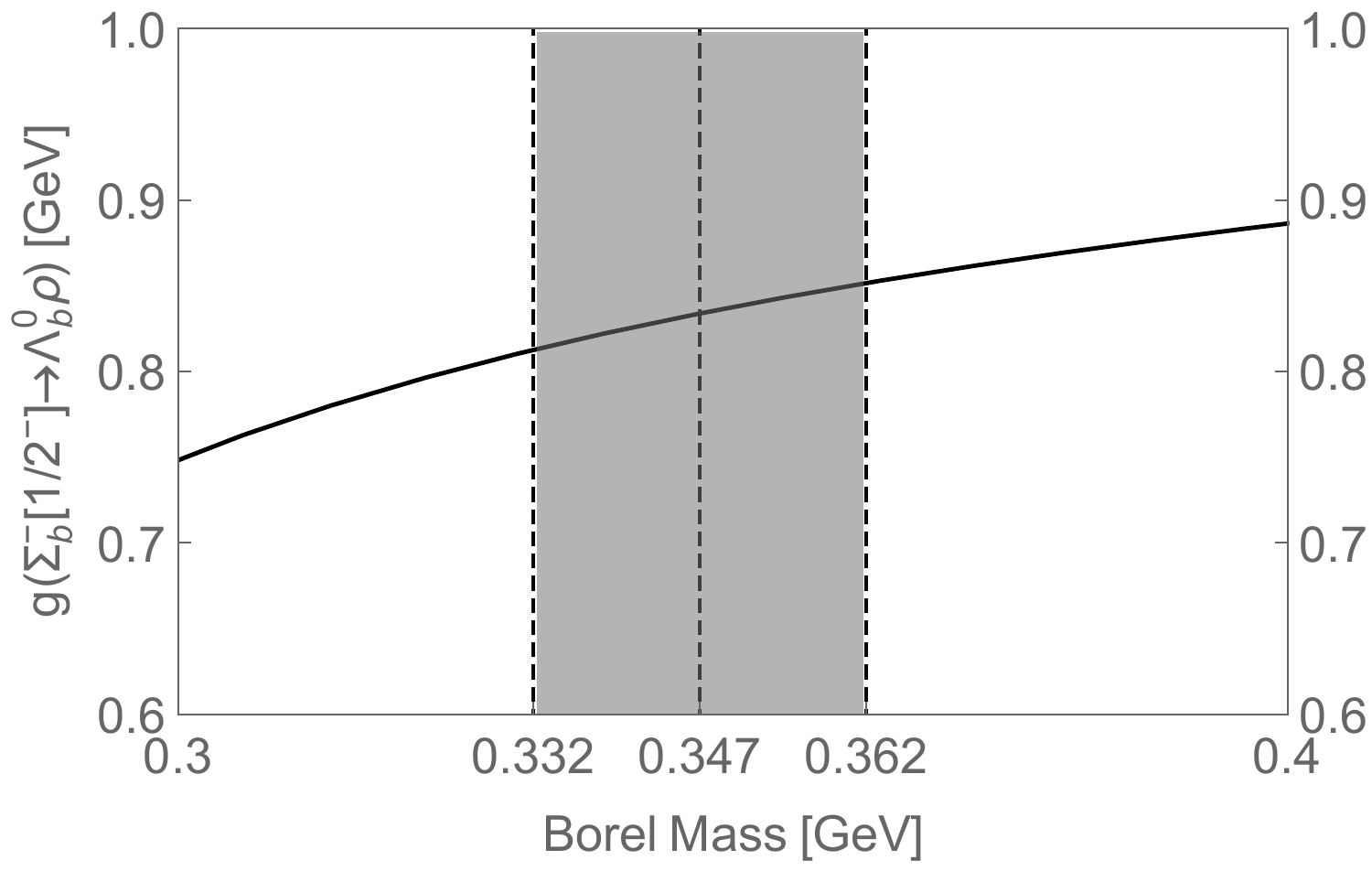}}}~~~
\subfigure[]{
\scalebox{0.37}{\includegraphics{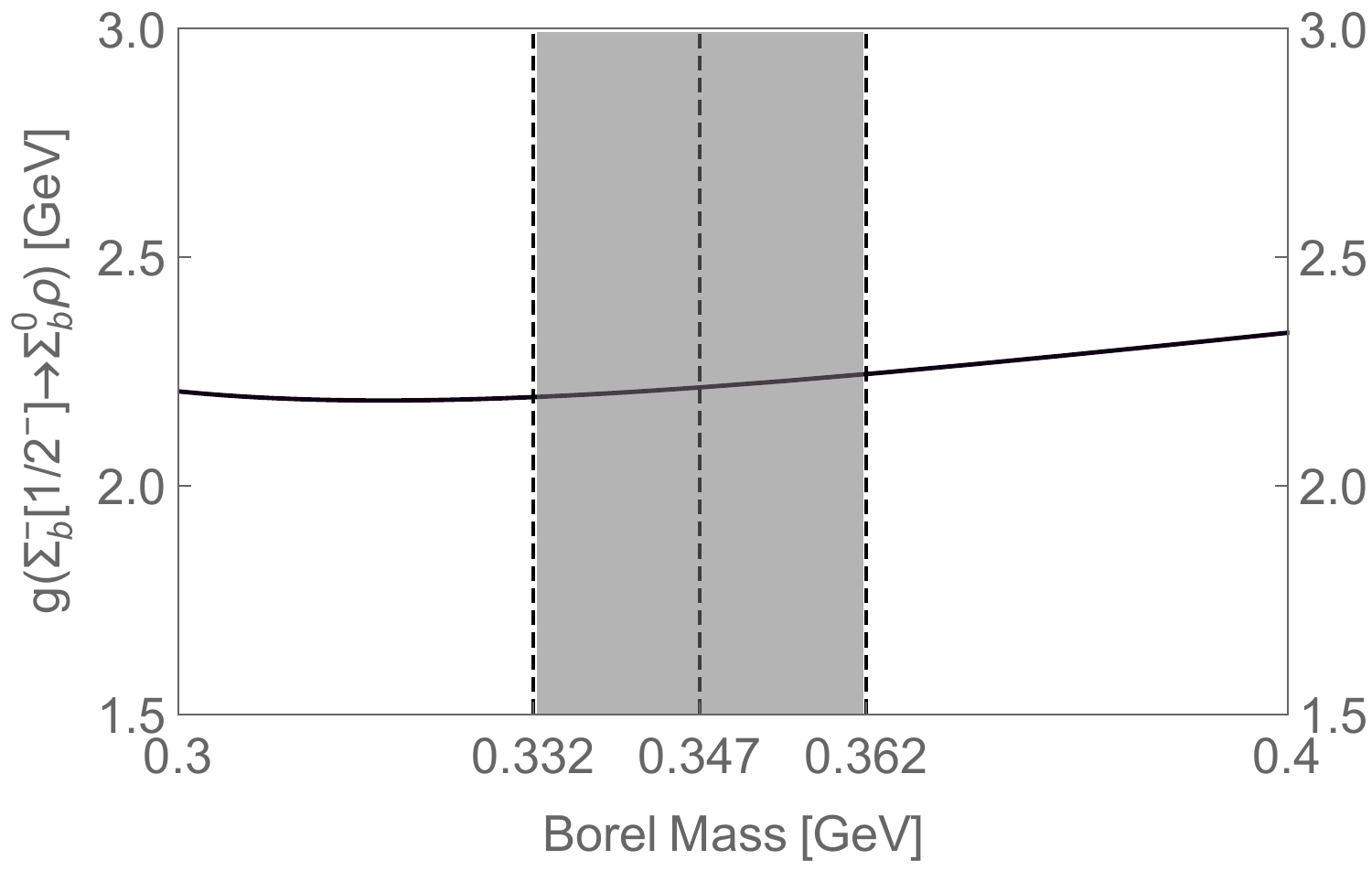}}}
\subfigure[]{
\scalebox{0.37}{\includegraphics{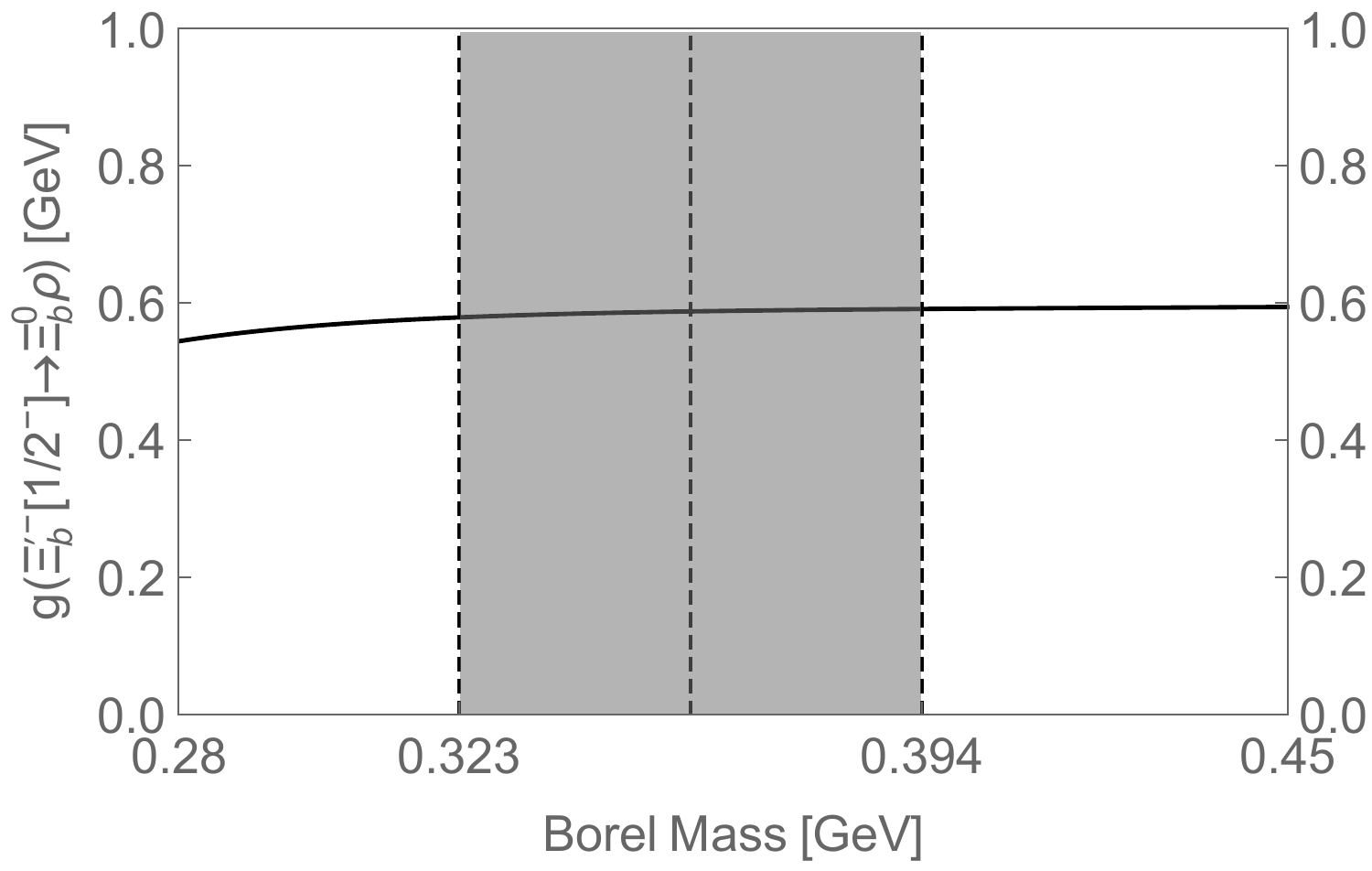}}}~~~
\subfigure[]{
\scalebox{0.37}{\includegraphics{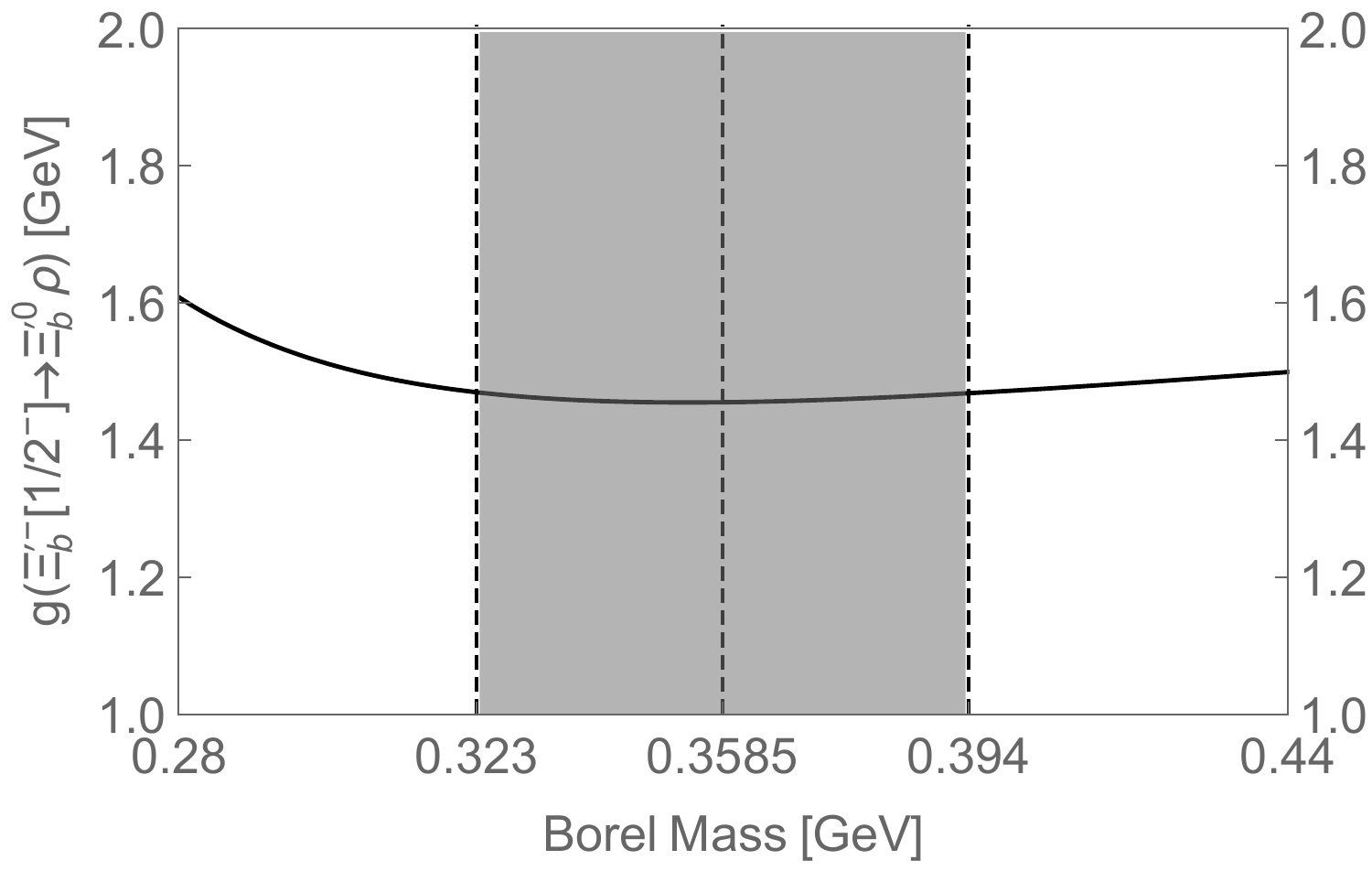}}}
\subfigure[]{
\scalebox{0.33}{\includegraphics{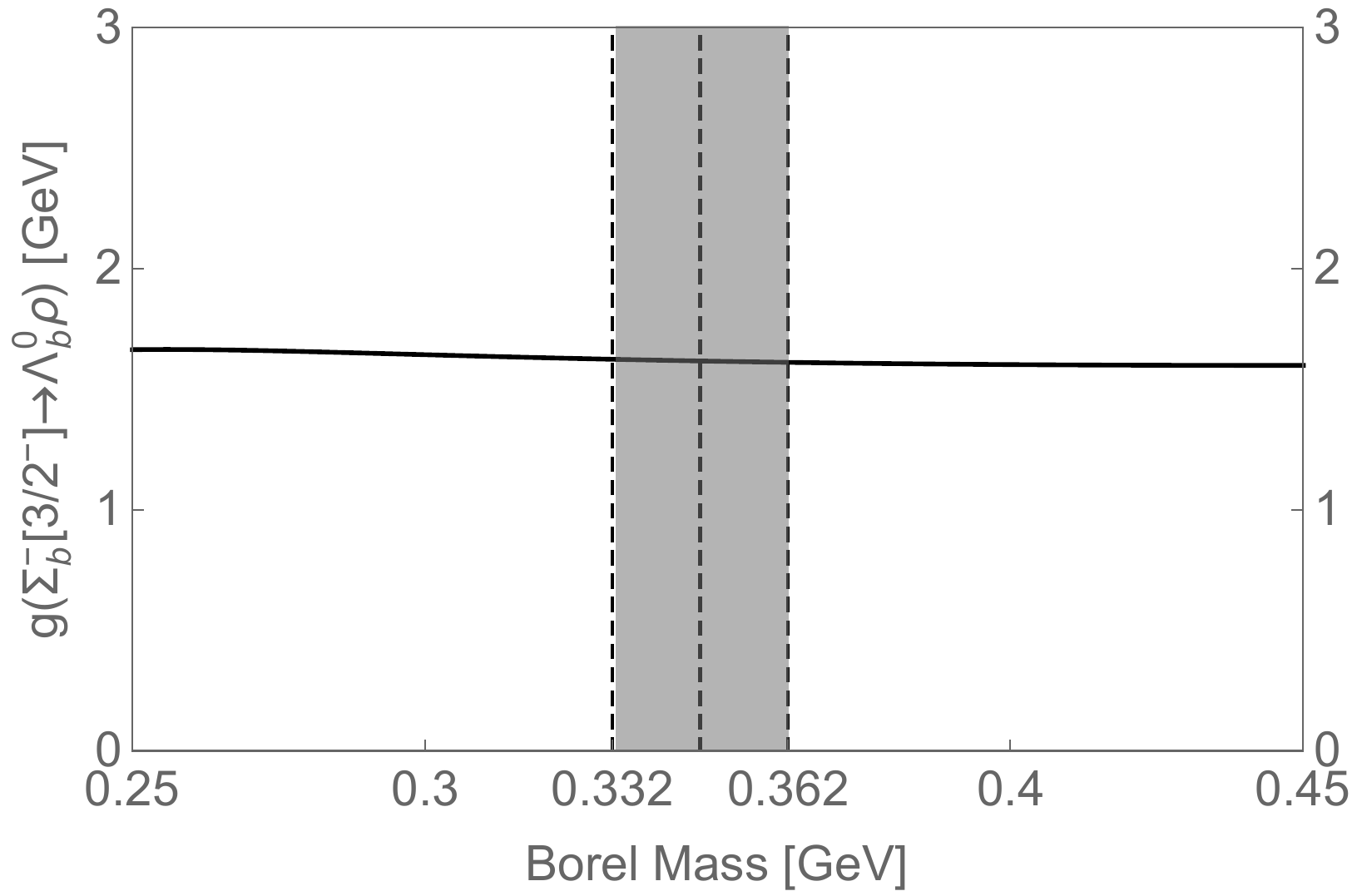}}}~~~
\subfigure[]{
\scalebox{0.37}{\includegraphics{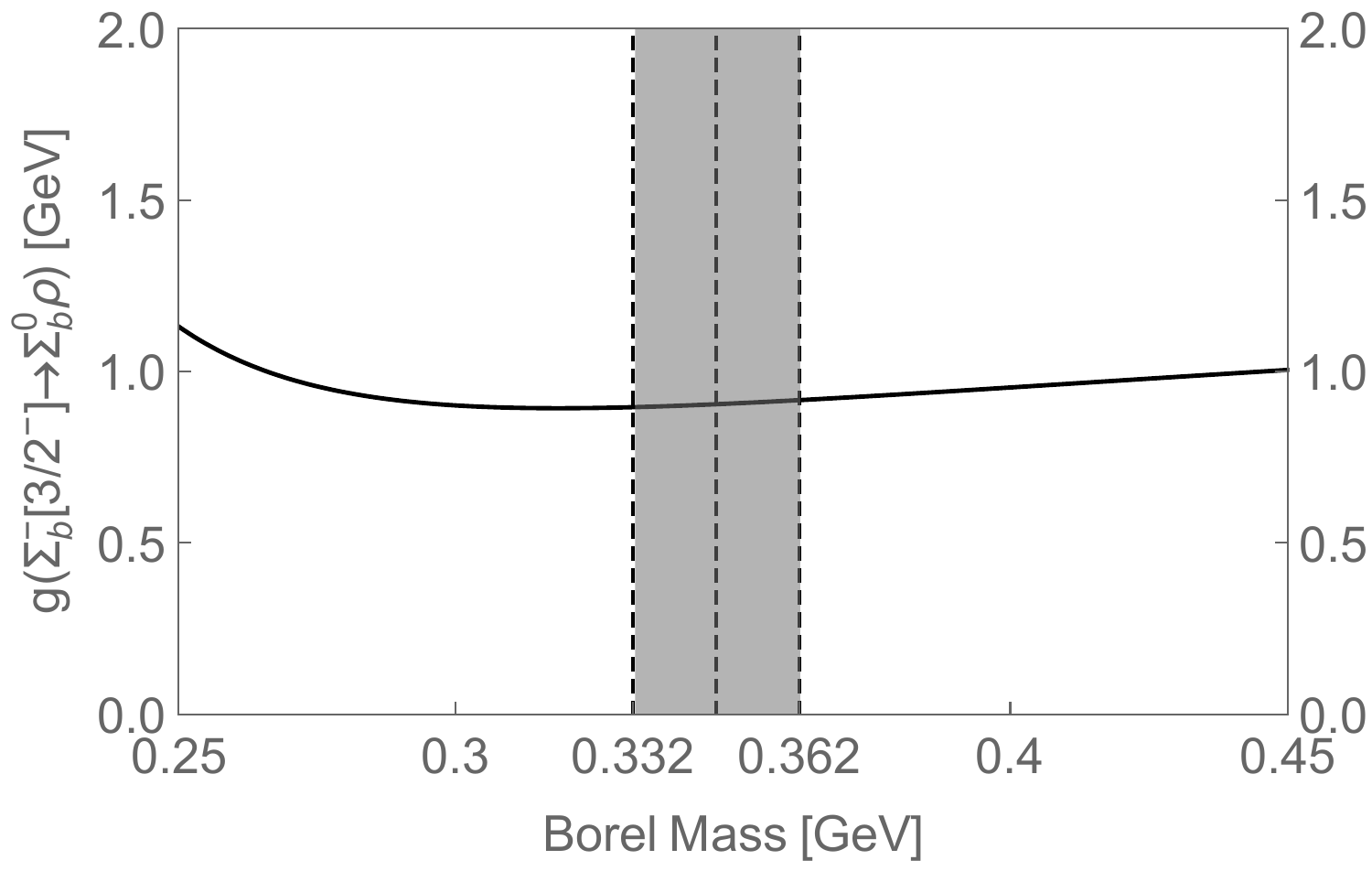}}}
\subfigure[]{
\scalebox{0.35}{\includegraphics{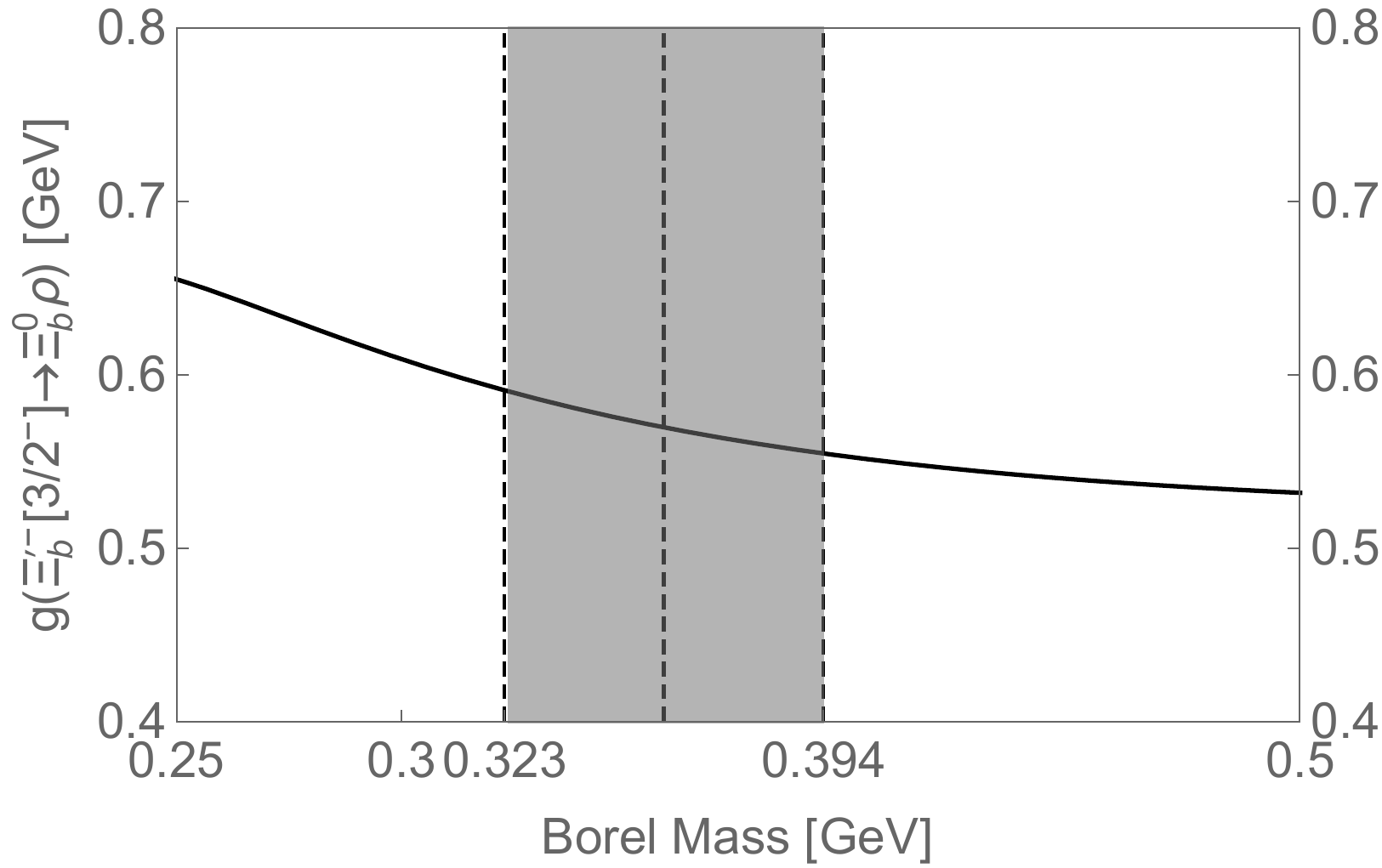}}}~~~
\subfigure[]{
\scalebox{0.35}{\includegraphics{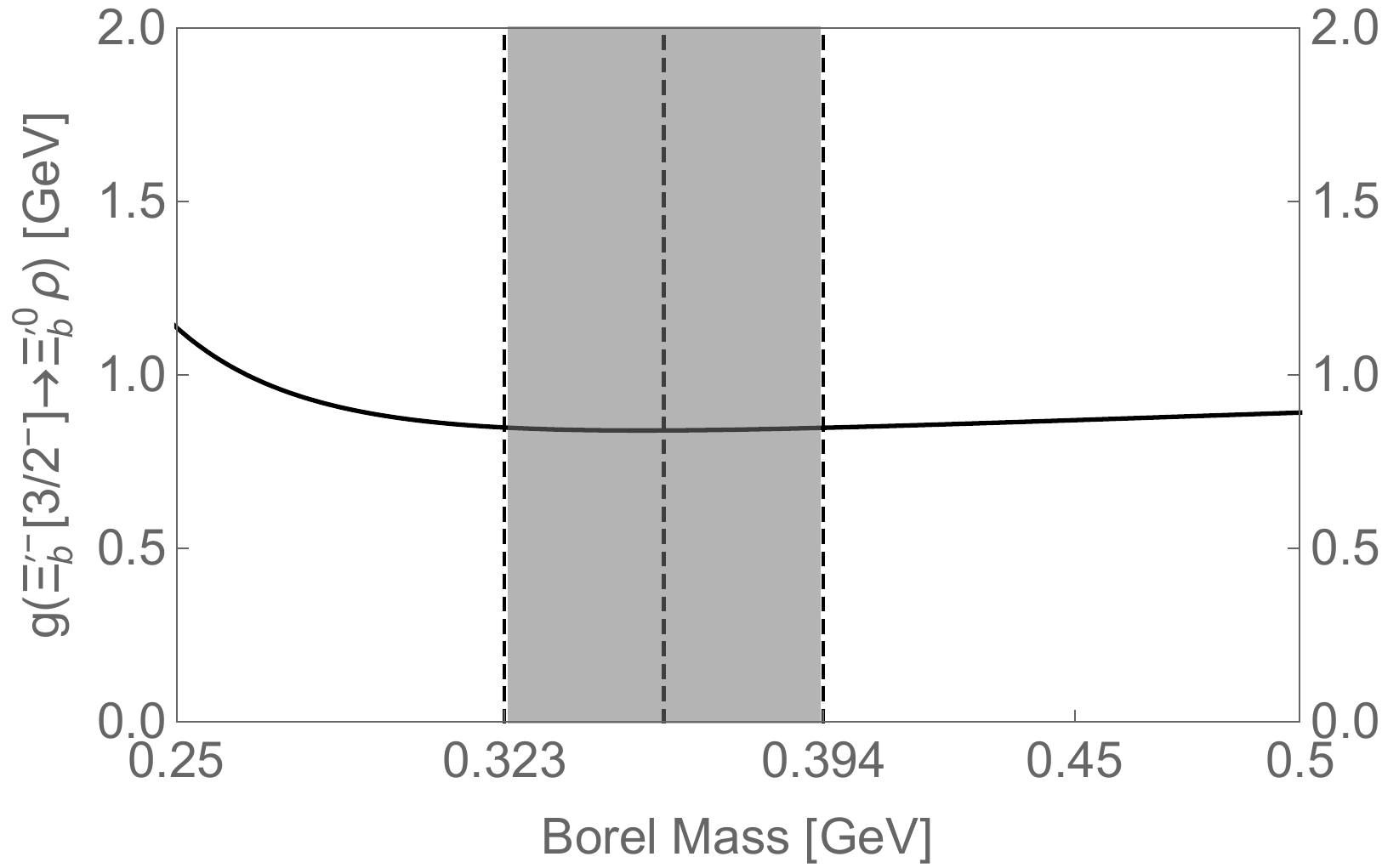}}}
\end{center}
\caption{The coupling constants (a) $g_{\Sigma_b^-[{1\over2}^-] \rightarrow \Lambda_b^0 \rho^-}$, (b) $g_{\Sigma_b^-[{1\over2}^-] \rightarrow \Sigma_b^0\rho^-}$, (c) $g_{\Xi_b^{\prime-}[{1\over2}^-] \rightarrow \Xi_b^0 \rho^-}$, (d) $g_{\Xi_b^{\prime-}[{1\over2}^-] \rightarrow \Xi_b^{\prime0} \rho^-}$, (e) $g_{\Sigma_b^-[{3\over2}^-] \rightarrow \Lambda_b^0 \rho^-}$, (f) $g_{\Sigma_b^-[{3\over2}^-] \rightarrow \Sigma_b^0 \rho^-}$, (g) $g_{\Xi_b^{\prime-}[{3\over2}^-] \rightarrow \Xi_b^0 \rho^-}$, and (h) $g_{\Xi_b^{\prime-}[{3\over2}^-] \rightarrow \Xi_b^{\prime0} \rho^-}$ as functions of the Borel mass $T$. Here the baryons $\Sigma_b({1\over2}^-/{3\over2}^-)$ and $\Xi^\prime_b({1\over2}^-/{3\over2}^-)$ belong to the bottom baryon doublet $[\mathbf{6}_F, 1, 1, \lambda]$, and the working regions for $T$ have been evaluated in mass sum rules~\cite{Chen:2015kpa,Mao:2015gya,Chen:2017sci,Cui:2019dzj} and summarized in Table~\ref{tab:pwaveparameter}.
\label{fig:611lambda}}
\end{figure*}

There are altogether twenty-four non-vanishing decay channels, whose coupling constants are extracted to be
\begin{eqnarray}
\nonumber &(a1)& g_{\Sigma_b[{1\over2}^-]\to \Lambda_b[{1\over2}^+] \rho} = 0.83 \, ,
\\
\nonumber &(a2)& g_{\Sigma_b[{1\over2}^-]\to \Sigma_b[{1\over2}^+] \rho} = 2.21 \, ,
\\
\nonumber &(a3)& g_{\Sigma_b[{1\over2}^-]\to \Sigma_b^{*}[{3\over2}^+] \rho}= 1.28 \, ,
\\
\nonumber &(b1)& g_{\Xi_b^{\prime}[{1\over2}^-]\to \Xi_b[{1\over2}^+] \rho} = 0.59 \, ,
\\
\nonumber &(b2)& g_{\Xi_b^{\prime}[{1\over2}^-]\to \Lambda_b[{1\over2}^+] K^{*}}= 1.60 \, ,
\\
\nonumber &(b3)& g_{\Xi_b^{\prime}[{1\over2}^-]\to \Xi_b^{\prime}[{1\over2}^+] \rho } = 1.45\, ,
\\
\nonumber &(b4)& g_{\Xi_b^{\prime}[{1\over2}^-]\to \Sigma_b[{1\over2}^+]K^{*}}= 0.13 \, ,
\\
\nonumber &(b5)& g_{\Xi_b^{\prime}[{1\over2}^-]\to \Xi_b^{*}[{3\over2}^+]\rho}= 0.84 \, ,
\\
\nonumber &(b6)& g_{\Xi_b^{\prime}[{1\over2}^-]\to\Sigma_b^{*}[{3\over2}^+]K^{*}}= 0.08 \, ,
\\
\nonumber &(c1)& g_{\Omega_b[{1\over2}^-]\to \Xi_b[{1\over2}^+]K^{*}}= 2.25 \, ,
\\
\nonumber &(c2)& g_{\Omega_b[{1\over2}^-]\to \Xi_b^{\prime}[{1\over2}^+]K^{*}}= 0.17 \, ,
\\
\nonumber &(c3)& g_{\Omega_b[{1\over2}^-]\to \Xi_b^{*}[{3\over2}^+]K^{*}}= 0.10 \, ,
\\        &(d1)& g_{\Sigma_b[{3\over2}^-]\to\Lambda_b[{1\over2}^+]\rho}=1.62 \, ,
\label{coupling611lambda}
\\
\nonumber &(d2)& g_{\Sigma_b[{3\over2}^-]\to\Sigma_b[{1\over2}^+]\rho}=0.90 \, ,
\\
\nonumber &(d3)& g_{\Sigma_b[{3\over2}^-]\to\Sigma_b^{*}[{3\over2}^+]\rho}= 1.48 \, ,
\\
\nonumber &(e1)& g_{\Xi_b^{\prime}[{3\over2}^-]\to \Xi_b[{1\over2}^+] \rho}=0.57 \, ,
\\
\nonumber &(e2)& g_{\Xi_b^{\prime}[{3\over2}^-]\to\Lambda_b[{1\over2}^+]K^{*}}=1.31 \, ,
\\
\nonumber &(e3)& g_{\Xi_b^{\prime}[{3\over2}^-] \to \Xi_b^{\prime}[{1\over2}^+] \rho }=0.84 \, ,
\\
\nonumber &(e4)& g_{\Xi_b^{\prime}[{3\over2}^-]\to\Sigma_b[{1\over2}^+]K^{*}}=0.08 \, ,
\\
\nonumber &(e5)& g_{\Xi_b^{\prime}[{3\over2}^-]\to\Xi_b^{*}[{3\over2}^+]\rho}=0.97 \, ,
\\
\nonumber &(e6)& g_{\Xi_b^{\prime}[{3\over2}^-]\to\Sigma_b^{*}[{3\over2}^+]K^{*}}=0.09 \, ,
\\
\nonumber &(f1)& g_{\Omega_b[{3\over2}^-]\to\Xi_b[{1\over2}^+]K^{*}}=1.61 \, ,
\\
\nonumber &(f2)& g_{\Omega_b[{3\over2}^-]\to\Xi_b^{\prime}[{1\over2}^+]K^{*}}=0.10 \, ,
\\
\nonumber &(f3)& g_{\Omega_b[{3\over2}^-]\to\Xi_b^{*}[{3\over2}^+]K^{*}}=0.12 \, .
\end{eqnarray}
Then we compute the three-body decay widths, which are kinematically allowed:
\begin{eqnarray}
\nonumber &(a1)& \Gamma_{\Sigma_b[{1\over2}^-]\to \Lambda_b[{1\over2}^+] \rho\to\Lambda_b[{1\over2}^+]\pi\pi} = 207.4 {\rm~keV} \, ,
\\
\nonumber &(a2)& \Gamma_{\Sigma_b[{1\over2}^-]\to \Sigma_b[{1\over2}^+] \rho\to\Sigma_b[{1\over2}^+]\pi\pi} = 5.7 \times 10^{-2} {\rm~keV} \, ,
\\
\nonumber &(b1)& \Gamma_{\Xi_b^{\prime}[{1\over2}^-]\to \Xi_b[{1\over2}^+] \rho\to\Xi_b[{1\over2}^+]\pi\pi} = 61.0 {\rm~keV} \, ,
\\
\nonumber &(b3)& \Gamma_{\Xi_b^{\prime}[{1\over2}^-]\to \Xi_b^{\prime}[{1\over2}^+] \rho \to \Xi_b^{\prime}[{1\over2}^+] \pi \pi}=0.2 {\rm~keV} \, ,
\\ &(d1)& \Gamma_{\Sigma_b[{3\over2}^-]\to\Lambda_b[{1\over2}^+]\rho\to\Lambda_b[{1\over2}^+]\pi\pi} = 261.1{\rm~keV} \, ,
\label{width611lambda}
\\
\nonumber &(d2)& \Gamma_{\Sigma_b[{3\over2}^-]\to\Sigma_b[{1\over2}^+]\rho\to\Sigma_b[{1\over2}^+]\pi\pi}=3.2\times10^{-3}  {\rm~keV } \, ,
\\
\nonumber &(e1)& \Gamma_{\Xi_b^{\prime}[{3\over2}^-]\to \Xi_b[{1\over2}^+] \rho\to\Xi_b[{1\over2}^+]\pi\pi}=18.8  {\rm~keV} \, ,
\\
\nonumber &(e3)& \Gamma_{\Xi_b^{\prime}[{3\over2}^-] \to \Xi_b^{\prime}[{1\over2}^+] \rho \to \Xi_b^{\prime}[{1 \over 2}^+] \pi\pi}=2.1\times10^{-2} {\rm~MeV} \, .
\end{eqnarray}
We summarize these results in Table~\ref{tab:decay611lambda}. For completeness, we also show the coupling constants as functions of the Borel mass $T$ in Fig.~\ref{fig:611lambda}.

\subsection{The bottom baryon doublet $[\mathbf{6}_F,2,1,\lambda]$}

\begin{table*}[hbt]
\begin{center}
\renewcommand{\arraystretch}{1.5}
\caption{$S$-wave decays of $P$-wave bottom baryons belonging to the doublet $[\mathbf{6}_F, 2, 1, \lambda]$ into ground-state bottom baryons and vector mesons.}
\begin{tabular}{ c | c | c | c}
\hline\hline
 ~~~Decay channels~~~ & ~Coupling constant $g$ ~ & Partial width  & Total width
\\ \hline\hline
(d2) $\Sigma_b({3\over2}^-)\to \Sigma_b({1\over2}^+) \rho\to\Sigma_b({1\over2}^+)\pi\pi$ & $5.90~{^{+3.56}_{-3.08}}$ & $0.14 ~{^{+0.19}_{-0.12}}${\rm~keV} &$0.14~{^{+0.19}_{-0.12}}${\rm~keV}
\\ \hline
(e3) $\Xi_b^{\prime}({3\over2}^-) \to \Xi_b^{\prime}({1\over2}^+) \rho \to \Xi_b^{\prime}({1 \over 2}^+) \pi\pi$ & $4.23~{^{+2.42}_{-2.13}}$ & $0.53~{^{+0.68}_{-0.46}}${\rm~keV}&$0.53~{^{+0.68}_{-0.46}}${\rm~keV}
\\ \hline
(g1) $\Sigma_b({5\over2}^-)\to\Sigma_b^{*}({3\over2}^+)\rho\to\Sigma_b^{*}({3\over2}^+)\pi\pi$ &$3.78~{^{+2.01}_{-1.65}}$&${\left(3~{^{+4}_{-2}}\right)}\times10^{-6}${\rm~keV}&${\left(3~{^{+4}_{-2}}\right)}\times10^{-6}${\rm~keV}
\\ \hline
(h2) $\Xi_b^\prime({5\over2}^-)\to\Xi_b^{*}({3\over2}^+)\rho\to\Xi_b^{*}({3\over2}^+)\pi\pi$ &$4.09~{^{+2.08}_{-1.72}}$&$0.29~{^{+0.32}_{-0.23}}${\rm~keV}&$0.29~{^{+0.32}_{-0.23}}${\rm~keV}
\\ \hline\hline
\end{tabular}
\label{tab:decay621lambda}
\end{center}
\end{table*}

The bottom baryon doublet $[\mathbf{6}_F, 2, 1, \lambda]$ consists of six members: $\Sigma_b({3\over2}^-/{5\over2}^-)$, $\Xi^\prime_b({3\over2}^-/{5\over2}^-)$, and $\Omega_b({3\over2}^-/{5\over2}^-)$. We use the method of light-cone sum rules within HQET to study their decays into ground-state bottom baryons and vector mesons.

\begin{figure*}[htb]
\begin{center}
\subfigure[]{
\scalebox{0.35}{\includegraphics{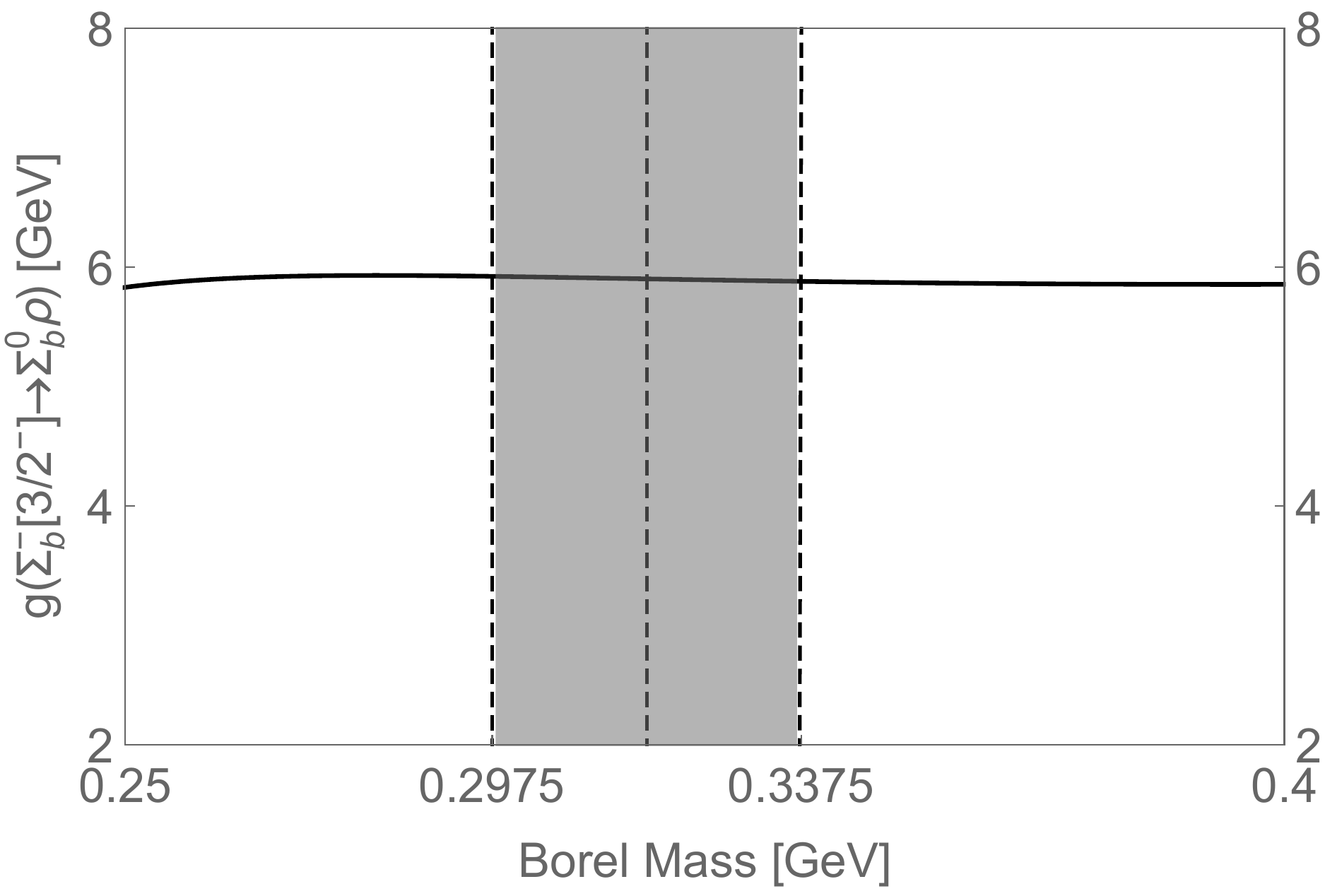}}}
\subfigure[]{
\scalebox{0.35}{\includegraphics{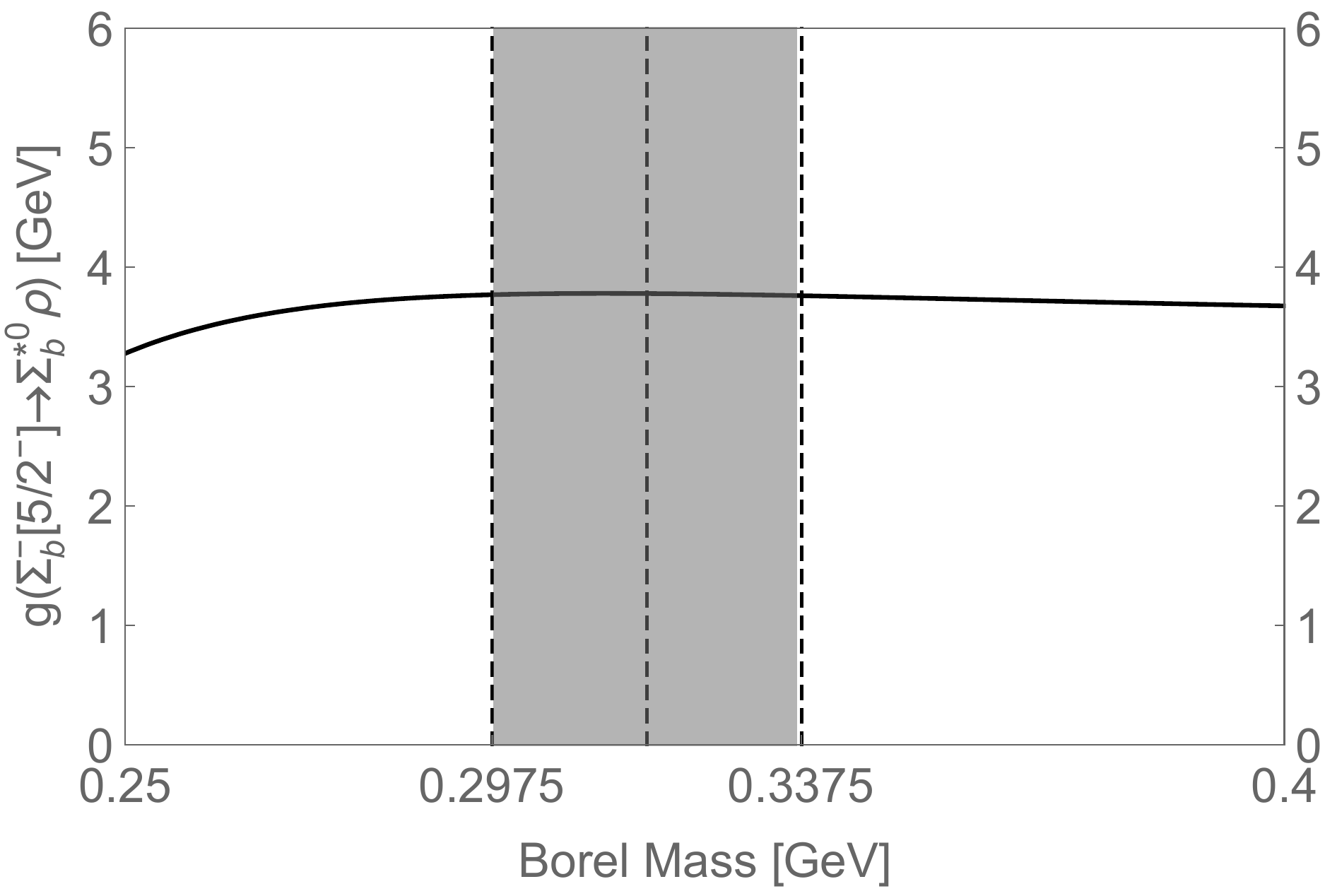}}}
\subfigure[]{
\scalebox{0.35}{\includegraphics{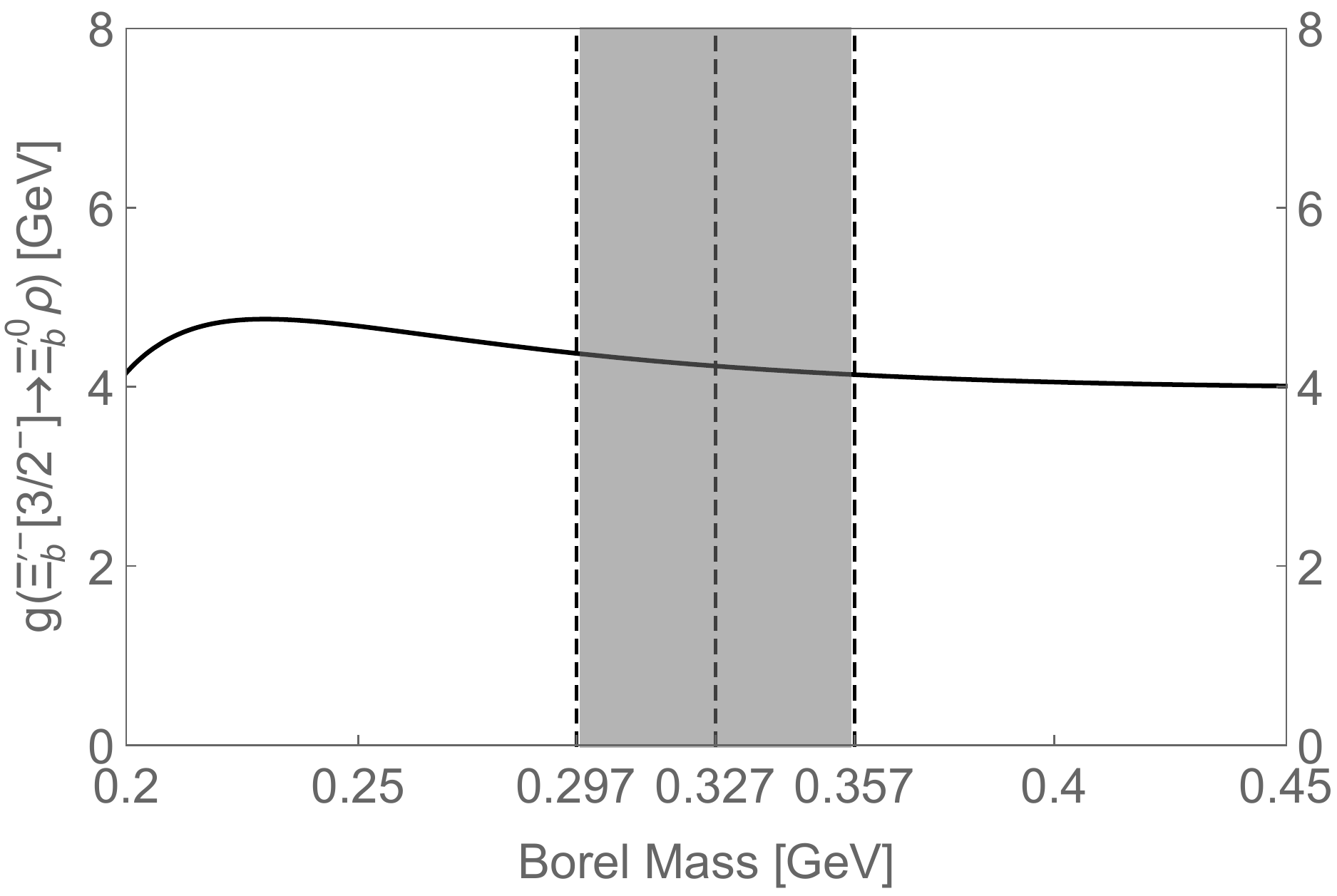}}}
\subfigure[]{
\scalebox{0.35}{\includegraphics{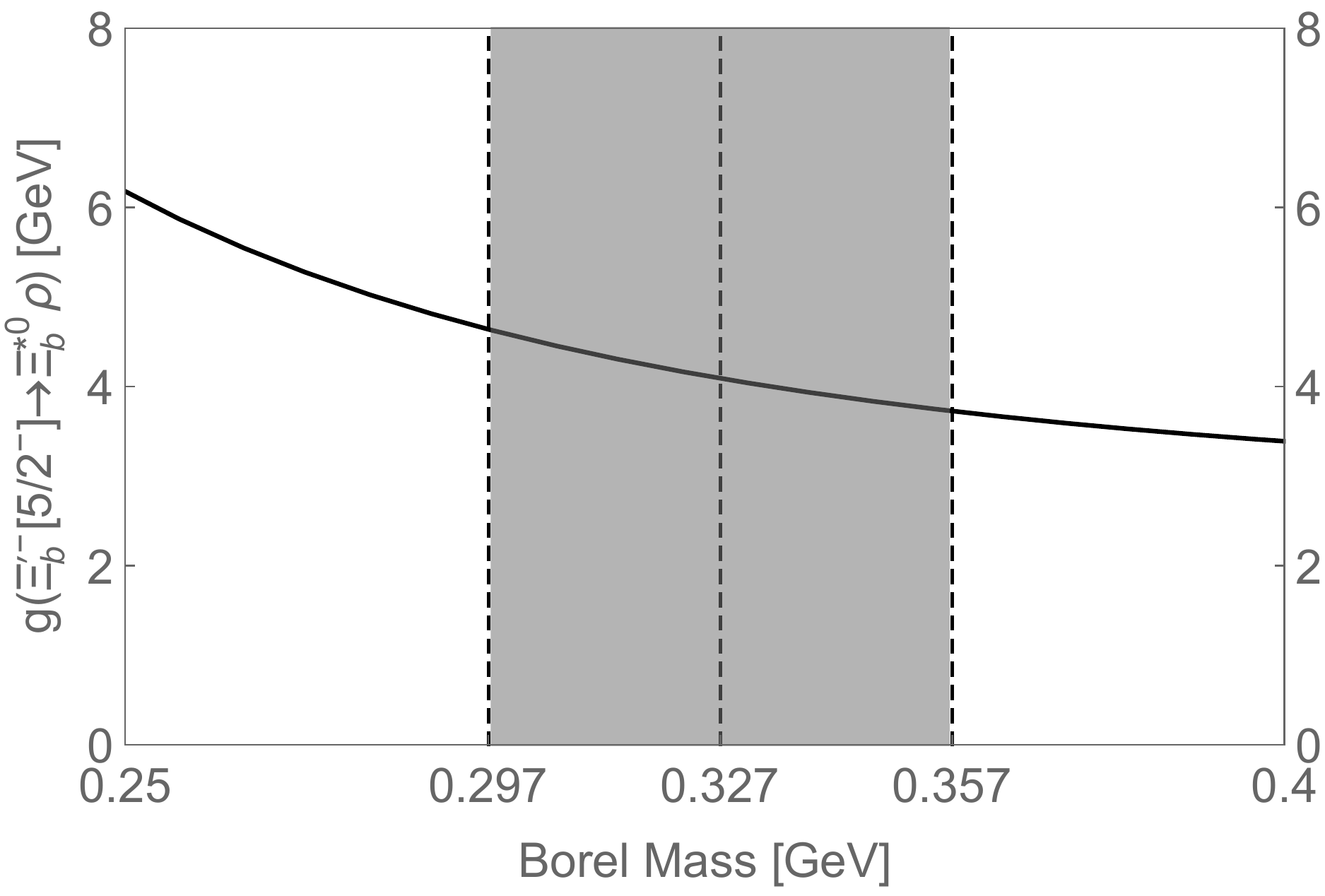}}}
\end{center}
\caption{The coupling constants (a) $g_{\Sigma_b^-[{3\over2}^-] \rightarrow \Sigma_b^0 \rho^-}$, (b) $g_{\Sigma_b^-[{5\over2}^-] \rightarrow \Sigma_b^{*0}\rho^-}$, (c) $g_{\Xi_b^{\prime-}[{3\over2}^-] \rightarrow \Xi_b^{\prime0} \rho^-}$, and (d) $g_{\Xi_b^{\prime-}[{5\over2}^-] \rightarrow \Xi_b^{*0} \rho^-}$ as functions of the Borel mass $T$. Here the baryons $\Sigma_b({3\over2}^-/{5\over2}^-)$ and $\Xi^\prime_b({3\over2}^-/{5\over2}^-)$ belong to the bottom baryon doublet $[\mathbf{6}_F, 2, 1, \lambda]$, and the working regions for $T$ have been evaluated in mass sum rules~\cite{Chen:2015kpa,Mao:2015gya,Chen:2017sci,Cui:2019dzj} and summarized in Table~\ref{tab:pwaveparameter}.
\label{fig:621lambda}}
\end{figure*}

There are altogether twelve non-vanishing decay channels, whose coupling constants are extracted to be
\begin{eqnarray}
\nonumber &(d2)& g_{\Sigma_b[{3\over2}^-]\to\Sigma_b[{1\over2}^+]\rho}=5.90~{^{+3.56}_{-3.08}} \, ,
\\
\nonumber &(d3)& g_{\Sigma_b[{3\over2}^-]\to\Sigma_b^{*}[{3\over2}^+]\rho}= 0.69~{^{+0.41}_{-0.35}} \, ,
\\
\nonumber &(e3)& g_{\Xi_b^{\prime}[{3\over2}^-] \to \Xi_b^{\prime}[{1\over2}^+] \rho }=4.23~{^{+2.42}_{-2.13}} \, ,
\\
\nonumber &(e4)& g_{\Xi_b^{\prime}[{3\over2}^-]\to\Sigma_b[{1\over2}^+]K^{*}}=3.17~{^{+2.37}_{-2.20}} \, ,
\\
\nonumber &(e5)& g_{\Xi_b^{\prime}[{3\over2}^-]\to\Xi_b^{*}[{3\over2}^+]\rho}=0.50~{^{+0.28}_{-0.24}} \, ,
\\
\nonumber &(e6)& g_{\Xi_b^{\prime}[{3\over2}^-]\to\Sigma_b^{*}[{3\over2}^+]K^{*}}=0.40~{^{+0.28}_{-0.26}} \, ,
\\
&(f2)& g_{\Omega_b[{3\over2}^-]\to\Xi_b^{\prime}[{1\over2}^+]K^{*}}=4.56~{^{+3.19}_{-2.97}} \, ,
\label{coupling621lambda}
\\
\nonumber &(f3)& g_{\Omega_b[{3\over2}^-]\to\Xi_b^{*}[{3\over2}^+]K^{*}}=0.60~{^{+0.38}_{-0.35}} \, ,
\\
\nonumber &(g1)& g_{\Sigma_b[{5\over2}^-]\to\Sigma_b^{*}[{3\over2}^+]\rho}= 3.78~{^{+2.01}_{-1.65}} \, ,
\\
\nonumber &(h1)& g_{\Xi_b^{\prime}[{5\over2}^-]\to\Xi_b^{*}[{3\over2}^+]\rho}=4.09~{^{+2.08}_{-1.72}} \, ,
\\
\nonumber &(h2)& g_{\Xi_b^{\prime}[{5\over2}^-]\to\Sigma_b^{*}[{3\over2}^+]K^{*}}=2.14~{^{+1.33}_{-1.18}} \, ,
\\
\nonumber &(i1)& g_{\Omega_b[{5\over2}^-]\to\Xi_b^{*}[{3\over2}^+]K^{*}}=6.36~{^{+3.75}_{-2.72}} \, .
\end{eqnarray}
Then we compute the three-body decay widths, which are kinematically allowed:
\begin{eqnarray}
\nonumber &(d2)& \Gamma_{\Sigma_b[{3\over2}^-]\to \Sigma_b[{1\over2}^+] \rho\to\Sigma_b[{1\over2}^+]\pi\pi} = 0.14~{^{+0.19}_{-0.12}} {\rm~keV} \, ,
\\
&(e3)& \Gamma_{\Xi_b^{\prime}[{3\over2}^-] \to \Xi_b^{\prime}[{1\over2}^+] \rho \to \Xi_b^{\prime}[{1 \over 2}^+] \pi\pi}=0.53~{^{+0.68}_{-0.46}} {\rm~keV} \, ,
\label{width621ambda}
\\
\nonumber
&(g1)& \Gamma_{\Sigma_b[{5\over2}^-]\to\Sigma_b^{*}[{3\over2}^+]\rho\to\Sigma_b^{*}[{3\over2}^+]\pi\pi} = \left(3~{^{+4}_{-2}}\right)\times10^{-6} {\rm~keV} \, ,
\\
\nonumber &(h2)& \Gamma_{\Xi_b^{\prime}[{5\over2}^-]\to\Xi_b^{*}[{3\over2}^+]\rho\to\Xi_b^{*}[{3\over2}^+]\pi\pi}=0.29~{^{+0.32}_{-0.23}}{\rm~keV } \, .
\end{eqnarray}
We summarize these results in Table~\ref{tab:decay621lambda}. For completeness, we also show the coupling constants as functions of the Borel mass $T$ in Fig.~\ref{fig:621lambda}.

%
\section{Summary and Discussions}\label{sec:summary}
%

To summarize this paper, we have used the method of light-cone sum rules within heavy quark effective theory to study decay properties of $P$-wave bottom baryons belonging to the flavor $\mathbf{6}_F$ representation.
We have studied their $S$-wave decays into ground-state bottom baryons and vector mesons. The possible decay channels are given in Eqs.~(\ref{eq:couple1}-\ref{eq:couple28}), and the extracted decay widths are listed 
in Tables~\ref{tab:decay610rho}, \ref{tab:decay601lambda}, \ref{tab:decay611lambda}, and \ref{tab:decay621lambda}.
These results are obtained separately for the four bottom baryon multiplets of flavor $\mathbf{6}_F$: $[\mathbf{6}_F, 1, 0, \rho]$, $[\mathbf{6}_F, 0, 1, \lambda]$, $[\mathbf{6}_F, 1, 1, \lambda]$, and $[\mathbf{6}_F, 2, 1, \lambda]$.

In Ref.~\cite{Cui:2019dzj} we have studied the mass spectrum and pionic decay properties of the $\Sigma_{b}(6097)$ and $\Xi_{b}(6227)$~\cite{Aaij:2018yqz,Aaij:2018tnn}. Our results suggest that they can be well interpreted as $P$-wave bottom baryons of $J^P = 3/2^-$, belonging to the bottom baryon doublet $[\mathbf{6}_F, 2, 1, \lambda]$. This doublet contains altogether six bottom baryons, $\Sigma_b({3\over2}^-/{5\over2}^-)$, $\Xi^\prime_b({3\over2}^-/{5\over2}^-)$, and $\Omega_b({3\over2}^-/{5\over2}^-)$. In the present study we further investigate their $S$-wave decays into ground-state bottom baryons and vector mesons, and extract:
\begin{eqnarray*}
\nonumber &(d2)& \Gamma_{\Sigma_b[{3\over2}^-]\to \Sigma_b[{1\over2}^+] \rho\to\Sigma_b[{1\over2}^+]\pi\pi} = 0.14~{^{+0.19}_{-0.12}} {\rm~keV} \, ,
\\
&(e3)& \Gamma_{\Xi_b^{\prime}[{3\over2}^-] \to \Xi_b^{\prime}[{1\over2}^+] \rho \to \Xi_b^{\prime}[{1 \over 2}^+] \pi\pi}=0.53~{^{+0.68}_{-0.46}} {\rm~keV} \, ,
\\
\nonumber
&(g1)& \Gamma_{\Sigma_b[{5\over2}^-]\to\Sigma_b^{*}[{3\over2}^+]\rho\to\Sigma_b^{*}[{3\over2}^+]\pi\pi} = \left(3~{^{+4}_{-2}}\right)\times10^{-6} {\rm~keV} \, ,
\\
\nonumber &(h2)& \Gamma_{\Xi_b^{\prime}[{5\over2}^-]\to\Xi_b^{*}[{3\over2}^+]\rho\to\Xi_b^{*}[{3\over2}^+]\pi\pi}=0.29~{^{+0.32}_{-0.23}}{\rm~keV } \, .
\end{eqnarray*}
Hence, these three $J^P = 5/2^-$ states are probably quite narrow, because their $S$-wave decays into ground-state bottom baryons and pseudoscalar mesons can not happen, and widths of the following $D$-wave decays are also calculated to be zero in Ref.~\cite{Cui:2019dzj}:
\begin{eqnarray}
\nonumber &(w^\prime)& \Gamma_{\Sigma_b^{-}({5/2}^-) \rightarrow \Lambda_b^{0} \pi^-} = 0 \, ,
\\
\nonumber &(x^\prime)& \Gamma_{\Xi_b^{\prime-}({5/2}^-) \rightarrow \Xi_b^{0} \pi^-} = 0 \, ,
\\
\nonumber &(y^\prime)& \Gamma_{\Xi_b^{\prime-}({5/2}^-) \rightarrow \Lambda_b^{0} K^-} = 0 \, ,
\\
\nonumber &(z^\prime)& \Gamma_{\Omega_b^{-}({5/2}^-) \rightarrow \Xi_b^{0} K^-} = 0 \, .
\end{eqnarray}

We suggest the LHCb and Belle/Belle-II experiments to search for these three narrow states. Especially, we propose to search for the $\Xi_b({5/2}^-)$, that is the $J^P = 5/2^-$ partner state of the $\Xi_{b}(6227)$, in the $\Xi_b({5/2}^-) \to \Xi_b^{*}\rho \to \Xi_b^{*}\pi\pi$ decay process. Its mass is $12 \pm 5$~MeV larger than that of the $\Xi_{b}(6227)$.

To end this work, we note that in the present study we have studied $S$-wave decays of flavor $\mathbf{6}_F$ $P$-wave heavy baryons into ground-state heavy baryons and vector mesons, which is actually a complement to Ref.~\cite{Chen:2017sci}, where we studied $S$-wave decays of flavor $\mathbf{\bar 3}_F$ $P$-wave heavy baryons into ground-state heavy baryons together with pseudoscalar and vector mesons, and $S$-wave decays of flavor $\mathbf{6}_F$ $P$-wave heavy baryons into ground-state heavy baryons together with pseudoscalar mesons. To make a complete QCD sum rule studies of $P$-wave heavy baryons within HQET, we still need to systematically study their $D$-wave and radiative decay properties, which is currently under investigation.

\begin{acknowledgements}
This project is supported by
the National Natural Science Foundation of China under Grant No.~11722540,
the Fundamental Research Funds for the Central Universities,
Grants-in-Aid for Scientific Research (No.~JP17K05441 (C)),
Grants-in-Aid for Scientific Research on Innovative Areas (No.~18H05407),
and
the Foundation for Young Talents in College of Anhui Province (Grant No.~gxyq2018103).
\end{acknowledgements}

\appendix

%
\section{Parameters of $S/P$-wave bottom baryons and $S$-wave mesons}
\label{sec:sbottom}
%

We list masses of ground-state bottom baryons used in the present study, taken from PDG~\cite{pdg}:
\begin{eqnarray}
   \nonumber        \Lambda_{b}(1/2^+)  ~:~ m&=&5619.60 \mbox{ MeV} \, ,
\\ \nonumber        \Xi_{b}(1/2^+)  ~:~ m&=&5793.20 \mbox{ MeV} \, ,
\\ \nonumber        \Sigma_{b}(1/2^+)    ~:~ m&=&5813.4 \mbox{ MeV} \, ,
\\ \nonumber        \Sigma_{b}^{*}(3/2^+)    ~:~ m&=&5833.6 \mbox{ MeV} \, , \,
\\ \nonumber        \Xi_{b}^{\prime}(1/2^+)  ~:~ m&=&5935.02 \mbox{ MeV} \, ,
\\ \nonumber        \Xi_{b}^{*}(3/2^+)  ~:~ m&=&5952.6 \mbox{ MeV} \, , \,
\\ \nonumber        \Omega_b(1/2^+)  ~:~ m&=&6046.1 \mbox{ MeV}\, ,
\\ \nonumber        \Omega_b^*(3/2^+)  ~:~ m&=&6063 \mbox{ MeV} \, .
\end{eqnarray}
Their QCD sum rule parameters can be found in Refs.~\cite{Liu:2007fg,Chen:2017sci,Cui:2019dzj}.

We list masses of $P$-wave bottom baryons used in the present study, taken from the LHCb experiments~\cite{Aaij:2018yqz,Aaij:2018tnn} as well as our previous QCD sum rule studies~\cite{Chen:2015kpa,Mao:2015gya,Cui:2019dzj}:
\begin{eqnarray}
\nonumber M_{\Sigma_b(1/2^-)} = M_{\Sigma_b(3/2^-)} &=& 6096.9~{\rm MeV}\, ,
\\ \nonumber M_{\Sigma_b(5/2^-)} -  M_{\Sigma_b(3/2^-)} &=& 13~{\rm MeV} \, ,
\\ \nonumber M_{\Xi_b^{\prime}(1/2^-)} = M_{\Xi_b^{\prime}(3/2^-)} &=& 6226.9~{\rm MeV}\, ,
\\ \nonumber M_{\Xi^\prime_b(5/2^-)} -  M_{\Xi^\prime_b(3/2^-)} &=& 12~{\rm MeV} \, ,
\\ \nonumber M_{\Omega_b(1/2^-)} = M_{\Omega_b(3/2^-)} &=& 6460~{\rm MeV} \, ,
\\ M_{\Omega_b(5/2^-)} -  M_{\Omega_b(3/2^-)} &=& 11~{\rm MeV} \, .
\end{eqnarray}
Their QCD sum rule parameters can be found in Table~\ref{tab:pwaveparameter}.
\begin{table*}[hbt]
\begin{center}
\renewcommand{\arraystretch}{1.4}
\caption{Parameters of $P$-wave bottom baryons belonging to the bottom baryon multiplets $[\mathbf{6}_F, 1, 0, \rho]$, $[\mathbf{6}_F, 0, 1, \lambda]$, $[\mathbf{6}_F, 1, 1, \lambda]$ and $[\mathbf{6}_F, 2, 1, \lambda]$. Detailed discussions can be found in Refs.~\cite{Chen:2015kpa,Mao:2015gya,Cui:2019dzj}.}
\begin{tabular}{c | c | c | c | c | c c | c | c}
\hline\hline
\multirow{2}{*}{Multiplets} & \multirow{2}{*}{B} & $\omega_c$ & Working region & $\overline{\Lambda}$ & Baryons & Mass & ~Difference~ & $f$
\\                                              &  & (GeV)      & (GeV)                & (GeV)                              & ($j^P$)       & (GeV)      & (MeV)        & (GeV$^{4}$)
\\ \hline\hline
\multirow{6}{*}{$[\mathbf{6}_F, 1, 0, \rho]$}
& \multirow{2}{*}{$\Sigma_b$} & \multirow{2}{*}{1.87} & \multirow{2}{*}{$0.31< T < 0.34$} & \multirow{2}{*}{$1.35 \pm 0.09$} & $\Sigma_b(1/2^-)$ & $6.10 \pm 0.11$ & \multirow{2}{*}{$3 \pm 1$} & $0.087 \pm 0.018$
\\ \cline{6-7}\cline{9-9}
& & & & & $\Sigma_b(3/2^-)$ & $6.10 \pm 0.10$ & &$0.050 \pm 0.011$
\\ \cline{2-9}
& \multirow{2}{*}{$\Xi^\prime_b$} & \multirow{2}{*}{2.02} & \multirow{2}{*}{$0.29< T < 0.36$} & \multirow{2}{*}{$1.49 \pm 0.09$} & $\Xi^\prime_b(1/2^-)$ & $6.24 \pm 0.11$ & \multirow{2}{*}{$3 \pm 1$} & $0.080 \pm 0.016$
\\ \cline{6-7}\cline{9-9}
& & & & & $\Xi^\prime_b(3/2^-)$ & $6.24 \pm 0.11$ & &$0.046 \pm 0.009$
\\ \cline{2-9}
& \multirow{2}{*}{$\Omega_b$} & \multirow{2}{*}{2.17} & \multirow{2}{*}{$0.33< T < 0.38$} & \multirow{2}{*}{$1.67 \pm 0.09$} & $\Omega_b(1/2^-)$ & $6.42 \pm 0.11$ & \multirow{2}{*}{$3 \pm 1$} & $0.155 \pm 0.030$
\\ \cline{6-7}\cline{9-9}
& & & & & $\Omega_b(3/2^-)$ & $6.42 \pm 0.11$ & &$0.090 \pm 0.017$
\\ \hline
\multirow{3}{*}{$[\mathbf{6}_F, 0, 1, \lambda]$} & $\Sigma_b$ & 1.75 & $0.30< T < 0.33$ & $1.29 \pm 0.08$ & $\Sigma_b(1/2^-)$ & $6.09 \pm 0.10$ & -- & $0.085 \pm 0.017$
\\ \cline{2-9}
                                                 & $\Xi^\prime_b$ & 1.90 & $0.30< T < 0.34$ & $1.44 \pm 0.08$ & $\Xi^\prime_b(1/2^-)$ & $6.25 \pm 0.10$ & -- & $0.077 \pm 0.016$
\\ \cline{2-9}
                                                 & $\Omega_b$ & 2.05 & $0.29< T < 0.35$ & $1.59 \pm 0.08$ & $\Omega_b(1/2^-)$ & $6.40 \pm 0.11$ & -- & $0.143 \pm 0.030$
\\ \hline
\multirow{6}{*}{$[\mathbf{6}_F, 1, 1, \lambda]$}
& \multirow{2}{*}{$\Sigma_b$} & \multirow{2}{*}{1.95} & \multirow{2}{*}{$0.33< T < 0.36$} & \multirow{2}{*}{$1.28 \pm 0.12$} & $\Sigma_b(1/2^-)$ & $6.10 \pm 0.13$ & \multirow{2}{*}{$7 \pm 3$} & $0.078 \pm 0.018$
\\ \cline{6-7}\cline{9-9}
& & & & & $\Sigma_b(3/2^-)$ & $6.10 \pm 0.13$ & &$0.045 \pm 0.010$
\\ \cline{2-9}
& \multirow{2}{*}{$\Xi^\prime_b$} & \multirow{2}{*}{2.10} & \multirow{2}{*}{$0.32< T < 0.39$} & \multirow{2}{*}{$1.47 \pm 0.11$} & $\Xi^\prime_b(1/2^-)$ & $6.33 \pm 0.13$ & \multirow{2}{*}{$5 \pm 3$} & $0.085 \pm 0.017$
\\ \cline{6-7}\cline{9-9}
& & & & & $\Xi^\prime_b(3/2^-)$ & $6.33 \pm 0.13$ & &$0.049 \pm 0.010$
\\ \cline{2-9}
& \multirow{2}{*}{$\Omega_b$} & \multirow{2}{*}{2.25} & \multirow{2}{*}{$0.31< T < 0.42$} & \multirow{2}{*}{$1.63 \pm 0.14$} & $\Omega_b(1/2^-)$ & $6.50 \pm 0.16$ & \multirow{2}{*}{$4 \pm 2$} & $0.171 \pm 0.036$
\\ \cline{6-7}\cline{9-9}
& & & & & $\Omega_b(3/2^-)$ & $6.51 \pm 0.16$ & &$0.099 \pm 0.021$
\\ \hline
\multirow{6}{*}{$[\mathbf{6}_F, 2, 1, \lambda]$}
& \multirow{2}{*}{$\Sigma_b$} & \multirow{2}{*}{1.84} & \multirow{2}{*}{$0.30< T < 0.34$} & \multirow{2}{*}{$1.29 \pm 0.09$} & $\Sigma_b(3/2^-)$ & $6.10 \pm 0.12$ & \multirow{2}{*}{$13 \pm 5$} & $0.102 \pm 0.022$
\\ \cline{6-7}\cline{9-9}
& & & & & $\Sigma_b(5/2^-)$ & $6.11 \pm 0.12$ & &$0.045 \pm 0.010$
\\ \cline{2-9}
& \multirow{2}{*}{$\Xi^\prime_b$} & \multirow{2}{*}{1.99} & \multirow{2}{*}{$0.30< T < 0.36$} & \multirow{2}{*}{$1.45 \pm 0.09$} & $\Xi^\prime_b(3/2^-)$ & $6.27 \pm 0.12$ & \multirow{2}{*}{$12 \pm 5$} & $0.099 \pm 0.021$
\\ \cline{6-7}\cline{9-9}
& & & & & $\Xi^\prime_b(5/2^-)$ & $6.29 \pm 0.11$ & &$0.044 \pm 0.009$
\\ \cline{2-9}
& \multirow{2}{*}{$\Omega_b$} & \multirow{2}{*}{2.14} & \multirow{2}{*}{$0.32< T < 0.38$} & \multirow{2}{*}{$1.62 \pm 0.09$} & $\Omega_b(3/2^-)$ & $6.46 \pm 0.12$ & \multirow{2}{*}{$11 \pm 5$} & $0.194 \pm 0.038$
\\ \cline{6-7}\cline{9-9}
& & & & & $\Omega_b(5/2^-)$ & $6.47 \pm 0.12$ & &$0.087 \pm 0.017$
\\ \hline \hline
\end{tabular}
\label{tab:pwaveparameter}
\end{center}
\end{table*}

We list masses and decay widths of the pseudoscalar and vector mesons used in the present study, taken from PDG~\cite{pdg}:
\begin{eqnarray}
 \pi(0^-) ~:~ m&=& 138.04 {\rm~MeV} \, ,
\\ \nonumber K(0^-) ~:~ m&=& 495.65 {\rm~MeV} \, ,
\\ \nonumber \rho(1^-) ~:~ m&=& 775.21 {\rm~MeV} \, ,\,
\\ \nonumber                                \Gamma&=& 148.2 {\rm~MeV} \, ,\,  {g}_{\rho \pi \pi} = 5.94 \, ,
\\  \nonumber         K^*(1^-) ~:~ m&=& 893.57 {\rm~MeV} \, ,\,
\\                                 \Gamma&=& 49.1 {\rm~MeV} \, ,\,  {g}_{K^* K \pi} = 3.20 \, ,
\end{eqnarray}
where the two coupling constants ${g}_{\rho \pi \pi}$ and ${g}_{K^* K \pi}$ are evaluated using the experimental decay widths of the $\rho$ and $K^*$~\cite{pdg} through the following Lagrangians
\begin{eqnarray}
\nonumber \mathcal{L}_{\rho \pi \pi} &=& {g}_{\rho \pi \pi} \times \left( \rho_\mu^0 \pi^+ \partial^\mu \pi^- - \rho_\mu^0 \pi^- \partial^\mu \pi^+ \right) + \cdots \, ,
\\ \nonumber \mathcal{L}_{K^* K \pi} &=& {g}_{K^* K \pi} K^{*+}_\mu \times \left( K^- \partial^\mu \pi^0 - \partial^\mu K^- \pi^0 \right) + \cdots \, .
\\
\end{eqnarray}

\section{Sum rule equations}
\label{sec:othersumrule}

In this appendix we show several examples of sum rule equations, which are used to extract $S$-wave decays of $P$-wave bottom baryons into ground-state bottom baryons and vector mesons.

\begin{widetext}
The sum rule for $\Xi_b^{\prime-}[{1\over2}^-]$ belonging to $[\mathbf{6}_F, 0 , 1, \lambda]$ is
\begin{eqnarray}
&& G_{\Xi_b^{\prime-}[{1\over2}^-] \rightarrow \Xi_b^{*0}\rho^-} (\omega, \omega^\prime)
= { g_{\Xi_b^{\prime-}[{1\over2}^-] \rightarrow \Xi_b^{*0}\rho^-} f_{\Xi_b^-[{1\over2}^-]} f_{\Xi_b^{*0}} \over (\bar \Lambda_{\Xi_b^-[{1\over2}^-]} - \omega^\prime) (\bar \Lambda_{\Xi_b^{*0}} - \omega)}
\\ \nonumber &=& \int_0^\infty dt \int_0^1 du e^{i(1-u)\omega^\prime t} e^{iu\omega t}\times 4\times \Big(\frac{if_\rho^{||} m_\rho}{6\pi^2 t^4}\phi_{2;\rho}^{||}(u)-\frac{2i f_\rho^{||} m_\rho}{3\pi^3 t^4}\phi_{3;\rho}^\perp(u)
\\ \nonumber &&-\frac{if_\rho^{||} m_\rho^3}{12\pi^2 t^4(v\cdot q)^2}\phi_{2;\rho}^{||}(u)+\frac{if_\rho^{||} m_\rho^3}{6\pi^2 t^4(v\cdot q)^2}\phi_{3;\rho}^{||}(u)-\frac{if_\rho^{||} m_\rho^3}{12\pi^2 t^4(v\cdot q)^2}\psi_{4;\rho}^{||}(u)+\frac{f_\rho^\perp m_s m_\rho^2}{12\pi^2 t^3 v\cdot q}\phi_{2;\rho}^\perp(u)
\\ \nonumber &&-\frac{f_\rho^\perp m_s m_\rho^2}{12\pi^2 t^3 v\cdot q}\psi_{4;\rho}^\perp(u)-\frac{f_\rho^{||} m_\rho v\cdot q}{12\pi^2 t^3}\psi_{3;\rho}^\perp(u)+\frac{if_\rho^{||} m_\rho^3}{96\pi^2 t^2}\phi_{4;\rho}^{||}(u)-\frac{if_\rho^\perp m_s m_\rho^2}{24\pi^2 t^2}\psi_{3;\rho}^{||}(u)
\\ \nonumber &&-\frac{if_\rho^\perp m_s u (v\cdot q)^2}{12\pi^2 t^2}\phi_{2;\rho}^\perp(u)+\frac{i f_\rho^{||} m_\rho u (v\cdot q)^2}{24\pi^2 t^2}\psi_{3;\rho}^\perp(u)+\frac{f_\rho^\perp m_\rho^2}{36 t v\cdot q}\langle\bar s s\rangle\phi_{2;\rho}^\perp (u)-\frac{f_\rho^\perp m_\rho^2}{36 v\cdot q}\langle\bar s s\rangle\psi_{4;\rho}^\perp(u)
\\ \nonumber &&-\frac{i f_\rho^{||}m_s m_\rho^3}{288(v\cdot q)^2}\langle\bar s s\rangle\phi_{2;\rho}^{||}(u)+\frac{i f_\rho^{||}m_s m_\rho^3}{144(v\cdot q)^2}\langle\bar s s\rangle\phi_{3;\rho}^\perp(u)-\frac{i f_\rho^{||} m_s m_\rho^3}{288(v\cdot q)^2}\psi_{4;\rho}^{||}(u)+\frac{i f_\rho^{||} m_s m_\rho}{144}\langle\bar s s\rangle\phi_{2;\rho}^{||}(u)
\\ \nonumber &&-\frac{i f_\rho^{||}m_s m_\rho}{36}\langle\bar s s\rangle\phi_{3;\rho}^\perp(u)-\frac{i f_\rho^\perp m_\rho^2}{72}\langle\bar s s\rangle\psi_{3;\rho}^{||}(u)-\frac{i f_\rho^\perp u(v\cdot q)^2}{36}\langle\bar s s\rangle\phi_{2;\rho}^\perp(u)-\frac{i f_\rho^\perp m_s m_\rho^2 u(v\cdot q)^2}{192\pi^2}\phi_{4;\rho}^\perp(u)
\\ \nonumber &&+\frac{f_\rho^\perp m_\rho^2 t}{576 v\cdot q}\langle g_s\bar s \sigma  G s\rangle\phi_{2;\rho}^\perp(u)-\frac{f_\rho^\perp m_\rho^2 t}{576 v\cdot q}\langle g_s\bar s \sigma  G s\rangle\psi_{4;\rho}^\perp(u)-\frac{f_\rho^{||} m_s m_\rho t v\cdot q}{288}\langle \bar s s\rangle\psi_{3;\rho}^\perp(u)+\frac{i f_\rho^{||} m_s m_\rho^3 t^2}{2304}\langle\bar s s\rangle\phi_{4;\rho}^{||}(u)
\\ \nonumber &&-\frac{i f_\rho^\perp m_\rho^2 t^2}{1152}\langle g_s\bar s \sigma  G s\rangle\psi_{3;\rho}^{||}(u)-\frac{i f_\rho^\perp u t^2 (v\cdot q)^2}{576}\langle g_s\bar s \sigma  G s\rangle\phi_{2;\rho}^\perp(u)-\frac{i f_\rho^\perp m_\rho^2 u t^2(v\cdot q)^2}{576}\langle\bar s s\rangle\phi_{4;\rho}^\perp(u)
\\ \nonumber &&+\frac{i f_\rho^{||}m_s m_\rho u t^2(v\cdot q)^2}{576}\langle\bar s s\rangle\psi_{3;\rho}^\perp(u)+\frac{f_\rho^\perp m_\rho^2 t^3 v\cdot q}{2304}\langle g_s\bar s \sigma  G s\rangle\phi_{4;\rho}^\perp(u)-\frac{i f_\rho^\perp m_\rho^2 u t^4(v\cdot q)^2}{9216}\langle g_s\bar s \sigma  G s\rangle\phi_{4;\rho}^\perp(u)\Big)
\\ \nonumber &&-\int_0^\infty dt \int_0^1 du \int \mathcal{D}\underline{\alpha}e^{i\omega^\prime t(\alpha_2+u\alpha_3)}e^{i\omega t(1-\alpha_2-u\alpha_3)}\times{1\over2}\times\Big(-\frac{i f_\rho^{||} m_\rho^3}{24 t^2}\Psi_{4;\rho}^{||}(\underline{\alpha})+\frac{i f_\rho^{||} m_\rho^3}{24\pi^2 t^2}\widetilde \Psi_{4;\rho}^{||}(\underline{\alpha})
\\ \nonumber &&-\frac{i f_\rho^{||}m_\rho^3 u}{12\pi^2 t^2}\Psi_{4;\rho}^{||}(\underline{\alpha})-\frac{i f_\rho^{||} m_\rho(v\cdot q)^2}{12\pi^2 t^2}\Phi_{3;\rho}^{||}(\underline{\alpha})-\frac{i f_\rho^{||} m_\rho(v\cdot q)^2}{12\pi^2 t^2}\widetilde \Phi_{3;\rho}^{||}(\underline{\alpha})+\frac{i f_\rho^{||}m_\rho u (v\cdot q)^2}{12\pi^2 t^2}\Phi_{3;\rho}^{||}(\underline{\alpha})
\\ \nonumber &&+\frac{i f_\rho^{||} m_\rho u (v\cdot q)^2}{12\pi^2 t^2}\widetilde \Phi_{3;\rho}^{||}(\underline{\alpha})+\frac{f_\rho^{||} m_\rho u (v\cdot q)^3}{12\pi^2 t}\Phi_{3;\rho}^{||}(\underline{\alpha})+\frac{f_\rho^{||} m_\rho u (v\cdot q)^3}{12\pi^2 t}\widetilde \Phi_{3;\rho}^{||}(\underline{\alpha})+\frac{f_\rho^{||} m_\rho u\alpha_2(v\cdot q)^3}{12\pi^2 t}\Phi_{3;\rho}^{||}(\underline{\alpha})
\\ \nonumber &&-\frac{f_\rho^{||} m_\rho u\alpha_2(v\cdot q)^3}{12\pi^2 t}\widetilde \Phi_{3;\rho}^{||}(\underline{\alpha})-\frac{f_\rho^{||} m_\rho u^2 \alpha_3(v\cdot q)^3}{12\pi^2 t}\Phi_{3;\rho}^{||}(\underline{\alpha})-\frac{f_\rho^{||} m_\rho u^2\alpha_3(v\cdot q)^3}{12\pi^2 t}\widetilde \Phi_{3;\rho}^{||}(\underline{\alpha})\Big) \, .
\end{eqnarray}
The sum rule for $\Xi_b^{\prime-}[{1\over2}^-]$ belonging to $[\mathbf{6}_F, 1, 0, \rho]$ is
\begin{eqnarray}
&& G_{\Xi_b^{\prime-}[{1\over2}^-] \rightarrow \Sigma_b^0 K^{*}} (\omega, \omega^\prime)
= { g_{\Xi_b^{\prime-}[{1\over2}^-] \rightarrow \Sigma_b^0 K^{*}} f_{\Xi_b^{\prime-}[{1\over2}^-]} f_{\Sigma_b^0} \over (\bar \Lambda_{\Xi_b^{\prime-}[{1\over2}^-]} - \omega^\prime) (\bar \Lambda_{\Sigma_b^0} - \omega)}
\\ \nonumber &=& \int_0^\infty dt \int_0^1 du e^{i(1-u)\omega^\prime t}e^{iu\omega t}\times4\times \Big (\frac{i f_{K^*}^\perp v\cdot q}{2\pi^2 t^4}\phi_{2;K^*}^\perp(u)-\frac{i f_{k^*}^\perp m_{K^*}^2}{2\pi^2 t^4 v\cdot q}\phi_{2;K^*}^\perp(u)
\\ \nonumber &&+\frac{i f_{K^*}^\perp m_{K^*}^2}{2\pi^2 t^4 v\cdot q}\psi_{4;K^*}^\perp(u)-\frac{f_{K^*}^\perp u(v\cdot q)^2}{4\pi^2 t^3}\phi_{2;K^*}^\perp(u)+\frac{if_{K^*}^\perp m_{K^*}^2 v\cdot q}{32\pi^2 t^2}\phi_{4;K^*}^\perp(u)-\frac{f_{K^*}^\perp m_{K^*}^2 u(v\cdot q)^2}{64\pi^2 t}\phi_{4;K^*}^\perp(u)
\\ \nonumber &&-\frac{i f_{K^*}^{||} m_{K^*} v\cdot q}{48}\langle\bar q q\rangle\psi_{3;K^*}^\perp(u)-\frac{f_{K^*}^{||} m_{K^*} u t(v\cdot q)^2}{96}\langle \bar q q\rangle\psi_{3;K^*}^\perp(u)-\frac{i f_{K^*}^{||}m_{K^*} t^2 v\cdot q}{768}\langle g_s\bar q \sigma  G q\rangle\psi_{3;K^*}^\perp(u)
\\ \nonumber &&-\frac{f_{K^*}^{||} m_{K^*} u t^3(v\cdot q)^2}{1536}\langle g_s\bar q\sigma  G q\rangle\psi_{3;K^*}^\perp(u)\Big)
\\ \nonumber &&-\int_0^\infty dt \int_0^1 du \int\mathcal{D}\underline{\alpha}e^{i\omega^\prime t(\alpha_2+u \alpha_3)}e^{i\omega t(1-\alpha_2-u \alpha_3)}\times{1\over2}\Big(-\frac{if_{K^*}^\perp v\cdot q}{8\pi^2 t^2}\Phi_{4;K^*}^{\perp(1)}(\underline{\alpha})+\frac{i f_{K^*}^\perp v\cdot q}{8\pi^2 t^2}\Phi_{4;K^*}^{\perp(2)}(\underline{\alpha})
\\ \nonumber &&-\frac{i f_{K^*}^\perp m_{K^*}^2 v\cdot q}{4\pi^2 t^2}\Phi_{4;K^*}^{\perp(3)}(\underline{\alpha})+\frac{i f_{K^*}^\perp m_{K^*}^2 v\cdot q}{4\pi^2 t^2}\Phi_{4;K^*}^{\perp(4)}(\underline{\alpha})+\frac{i f_{K^*}^\perp m_{K^*}^2 v\cdot q}{8\pi^2 t^2}\widetilde \Psi_{4;k^*}^\perp(\underline{\alpha})+\frac{i f_{K^*}m_{K^*}^2 v\cdot q}{8\pi^2 t^2}\Psi_{4;K^*}^\perp(\underline{\alpha})
\\ \nonumber &&+\frac{i f_{K^*}^\perp m_{K^*}^2 u v\cdot q}{8\pi^2 t^2}\Phi_{4;K^*}^{\perp(1)}(\underline{\alpha})-\frac{i f_{K^*}^\perp u v\cdot q}{8\pi^2 t^2}\Phi_{4;K^*}^{\perp(2)}(\underline{\alpha})-\frac{i f_{K^*} m_{K^*}^2 u v\cdot q}{4\pi^2 t^2}\widetilde \Psi_{4;K^*}^\perp(\underline{\alpha})+\frac{f_{K^*} m_{K^*}^2(v\cdot q)^2}{16\pi^2 t}\Phi_{4;K^*}^{\perp(1)}(\underline{\alpha})
\\ \nonumber &&-\frac{f_{K^*}^\perp m_{K^*}^2(v\cdot q)^2}{16\pi^2 t}\Phi_{4;K^*}^{\perp(2)}(\underline{\alpha})-\frac{f_{K^*}^\perp m_{K^*}^2(v\cdot q)^2}{16\pi^2 t}\Phi_{4;K^*}^{\perp(3)}(\underline{\alpha})+\frac{f_{K^*}^\perp m_{K^*}^2(v\cdot q)^2}{16\pi^2 t}\Phi_{4;K^*}^{\perp(4)}(\underline{\alpha})+\frac{f_{K^*}^\perp m_{K^*}^2(v\cdot q)^2}{16\pi^2 t}\widetilde \Psi_{4;K^*}^\perp(\underline{\alpha})
\\ \nonumber &&-\frac{f_{K^*}^\perp m_{K^*}^2(v\cdot q)^2}{16\pi^2 t}\Psi_{4;K^*}^\perp(\underline{\alpha})-\frac{f_{K^*}^\perp m_{K^*}^2(v\cdot q)^2}{8\pi^2 t}\Phi_{4;K^*}^{\perp(3)}(\underline{\alpha})+\frac{f_{K^*}^\perp m_{K^*}^2 u (v\cdot q)^2}{8\pi^2 t}\Phi_{4;K^*}^{\perp(4)}(\underline{\alpha})+\frac{f_{K^*}^\perp m_{K^*}^2 u(v\cdot q)^2}{8\pi^2 t}\widetilde \Psi_{4;K^*}^\perp(\underline{\alpha})
\\ \nonumber &&-\frac{f_{K^*}^\perp m_{K^*}^2 \alpha_2(v\cdot q)^2}{16\pi^2 t}\Phi_{4;K^*}^{\perp(1)}(\underline{\alpha})+\frac{f_{K^*}^\perp m_{K^*}^2 \alpha_2 (v\cdot q)^2}{16\pi^2 t}\Phi_{4;K^*}^{\perp(2)}(\underline{\alpha})+\frac{f_{K^*}^\perp m_{K^*}^2 \alpha_2(v\cdot q)^2}{16\pi^2 t}\Phi_{4;K^*}^{\perp(3)}(\underline{\alpha})
\\ \nonumber &&-\frac{f_{K^*}^\perp m_{K^*}^2\alpha_2(v\cdot q)^2}{16\pi^2 t}\Phi_{4;K^*}^{\perp(4)}(\underline{\alpha})-\frac{f_{K^*}^\perp m_{K^*}^2\alpha_2(v\cdot q)^2}{16\pi^2 t}\widetilde \Psi_{4;K^*}^\perp(\underline{\alpha})+\frac{f_{K^*}^\perp m_{K^*}^2 \alpha_2(v\cdot q)^2}{16\pi^2 t}\Psi_{4;K^*}^\perp(\underline{\alpha})
\\ \nonumber &&+\frac{f_{K^*}^\perp m_{K^*}^2 u\alpha_2(v\cdot q)^2}{8\pi^2 t}\Phi_{4;K^*}^{\perp(3)}(\underline{\alpha})-\frac{f_{K^*}^\perp m_{K^*}^2 u\alpha_2(v\cdot q)^2}{8\pi^2 t}\Phi_{4;K^*}^{\perp(4)}(\underline{\alpha})-\frac{f_{K^*}^\perp m_{K^*}^2 u\alpha_2(v\cdot q)^2}{8\pi^2 t}\widetilde \Psi_{4;K^*}^\perp(\underline{\alpha})
\\ \nonumber &&+\frac{f_{K^*}^\perp m_{K^*}^2 u^2\alpha_3(v\cdot q)^2}{8\pi^2 t}\Phi_{4;K^*}^{\perp(3)}(\underline{\alpha})-\frac{f_{K^*}^\perp m_{K^*}^2 u^2\alpha_3(v\cdot q)^2}{8\pi^2 t}\Phi_{4;K^*}^{\perp(4)}(\underline{\alpha})-\frac{f_{K^*}^\perp m_{K^*}^2 u^2\alpha_3(v\cdot q)^2}{8\pi^2}\widetilde \Psi_{4;K^*}^\perp(\underline{\alpha})
\\ \nonumber &&-\frac{f_{K^*}^\perp m_{K^*}^2 u\alpha_3(v\cdot q)^2}{16\pi^2 t}\Phi_{4;K^*}^{\perp(1)}(\underline{\alpha})+\frac{f_{K^*}^\perp m_{K^*}^2 u\alpha_3(v\cdot q)^2}{16\pi^2}\Phi_{4;K^*}^{\perp(2)}(\underline{\alpha})+\frac{f_{K^*}^\perp m_{K^*}^2 u\alpha_3(v\cdot q)^2}{16\pi^2 t}\Phi_{4;K^*}^{\perp(3)}(\underline{\alpha})
\\ \nonumber &&-\frac{f_{K^*}^\perp m_{K^*}^2 u\alpha_3(v\cdot q)^2}{16\pi^2 t}\Phi_{4;K^*}^{\perp(4)}-\frac{f_{K^*}^\perp m_{K^*}^2 u\alpha_3(v\cdot q)^2}{16\pi^2 t}\widetilde \Psi_{4;K^*}^\perp(\underline{\alpha})+\frac{f_{K^*}^\perp m_{K^*}^2 u\alpha_3(v\cdot q)^2}{16\pi^2 t}\Psi_{4;K^*}^\perp(\underline{\alpha})\Big) \, .
\end{eqnarray}
The sum rule for $\Sigma_b^-[{1\over2}^-]$ belonging to $[\mathbf{6}_F, 1, 1, \lambda]$ is
\begin{eqnarray}
&& G_{\Sigma_b^-[{1\over2}^-] \rightarrow \Sigma_b^0\rho^-} (\omega, \omega^\prime)
= { g_{\Sigma_b^-{1\over2}^-] \rightarrow \Sigma_b^0\rho^-} f_{\Sigma_b^-[{1\over2}^-]} f_{\Sigma_b^0} \over (\bar \Lambda_{\Sigma_b^-[{1\over2}^-]} - \omega^\prime) (\bar \Lambda_{\Sigma_b^0} - \omega)}
\\ \nonumber &=& \int_0^\infty dt \int_0^1 du e^{i (1-u) \omega^\prime t} e^{i u \omega t} \times 8 \times \Big (-\frac{f_\rho^{||} m_\rho}{2\pi^2 t^4}\phi_{2;\rho}^{||}(u)-\frac{f_\rho^{||} m_\rho}{2\pi^2 t^4}\phi_{3;\rho}^\perp(u)
\\ \nonumber &&+\frac{f_\rho^{||} m_\rho^3}{4\pi^2 t^4(v\cdot q)^2}\phi_{2;\rho}^{||}(u)-\frac{f_\rho^{||} m_\rho^3}{2\pi^2 t^4(v\cdot q)^2}\phi_{3;\rho}^\perp(u)+\frac{f_\rho^{||} m_\rho^3}{4\pi^2 t^2(v\cdot q)^2}\psi_{4;\rho}^{||}(u)-\frac{if_\rho^{||} m_\rho v\cdot q}{8\pi^2 t^3}\psi_{3;\rho}^\perp(u)
\\ \nonumber &&-\frac{f_\rho^{||} m_\rho u(v\cdot q)^2}{16\pi^2 t^2}\psi_{3;\rho}^\perp(u)-\frac{f_\rho^{||} m_\rho^3}{32\pi^2 t^2}\phi_{4;\rho}^{||}(u)+\frac{if_\rho^\perp m_\rho^2}{24 t v\cdot q}\langle\bar q q\rangle\phi_{2;\rho}^\perp(u)-\frac{if_\rho^\perp m_\rho^2}{24 t v\cdot q}\langle\bar q q\rangle\psi_{4;\rho}^\perp(u)
\\ \nonumber &&+\frac{f_\rho^\perp u(v\cdot q)^2}{24}\langle\bar q q\rangle\phi_{2;\rho}^\perp(u)+\frac{f_\rho^\perp m_\rho^2}{24}\langle\bar q q\rangle\psi_{3;\rho}^{||}(u)+\frac{if_\rho^\perp m_\rho^2 t}{384 v\cdot q}\langle g_s\bar q \sigma  G q\rangle\phi_{2;\rho}^\perp(u)-\frac{if_\rho^\perp m_\rho^2 t}{384 v\cdot q}\langle g_s\bar q \sigma  G q\rangle\psi_{4;\rho}^\perp(u)
\\ \nonumber &&+\frac{f_\rho^\perp u t^2(v\cdot q)^2}{384}\langle g_s\bar q \sigma  G q\rangle \phi_{2;\rho}^\perp(u)+\frac{f_\rho^\perp m_\rho^2 u t^2(v\cdot q)^2}{384}\langle\bar q q\rangle\phi_{4;\rho}^\perp(u)+\frac{f_\rho^\perp m_\rho^2 t^2}{384}\langle g_s\bar q \sigma  G q\rangle\psi_{3;\rho}^{||}(u)
\\ \nonumber &&+\frac{f_\rho^\perp m_\rho^2 u t^4 (v\cdot q)^2}{6144}\langle g_s\bar q \sigma  G q\rangle\phi_{4;\rho}^\perp(u)\Big)
\\ \nonumber &&-\int_0^\infty dt \int_0^1 du\int\mathcal{D}\underline{\alpha}e^{i\omega^\prime t(\alpha_2+u\alpha_3)}e^{i\omega t(1-\alpha_2-u\alpha_3)}\times\Big(\frac{f_\rho^{||} m_\rho^3}{8\pi^2 t^2}\Phi_{4;\rho}^{||}(\underline{\alpha})-\frac{f_\rho^{||} m_\rho^3}{8\pi^2 t^2}\widetilde\Phi_{4;\rho}^{||}(\underline{\alpha})
\\ \nonumber &&+\frac{f_\rho^{||} m_\rho^3 u}{4\pi^2 t^2}\Phi_{4;\rho}^{||}(\underline{\alpha})-\frac{f_\rho^{||} m_\rho(v\cdot q)^2}{8\pi^2 t^2}\Phi_{3;\rho}^{||}(\underline{\alpha})-\frac{f_\rho^{||} m_\rho(v\cdot q)^2}{8\pi^2 t^2}\widetilde\Phi_{3;\rho}^{||}(\underline{\alpha})
\\ \nonumber &&+\frac{3f_\rho^{||} m_\rho u(v\cdot q)^2}{8\pi^2 t^2}\Phi_{3;\rho}^{||}(\underline{\alpha})-\frac{f_\rho^{||} m_\rho u (v\cdot q)^2}{8\pi^2 t^2}\widetilde\Phi_{3;\rho}^{||}(\underline{\alpha})-\frac{i f_\rho^{||} m_\rho(v\cdot q)^3}{8\pi^2 t}\Phi_{3;\rho}^{||}(\underline{\alpha})
\\ \nonumber &&+\frac{if_\rho^{||} m_\rho\alpha_2(v\cdot q)^3}{8\pi^2 t}\Phi_{3;\rho}^{||}(\underline{\alpha})-\frac{if_\rho^{||}m_\rho\alpha_2(v\cdot q)^3}{8\pi^2 t}\widetilde\Phi-{3;\rho}^{||}(\underline{\alpha})+\frac{if_\rho^{||} m_\rho u\alpha_2(v\cdot q)^3}{8\pi^2 t}\Phi_{3;\rho}^{||}(\underline{\alpha})
\\ \nonumber &&-\frac{i f_\rho^{||} m_\rho u \alpha_2(v\cdot q)3}{8\pi^2 t}\widetilde\Phi_{3;\rho}^{||}(\underline{\alpha})+\frac{if_\rho^{||} m_\rho u^2\alpha_3(v\cdot q)^3}{8\pi^2 t}\Phi_{3;\rho}^{||}(\underline{\alpha})-\frac{i f_\rho^{||} m_\rho u^2\alpha_3(v\cdot q)^3}{8\pi^2 t}\widetilde\Phi_{3;\rho}^{||}(\underline{\alpha})
\\ \nonumber &&+\frac{if_\rho^{||} m_\rho u\alpha_3(v\cdot q)^3}{8\pi^2 t}\Phi_{3;\rho}^{||}(\underline{\alpha})-\frac{if_\rho^{||} m_\rho u\alpha_3(v\cdot q)^3}{8\pi^2 t}\widetilde\Phi_{3;\rho}^{||}(\underline{\alpha}) \, .
\end{eqnarray}
The sum rule for $\Omega_b^-[{1\over2}^-]$ belonging to $[\mathbf{6}_F, 2, 1, \lambda]$ is
\begin{eqnarray}
G_{\Omega_b^-[{1\over2}^-] \rightarrow \Xi_b^{0}K^*} (\omega, \omega^\prime)
= { g_{\Omega_b^-[{1\over2}^-] \rightarrow \Xi_b^0K^*} f_{\Omega_b^-[{1\over2}^-]} f_{\Xi_b^0} \over (\bar \Lambda_{\Omega_b^-[{1\over2}^-]} - \omega^\prime) (\bar \Lambda_{\Xi_b^0} - \omega)}
= 0 \, .
\end{eqnarray}
\end{widetext}

%

%

\end{document}